\documentclass[a4paper,11pt]{article}
\usepackage[english]{babel}
\usepackage{cite}
\usepackage{jheppub}

\usepackage{amsmath,amssymb,graphicx,textcomp,enumerate,alltt}

\newcommand{\Beta}{\mathrm{B}}
\newcommand{\mathd}{\mathrm{d}}
\newcommand{\nobracket}{}
\newcommand{\nocomma}{}
\newcommand{\noplus}{}
\newcommand{\tmem}[1]{{\em #1\/}}
\newcommand{\tmop}[1]{\ensuremath{\operatorname{#1}}}
\newcommand{\tmstrong}[1]{\textbf{#1}}
\newcommand{\tmtextbf}[1]{{\bfseries{#1}}}

\newcommand{\tmtexttt}[1]{{\ttfamily{#1}}}
\newcommand{\tmverbatim}[1]{{\ttfamily{#1}}}
\newenvironment{enumeratealphacap}{\begin{enumerate}[A.] }{\end{enumerate}}
\newenvironment{itemizedot}{\begin{itemize} }{\end{itemize}}

\title{On the Treatment of Resonances in Next-to-Leading Order Calculations
       Matched to a Parton Shower.}

\author[a]{Tom\'a\v{s} Je\v{z}o,} 
\author[b]{Paolo Nason}

\emailAdd{tomas.jezo@mib.infn.it}
\emailAdd{paolo.nason@mib.infn.it}

\affiliation[a]{
  Universit\`a di Milano-Bicocca and INFN, Sezione di Milano-Bicocca,
  Piazza della Scienza 3,\\ 20126 Milano, Italy
}

\affiliation[b]{
  INFN, Sezione di Milano-Bicocca, Piazza della Scienza 3, 20126 Milano, Italy
}

\abstract{
  In this work we present a new subtraction method for next-to-leading order
  calculations that is particularly convenient even when narrow resonances are
  present. The method is particularly suitable for the implementation of
  next-to-leading order
  calculations matched to parton shower generators. It allows at the same time
  for the inclusion of all finite width effects, including interferences, and
  for a consistent treatment of resonances in the shower approach, preserving the
  mass of resonances near their peak. We implement our method, in a fully
  general and automatic way, within the {\tt POWHEG BOX} framework, and illustrate
  it using as a test case the process of $p p \to \mu^+ \nu_\mu j_b j$, that is
  dominated by $t$-channel single top production.
}

\keywords{QCD, Hadronic Colliders, Monte Carlo simulations, NLO calculations}

\begin{document}
\maketitle
\flushbottom

\section{The problem}

At present, several methods exist for the computation of next-to-leading order
(NLO) corrections in the Standard Model. When strong and/or electromagnetic
interactions are present, these calculations must deal properly with collinear
and soft divergences, that must cancel when infrared insensitive (IR-safe)
observables are computed. The so-called subtraction methods are generally used
in order to deal with this problem. In essence, they work as follows. A
generic NLO cross section can be written symbolically as
\begin{equation}
  \mathd \sigma = \mathd \Phi_B (B (\Phi_B) + V (\Phi_B)) + \mathd \Phi_R R
  (\Phi_R), \label{eq:basicsigma}
\end{equation}
where $\Phi_B$ stands for the Born phase space and $\Phi_R$ is the real
emission phase space. $B (\Phi_B)$, $V(\Phi_B)$ and $R (\Phi_R)$
represent the Born, Virtual and Real cross section respectively.
The real emission
process corresponds to the Born process in association with an extra
parton.\footnote{For simplicity we discuss the QCD case. All what we do is
straightforwardly extended to the electrodynamics case.} \ For the purpose of
this example we assume that we do not have hadrons in the initial state, i.e.
we consider processes like $Z$ decays into hadrons. The expression in
eq.~(\ref{eq:basicsigma}), in order to make sense, must be evaluated with some
form of regularization for the soft and collinear singularities. Assuming that
we are using dimensional regularization, the phase space $\Phi_B$ and $\Phi_R$
are evaluated in $d = 4 - 2 \epsilon$ dimensions. The value of an
observable $\mathcal{O}$ is then given by
\begin{equation}
  \langle \mathcal{O} \rangle = \int \mathcal{O} \mathd \sigma = \int \mathd
  \Phi_B (B (\Phi_B) + V (\Phi_B))  \mathcal{O} (\Phi_B) + \int \mathd \Phi_R
  R (\Phi_R) \mathcal{O} (\Phi_R) . \label{eq:observableO}
\end{equation}
We can think of our observable $\mathcal{O}$ as the cross section in a given
histogram bin of some kinematic distribution. Again, in
eq.~(\ref{eq:observableO}) we assume that we have a $d$-dimensional definition for
our observable, with the appropriate 4-dimensional limit. If the observable is
IR-safe, soft and collinear singularities will cancel in
eq.~(\ref{eq:observableO}), yielding a finite result.

\subsection{Subtraction method}

In the subtraction method, one introduces a parametrization of the real phase
space of the form $\Phi_R = \Phi_R (\Phi_B, \Phi_{\tmop{rad}})$, with
\begin{equation}
  \mathd \Phi_R = \mathd \Phi_B \mathd \Phi_{\tmop{rad}},
\end{equation}
where $\Phi_{\tmop{rad}}$ has $d - 1$ dimensions, and parametrizes the
emission of the extra parton. The parametrization must have a smooth behaviour
in the soft and collinear limit. Thus, in the limit of soft
emission, the kinematics of all but the soft parton described by
$\Phi_R(\Phi_B, \Phi_{\tmop{rad}})$ must match the $\Phi_B$ kinematics. In the
collinear limit, the kinematics of the system obtained by replacing the two
collinear partons in $\Phi_R (\Phi_B, \Phi_{\tmop{rad}})$ with a single parton
with the appropriate flavour, having momentum equal to the sum of the momenta
of the collinear partons, must match the $\Phi_B$ kinematics. One also
introduces a simplified approximation to the real cross section, $R_s$, that
coincides with $R$ in the soft and collinear singular limits.
Eq.~(\ref{eq:observableO}) is rewritten as
\begin{eqnarray}
  \langle \mathcal{O} \rangle & = & \int \mathd \Phi_B \left[ B (\Phi_B) + V
  (\Phi_B) + \int \mathd \Phi_{\tmop{rad}} R_s (\Phi_B, \Phi_{\tmop{rad}})
  \right]  \mathcal{O} (\Phi_B) \\
  & + & \int \mathd \Phi_B \mathd \Phi_{\tmop{rad}} [R (\Phi_R (\Phi_B,
  \Phi_{\tmop{rad}})) \mathcal{O} (\Phi_R (\Phi_B, \Phi_{\tmop{rad}})) - R_s
  (\Phi_R (\Phi_B, \Phi_{\tmop{rad}})) \mathcal{O} (\Phi_B)] . \nonumber
  \label{eq:subtrmeth} 
\end{eqnarray}
eq.~(\ref{eq:subtrmeth}) is clearly identical to eq.~(\ref{eq:observableO}).
It has however the nice property that $1 / \epsilon$ divergences in the square
bracket of the first term on the right-hand side of the equation (arising in
the virtual term and in the $\mathd \Phi_{\tmop{rad}}$ integration of $R_s$)
cancel among each other. Furthermore, collinear and soft divergences cancel under
the integral sign in the square bracket of the second term. The second term
can thus be evaluated in 4 dimensions with numerical methods. The square
bracket in the first term can be evaluated analytically once and for all.
One usually defines
\begin{equation}
  V_{\tmop{sv}} (\Phi_B) = \lim_{\epsilon \rightarrow 0} \left[ V (\Phi_B) +
  \int \mathd \Phi_{\tmop{rad}} R_s (\Phi_B, \Phi_{\tmop{rad}}) \right],
\end{equation}
and the first term becomes the 4-dimensional expression
\begin{equation}
  \int \mathd \Phi_B [B (\Phi_B) + V_{\tmop{sv}} (\Phi_B)]  \mathcal{O}
  (\Phi_B),
\end{equation}
that can be computed numerically.

The development of the subtraction method started since the very early QCD
computations, already appearing in the bud in the calculation of the Drell-Yan
process of ref.~\cite{Altarelli:1979ub}. A more systematic use of it was
made in ref.~{\cite{Ellis:1980wv}}, in the context of $e^+ e^-$ annihilation
into hadrons. In ref.~{\cite{Kunszt:1989km}} the calculation of ref.~{\cite{Ellis:1980wv}}
was implemented as a parton level generator, such that
any given observable could be computed with it without any dedicated analytic
work, and was in fact used to compute a number of commonly used IR-safe
observables for QCD studies at LEP. Subsequently, the subtraction method
implemented in parton level generators was applied also for processes
initiated by hadrons~{\cite{Mele:1990bq}}, and it became common practice to
compute the $R_s$ term by using the collinear and the soft approximations in
$d$ dimensions (see for example {\cite{Mangano:1991jk}}).

More recently, fully general formulations of the subtraction method
have appeared.  The procedure of ref.~{\cite{Catani:1996vz}}, known
as the CS method, uses local subtraction terms for the $R_s$ cross
section. The formulation given in ref.~{\cite{Frixione:1995ms}}, known
as the FKS method, is instead based upon the more traditional phase
space parametrizations used in refs.~{\cite{Ellis:1980wv}}
and~{\cite{Mele:1990bq}}.

\subsection{The subtraction method and resonances}

When resonances are present, in the zero width limit, the cross section
factorizes into the product of production and decay terms. In these cases,
a standard subtraction method can be applied
independently to the production and decay processes. In fact this was done
in refs.~{\cite{Melnikov:2011qx}} and~{\cite{Campbell:2012uf}}
for top production and decay. Problems do arise,
however, if finite width effects are fully included, so that also interference
among radiation produced in production and decays, or among radiation
produced in the decay of different
resonances, is included.

\begin{figure}[tbh]
  \centering
  \includegraphics[width=6cm]{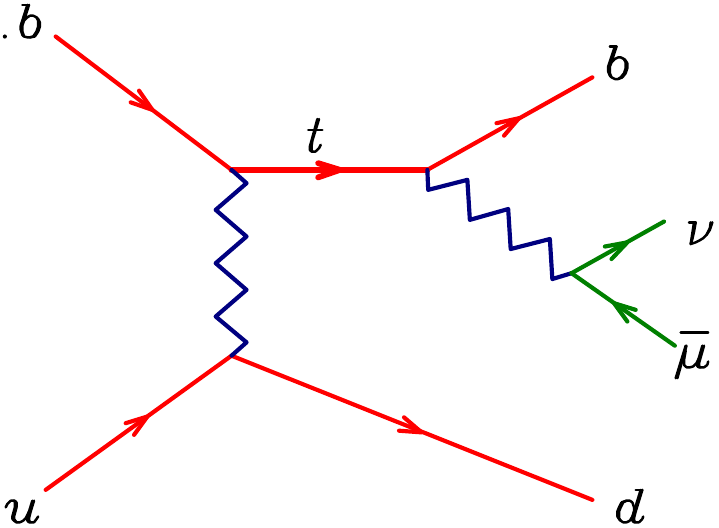}
  \caption{Single top $t$-channel production. \label{fig:ST}}
\end{figure}

On the one hand, the presence of a finite width regulates the singularity
associated with the resonance peak, so that, strictly speaking, a subtraction
method will formally lead to finite and consistent results. On the other hand,
taking the zero width limit, a standard subtraction method approach will lead
to divergent results. In order to illustrate this problem, we consider the
example of $t$-channel single top production and decay. One Born amplitude for
this process is illustrated in fig.~\ref{fig:ST}.
The final state is composed by a $b$ and $d$ quark, a muon
neutrino and an anti-muon. We assume for the sake of illustration a massless
$b$ quark. The system comprising the final state $b$ quark, the neutrino and
the anti-muon have an invariant mass close to the top mass, becoming identical
to the top mass in the zero width limit. Consider now the real contribution
obtained by adding gluon radiation to the final state. As illustrated
above, in generic subtraction methods, the subtraction counterterms are
obtained by factorizing the real phase space in terms of a Born phase space
times a radiation phase space. The subtraction term for the collinear
singularity corresponding to the final state gluon being collinear to the
final $b$ uses a Born phase space where the collinear $b g$ pair is merged
into a single $b$. The problem with the resonance is better illustrated in the
CS subtraction framework, where the kinematics of the subtraction term is
built as follows. Calling $k_{\oplus}$ the 4-momentum of the incoming $b$, and
$k_b$, $k_g$ the 4-momenta of the final $b$ and $g$ partons, one defines the
momentum of the $b$ quark in the underlying Born configuration as
\begin{equation}
  \bar{k}_b = k_b + k_g - k_{\oplus} \frac{(k_b + k_g)^2}{2 (k_b + k_g) \cdot
  k_{\oplus}},
\end{equation}
in such a way that $\bar{k}_b^2 = 0$. Furthermore, the incoming $b$ quark
momentum is redefined as
\begin{equation}
  \bar{k}_{\oplus} = k_{\oplus} - k_{\oplus} \frac{(k_b + k_g)^2}{2 (k_b +
  k_g) \cdot k_{\oplus}},
\end{equation}
so that the total 4-momentum is conserved. We see that in this way the
4-momentum of the top quark has been altered, near the collinear
limit, by an amount
\begin{equation}
  k_{\oplus} \frac{(k_b + k_g)^2}{2 (k_b + k_g) \cdot k_{\oplus}} \approx
  \frac{m_{b g}^2}{E_{b g}} .
\end{equation}
Since the CS procedure does not impose that the top 4-momentum
is the same in the real and subtraction terms\footnote{It instead imposes that the
incoming $b$ momentum minus the momenta of the final $b$ and of the radiated
gluon $g$ in the real term equals to the $b$ incoming momentum minus the final
$b$ momentum in the subtraction term, and all other momenta remain the same.}
it will turn out that the top virtualities will
differ there by an amount of order $m_{b g}^2 / E_{b g}$. The collinear
singularity in the real and subtraction terms will thus match only if
\begin{equation}
  m_{b g}^2 \ll \Gamma_t E_{b g}, \label{eq:collimgamma}
\end{equation}
where $\Gamma_t$ is the top width. It is easy to see that this is also true in
other subtraction methods. For example, in the one used in the
\tmverbatim{POWHEG BOX}, the momentum of the collinear counterterm is built by
setting the 3-momentum of the $b$ quark parallel to the sum of the 3-momenta of
the $b$ and $g$ particles in the partonic CM frame. Furthermore, the momentum
of the $d \bar{{\mu}} \nu$ system is boosted along the direction of the
merged $b$ quark in order to conserve 3-momentum, and the absolute value of
the $b$ quark momentum is chosen in such a way that the final state CM energy
is conserved. This procedure is designed to conserve the mass of the final
state system, and the mass of the system that recoils with respect to the
splitting partons, i.e. the $\bar{{\mu}} \nu d$ system, while the mass of
the top resonance is not conserved.

We thus expect that the collinear singularities present in the real and
subtraction terms will be exposed in the narrow width limit, spoiling the
convergence of the subtraction method. In fact, double logs of the resonance
width will arise in different regions of the real cross section, yielding to a
failure of convergence in the limit $\Gamma \rightarrow 0$. It is also clear
that, in order to overcome this problem, one must devise a subtraction method
such that the resonance mass is the same in the real and subtraction terms
when approaching the resonance peak even when the resonance is off-shell by
an amount greater than its width.

\subsection{NLO+PS and resonances}

If we plan to use an NLO calculation with an interface to a shower
generator (NLO+PS from now on), further problems arise due to the
resonance treatment.

In the \tmverbatim{MC@NLO} method~\cite{Frixione:2002ik}, one should
consider the recoil scheme used by the Shower Monte Carlo to build
radiation from a decaying resonance and construct the MC counterterms
accordingly.

In the \tmverbatim{POWHEG} method~\cite{Nason:2004rx,Frixione:2007vw,Alioli:2010xd},
one first computes the inclusive cross
section for the production of an event with a given underlying Born
configuration. Radiation is then generated according to a Sudakov form factor
with the following form:
\begin{equation}
  \Delta (p_T^2) = \exp \left[ - \int \frac{R (\Phi_B, \Phi_{\tmop{rad}})}{B
  (\Phi_B)} \theta (k_T (\Phi_{\tmop{rad}}) - p_T) \mathd \Phi_{\tmop{rad}}
  \right] .
\end{equation}
The mapping of the real phase space into a product of an underlying Born
times a radiation phase space is the same used in the NLO subtraction
procedure. In general, it will not preserve resonance masses, so that in the
$R / B$ ratio, unless the condition~(\ref{eq:collimgamma}) is met, the
numerator and the denominator will not be on the resonance peak at the same
time. In case when $R$ is on peak and $B$ is not, this will yield large ratios
that badly violate the collinear approximation.

A further problem arises when interfacing the NLO+PS calculation to a Shower
generator, in order to generate the next-to-hardest radiation. Shower Monte
Carlo's should be instructed to preserve the mass of the resonances.
Thus, radiation should have a resonance assignment. This is generally
not available in processes that include interference among radiation
generated by different resonances, or by a resonance and the production
process itself.

\section{The method}
In order to solve the problems mentioned above, we should separate all
contributions to the cross section into terms with definite resonance
structure. Each term individually should have resonance peaks only in a
single, well defined, resonance cascade chain. The mapping into an underlying
Born configuration should be defined for these terms in such a way that the
resonance masses are preserved. Thus, when looking for parton pairs that can
give rise to a collinear singularity, one should only consider pairs arising
from the same resonance decay, or directly from the production process.

In the \tmverbatim{POWHEG BOX} framework, a subtraction method that preserves
the resonance masses is already implemented, but it is presently available
only for calculations performed in the zero width approximation. In these
cases, only one resonance decay chain is possible, and the real emission
contributions are separated according to the resonance that originates the
radiation. The method is discussed in detail in ref.~{\cite{Campbell:2014kua}}.
In essence, with this method, the subtraction
procedure for initial state radiation is the same one used in
ref.~{\cite{Frixione:2007vw}} (the FNO paper from now on).
For final state radiation arising from the production process
the subtraction procedure is also the same one discussed in Section 5.2 of FNO.
This procedure is such that the
mapping of the real to the underlying Born configuration does not change the
four momentum of the final state. In case of radiation from the decay of a
resonance, the subtraction procedure is essentially the same, except that it
is applied in the resonance frame, and thus does not alter the resonance four
momentum and the momenta of all particles that do not have the resonance as an
ancestor.

In the general context when finite width effects are to be included, more than
one resonance cascade chain (from now on ``resonance history'') may be present,
and interference between amplitudes with
different resonance histories must also be included. We thus need to perform a
separation of the cross section into a sum of contributions, each one of them
dominating only for a single resonance history. For each of these contributions
we should apply the resonance aware subtraction method of ref.~{\cite{Campbell:2014kua}}.

In the following we will describe in great detail the procedure adopted for
the separation of the cross section contributions into terms with a definite
resonance structure. We will discuss the procedure for the terms that have the
Born kinematics (i.e. the Born, Virtual and Collinear Remnant terms), and for
real terms. For the latter, subtraction terms having the Born kinematics are
also present. We will require that in
the collinear and soft limits the separation of the
contributions associated with given resonance histories in the real term
smoothly matches the corresponding separation in the Born kinematics.

\subsection{The Born resonance histories}

We need to single out contributions from the Born term corresponding to
several different resonance histories of the final state. Each resonance
history corresponds to a \tmtextbf{tree graph}, where the \tmtextbf{leaves} of
the tree are the final state particle, and the \tmtextbf{intermediate nodes}
are the resonances. In our case, we include in the tree also the two initial
state particles, before the root node.

\begin{center}
\includegraphics{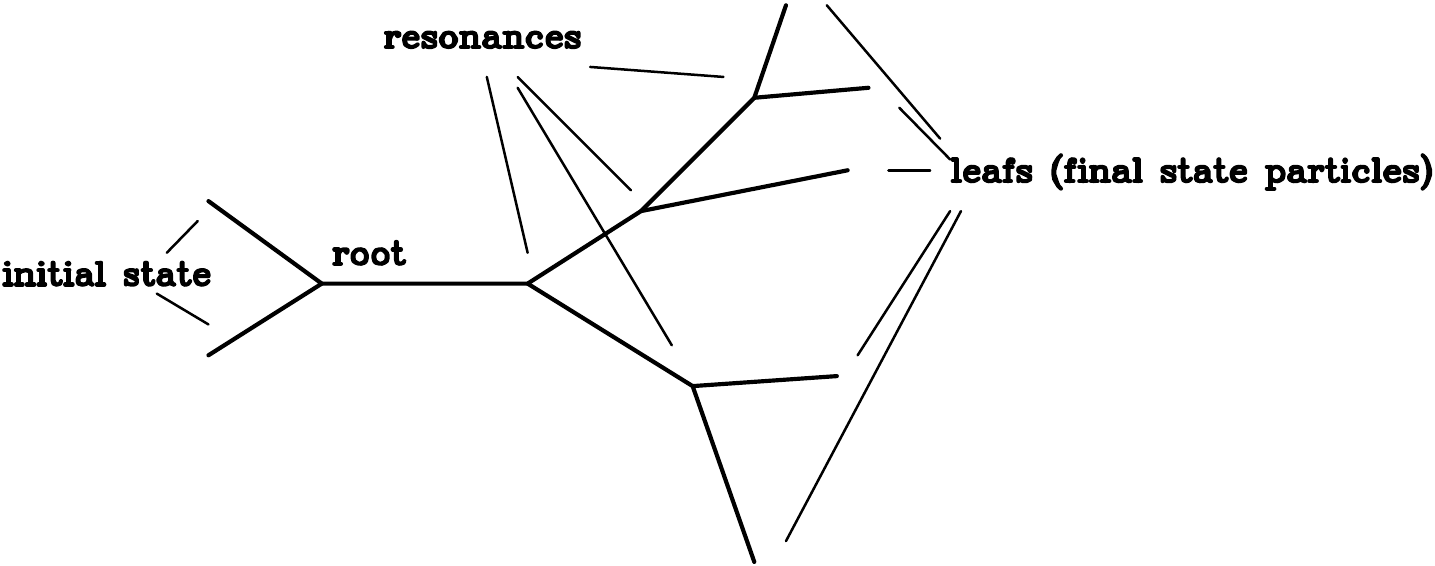}
\end{center}

The root of the tree does not correspond necessarily to any real resonance.
For uniformity of treatment, we will however associate to the root a
fictitious resonance, and we will refer to it as the ``production
resonance''.

For each given initial and final flavour configurations, we have several
possible resonance histories. We will denote with $F_b$ the initial and final
flavour structure of the Born process, irrespective of the internal nodes of the resonance
history. We will instead denote with $f_b$ the
flavour structure including the resonances decay
cascade. We will also refer to it as the \tmtextbf{resonance history}. Summarizing, we
will refer to $F_b$ as the \tmtextbf{bare flavour structure} of the process,
and to $f_b$ as the \tmtextbf{full} {\tmstrong{flavour}}
{\tmstrong{structure}}, or simply as the {\tmstrong{flavour structure}}.

The Born contributions will be labeled as $B_{F_b}$. Thus, $B_{F_b}$ is the
square of the amplitude for the production of the final state $F_b$,
including all possible resonance histories allowed for the process. We
separate the Born contribution in the following way:
\begin{equation}
  B_{F_b} = \sum_{f_b \in T (F_b)} B_{f_b}, \hspace{2em} B_{f_b} = \Pi_{f_b}
  B_{F_b},
\end{equation}
where $T (F)$ is the set of all trees having the same bare flavour structure
$F$. The factors $\Pi_{f_b}$ have the property
\begin{equation}
  \sum_{f_b \in T (F_b)} \Pi_{f_b} = 1 . \label{eq:pifbsum}
\end{equation}
Furthermore, they must be such that $\Pi_{f_b} B_{F_b}$ must have resonance
peaks compatible with the resonance history of $f_b$. One possible definition
for the $\Pi_{f_b}$ is the following. With each resonance $i$ in the resonance
history, we associate the factor
\begin{equation}
  \frac{M_i^4}{(s_i - M_i^2)^2 + \Gamma_i^2 M_i^2},
\end{equation}
and define
\begin{equation}
  P^{f_b} = \prod_{i \in \tmop{Nd} (f_b) }  \frac{M_i^4}{(s_i - M_i^2)^2 +
  \Gamma_i^2 M_i^2},
\end{equation}
where $s_i$, $M_i$ and $\Gamma_i$ are respectively the invariant mass of the decay product system,
the mass of the resonance and its width. By
$\tmop{Nd} (f_b)$ we denote the set of all nodes of the resonance tree for
$f_b$ (excluding the root). We then define
\begin{equation}
  \Pi_{f_b} = \frac{P^{f_b}}{\sum_{f_b' \in T (F_b (f_b))} P^{f_b'}},
\end{equation}
where we have introduced the notation $F_b (f_b)$ to denote the bare flavour
structure associated to a given full flavour structure $f_b$. This definition
clearly satisfies the property (\ref{eq:pifbsum}). Thus $B_{f_b}$
exhibits resonance peaks only in correspondence with resonances in its own
resonance history. In fact the $P^{f'_b}$ factors for all alternative
resonance histories in the denominator of $\Pi_{f_b}$
cancel the resonance peaks due to alternative
resonance histories in $B_{F_b}$. Only the peaks compatible with the $f_b$
resonance structure, that have a corresponding enhancement factor in the
numerator, will remain.

It is worth pointing out that our definition of the $\Pi$ factor is certainly
not unique. In particular, there is an alternative possibility
that is easily implemented if one
has access to the individual sub-amplitudes contributing to the total amplitude
characterized by $F_b$:
\begin{equation}
  B_{F_b} = \Bigl|^{} \sum_i \mathcal{A}_i  \Bigr|^2 .
\end{equation}
The structure of each sub-amplitude represents in this case a resonance
history, so that we can create a correspondence $i \leftrightarrow f_b$, and
define
\begin{equation}
  P^{f_b} = | \mathcal{A}_{f_b} |^2 . 
\end{equation}
This possibility may prove convenient with current numerical matrix elements
programs, where the numerical calculation of the individual amplitude is a
necessary step for the computation of the full matrix element. Since this
procedure is gauge dependent, care should be taken in the choice of an
appropriate gauge.

\subsubsection{Implementation of the Born resonance histories in the
{\tmstrong{\tmverbatim{POWHEG BOX}}}}

The internal implementation of the Born flavour structure can be inherited
from the present Born level structure in the \tmverbatim{POWHEG-BOX-V2},
starting with the extension of ref.~{\cite{Campbell:2014kua}} for the
inclusion of narrow width resonances. In this implementation, the full flavour
structure of a Born term is represented by two arrays,
\tmverbatim{flst\_born(j,iborn)} and \tmverbatim{flst\_bornres(j,iborn)},
where the index \tmverbatim{iborn} labels the particular Born full flavour
structure $f_b$. The \tmverbatim{j} index labels the external leg and the
internal resonances, with 1 and 2 representing the incoming legs, and the
(integer) value of the \tmverbatim{flst\_born} array represents the
corresponding flavour code (that coincides with the PDG code, except for
gluons, that are labeled 0). The \tmverbatim{flst\_bornres(j,iborn)} integer
array represents the resonance pointers, so that the whole resonance structure
can be reconstructed. For example, for the case of the full flavour structure
corresponding to the process $g g \rightarrow (t \rightarrow (W^+ \rightarrow
e^+ \nu_e) b) ( \bar{t} \rightarrow (W^- \rightarrow {\mu}^-
\bar{\nu}_{{\mu}})  \bar{b} )$, we have

{\noindent}\tmverbatim{flst\_born(1:12,iborn) \ \ \ = [ 0, 0, 6, -6, 24, -24,
-11, 12, 13, -14, 5, -5]}

{\noindent}\tmverbatim{flst\_bornres(1:12,iborn) = [ 0, 0, 0, \ 0, \ 3, \ \ 4,
\ \ 5, \ 5, \ 6, \ \ 6, 3, \ 4]}.

{\noindent}We see that the resonance pointer list contains zero for particles
generated at the production stage (in \verb!POWHEG! we represent the fictitious
production stage resonance as having index 0), while for particles produced in
resonance decays the corresponding entry is the position of mother
resonance in the list.

At variance with the V2 implementation, in the present case we must be
prepared to assume that not all Born flavour configurations have the same
resonance history and the same number of resonances, so we must admit Born
flavour lists of different length. We thus introduce an array
\tmtexttt{flst\_bornlength(iborn)}, carrying the length of the flavour list
for the Born $f_b$ labeled by the \tmtexttt{iborn} index. The entries of this
array are set in the user process \tmtexttt{init\_processes} subroutine.

The \verb!POWHEG BOX!
integration program  (the \tmtexttt{mint} integrator
{\cite{Nason:2007vt}}) was updated in order to deal with the
resonance histories. Since several resonance histories may be present,
the \tmtexttt{mint} integrator was also updated to be able to deal with
a discrete (summation) variable. It now computes a multidimensional integral in
a unit hypercube and the summation over a discrete index. The discrete index
is used to label each resonance history. The phase space generator examines
the value of this discrete index, identifies the corresponding resonance
history, and chooses automatically a phase space parametrization that
performs importance sampling over the resonance regions, generating the
resonance virtualities with an appropriate Breit-Wigner distribution.

\subsection{The real resonance histories and singular regions}

In the case of real graphs, we have more resonance histories, because we have
one more final state particle that can belong to resonances. In analogy with
the Born case, we introduce $F_r$ and $f_r$ as before, labeling the bare and
full flavour structure for a real graph. We will now introduce a label
$\alpha_r$, that labels a singular region\footnote{We assume throughout that
the reader is familiar with the notation introduced of the FNO
paper~\cite{Frixione:2007vw}. A singular region corresponds to a configuration
where two final state particle become collinear, or a final state particle
becomes collinear to an initial particle.} compatible (in a sense that
we will specify in the following) with a given $f_r$
\begin{equation}
  \alpha_r \in \tmop{Sr} (f_r) .
\end{equation}
Also here we will use the notation $f_r (\alpha_r)$ and $F_r (\alpha_r)$ to
denote the full and bare flavour structure associated with a given singular
region.

We only consider singular regions that are compatible with the given
resonance history in the following sense: {\em the particles that become
collinear should be siblings, i.e. should arise directly from the decay of the same
resonance or from the root (if they are directly produced in the hard
reaction)}.

We now perform the separation of the real cross section with a
given bare flavour structure into singular region contributions:
\begin{equation}
  R_{\alpha_r} = \frac{P^{f_r (\alpha_r)} d^{- 1} (\alpha_r)}{\sum_{f_r'
  \nocomma \in T (F_r (\alpha_r))} P^{f'_r}  \sum_{\alpha'_r \in \tmop{Sr}
  (f'_r)} d^{- 1} (\alpha_r')} R_{F_r (\alpha_r)}, \label{eq:realpartition}
\end{equation}
where $F_r (\alpha_r)$ stands for the bare flavour structure associated with
$\alpha_r$.
We require that the real weights $P^{f_r (\alpha_r)}$ are
compatible with the Born weights, in the sense that, in the soft or collinear
limit, the $P^{f_r (\alpha_r)}$ must approach smoothly a $P^{f_b}$ factor of
the corresponding underlying Born. This is certainly the case if they are
defined as in the Born case.

We notice that in the standard \verb!POWHEG! scheme, the real contribution to a given
region is enhanced if the collinear pair has a smaller transverse momentum than
all other possible collinear pairings. In the present scheme, the relative
transverse momentum of the pair is no longer the only element that decides
about the partition of singular regions.
As an example, consider three final state partons $i$, $j$ and $k$.
The cross section is parted among the $i,j$ and $i,k$ singular regions,
depending upon how small are the relative transverse momenta in the two cases,
and how far from the resonance peaks are the resonances containing respectively
the $i,j$ and $i,k$ partons.

The $d^{- 1}$ factors used in the \tmverbatim{POWHEG BOX} have the form
\begin{eqnarray}  \label{eq:didef}
  d_i & = & [E_i^2 (1 - \cos^2 \theta_i)]^b, \\  \label{eq:dipmdef}
  d_i^{\pm} & = & [E_i^2 2 (1 \pm \cos \theta_i)]^b, \\ \label{eq:dijdef}
  d_{i j} & = & \left[ \frac{E_i^2 E^2_j}{(E_i + E_j)^2}  (1 - \cos \theta_{i
  j}) \right]^b, 
\end{eqnarray}
where $b$ is a positive constant parameter.
Eq.~(\ref{eq:didef}) is used for the collinear region characterized by parton $i$,
with energy $E_i$ and angle $\theta_i$ (relative to the beam)
in the partonic CM, becoming collinear to either incoming hadrons.
Eq.~(\ref{eq:dipmdef}) is again for initial
state collinear regions, but distinguishes among the two collinear directions.
Eq.~(\ref{eq:dijdef}) is used for the region characterised by final state partons
$i$ and $j$ becoming collinear. They are
commonly evaluated in the partonic rest frame. In the present case, however,
in case of final state singularities associated with the decay products of a
resonance, it seems more natural to compute them in the resonance rest frame.
They thus become dependent upon the full flavour structure $f_r$ of the real
contribution. It is however important for the following developments that the
$d_{i j}$ factors do not depend upon the resonance structure in the collinear
limit. This is in fact the case with our definition, since
\[ \lim_{i j} d_{i j} = \lim_{\tmop{coll}} \left[ \frac{E_i E_j}{(E_i +
   E_j)^2} k_i \cdot k_j \right]^b \Longrightarrow [z (1 - z) k_i \cdot k_j]^b
\]
where $\lim_{i j}$ denotes the limit for particles $i$ and $j$ becoming
collinear and $z$ is the energy fraction. The last expression is obviously
Lorentz invariant in the
collinear limit. Thanks to this property, it will turn out that the sum of all
$R_{\alpha_r}$ associated with the same underlying Born full flavour structure
factorizes in the collinear limit\footnote{Notice that with the notation $f_b
(\alpha_r) = f_b$ means: all $\alpha_r$ that leads to a full underlying Born
flavour equal to $f_b$. We thus use $f_b$ both as a function name and as a
variable, since in the present context this cannot generate confusion.}
\begin{equation}
  \lim_{i j} \sum_{\alpha_r \nocomma \nocomma}^{f_b (\alpha_r) = f_b}
  R_{\alpha_r} \propto B_{f_b} \times P_{i j} (z) .
\end{equation}
This follows from the fact that all (and only) the $\alpha_r$ associated with
particles $i, j$ becoming collinear dominate in this limit, and, being all
equal, they simplify out in the numerator and denominator of eq.~(\ref{eq:realpartition}).
We emphasize, however, that the $d_{ij}$ terms are not frame independent in the
soft limit. This is quite clear from eq.~(\ref{eq:dijdef}), that in the $E_i\to 0$
limit becomes
\begin{equation} \label{eq:dijsoftlim}
d_{ij}\approx \left[\frac{E_i}{E_j} k_i\cdot k_j \right]^b,
\end{equation}
that is clearly frame dependent.

As in the Born case, the scheme discussed here is not the only alternative for
the partition of the singular regions and of the resonance structure. Using
weights equal to the square of individual sub-amplitudes is still a valid
alternative, as long as one computes the amplitudes in a physical gauge, in
such a way that squared amplitudes also retain the full collinear singularity
structure. In this case one does not need to introduce the $d_{i j}$ factors,
since the squared amplitudes already have the appropriate singular behaviour
in the collinear limit. In order to further pursue this alternative, issues
related to the lack of gauge invariance of the individual amplitudes squared
should be addressed. In the present work we did not investigate this
alternative any further, since we prefer to assume that in general the
individual amplitude for the process may not be available.

\subsection{Example: electroweak $u \bar{u} \rightarrow u \bar{d} \bar{u} d$}
\begin{figure}[h]
  \centering
  \includegraphics{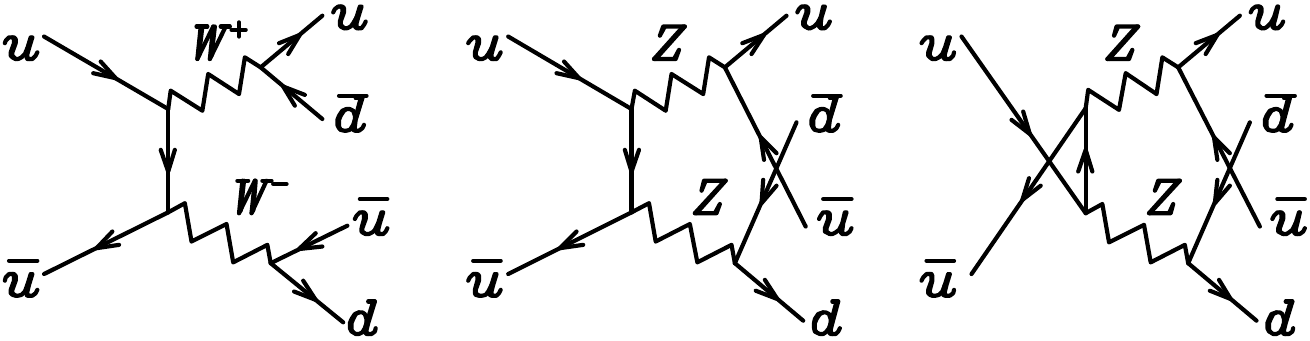}
  \caption{\label{fig:WWfeynman}
  Feynman diagrams for $u \bar{u} \rightarrow u \bar{d} \bar{u} d$.}
\end{figure}

\begin{figure}[h]
  \centering
  \includegraphics{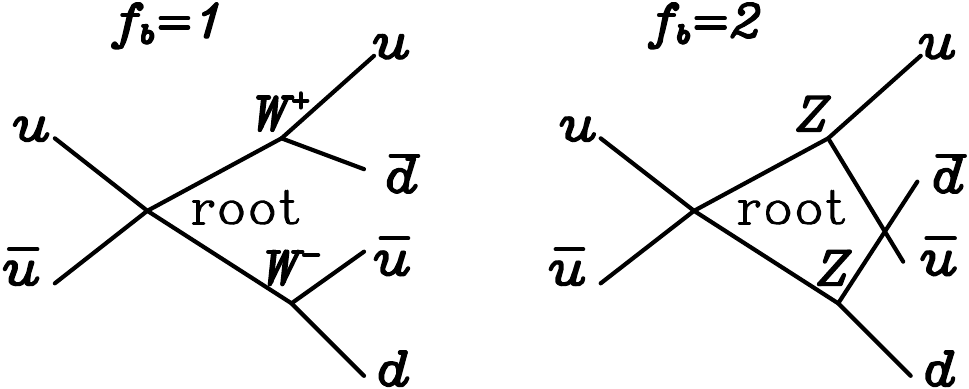}
  \caption{Trees for $u \bar{u} \rightarrow u \bar{d} \bar{u} d$.
  \label{fig:WWborntrees}}
\end{figure}

We illustrate the separation of the resonance structures in the process $u \bar{u}
\rightarrow u \bar{d} \bar{u} d$, considering only electroweak interactions.
In order to simplify the discussion, we will (wrongfully!) assume that only
the diagrams illustrated in fig.~\ref{fig:WWfeynman} contribute to
it. We remark that this process is chosen only for illustration purposes. We
are aware of the fact that it has no physical relevance and that we are
omitting other relevant resonance histories.
There is only one $F_b$, corresponding to the bare flavour structure $u
\bar{u} \rightarrow u \bar{d} \bar{u} d$. We have two $f_b$, represented in
fig.~\ref{fig:WWborntrees}, corresponding respectively to $u \bar{u}
\rightarrow (W^+ \rightarrow u \bar{d}) (W^- \rightarrow \bar{u} d)$ and $u
\bar{u} \rightarrow (Z \rightarrow u \bar{u}) (Z \rightarrow d \bar{d})$.

The $P$ factors for the two configurations are
\begin{eqnarray*}
  P^1_b & = & \frac{M_W^4}{(s_{34} - M_W^2)^2 + \Gamma_W^2 M_W^2} \times
  \frac{M_W^4}{(s_{56} - M_W^2)^2 + \Gamma_W^2 M_W^2}\,,\\
  P^2_b & = & \frac{M_Z^4}{(s_{35} - M_Z^2)^2 + \Gamma_Z^2 M_Z^2} \times
  \frac{M_Z^4}{(s_{46} - M_Z^2)^2 + \Gamma_Z^2 M_Z^2}\, .
\end{eqnarray*}
Notice that we have assigned the values 1 and 2 to the $f_b$ index of the two
flavour configurations depicted in the figure. Particles are labeled by an
integer, starting from the lower incoming line, and going through all other
particles clockwise. Summarizing, we have two (full) flavour structures for the given
bare flavour structure $u \bar{u} \rightarrow u \bar{d} \bar{u} d$. The
corresponding Born contributions will be given by
\[ B_1 = \frac{P^1_b B}{D_b}, \hspace{2em} B_2 = \frac{P^2_b B}{D_b}, \]
with
\[ D_b = P_b^1 + P_b^2 . \]
Notice that $B$ is the full Born contribution, given by the square of the sum
of the graphs in fig.~\ref{fig:WWfeynman}. However, $B_1$ will be dominated by the square of the
first graph, and $B_2$ by the second.

\begin{figure}[h]
  \centering
  \includegraphics{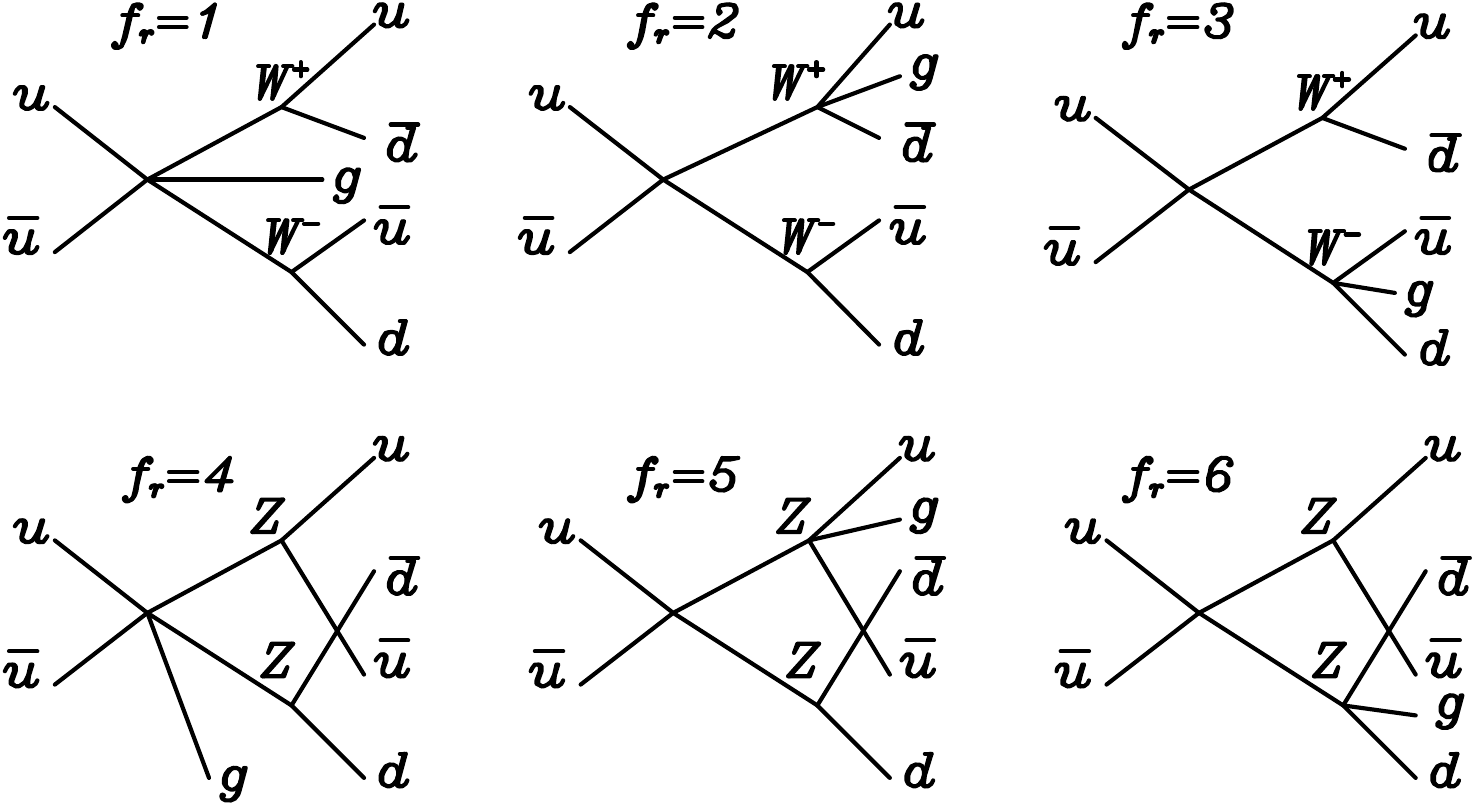}
  \caption{Trees for $u \bar{u} \rightarrow u \bar{d} \bar{u} d g$.\label{fig:WWrealtrees}}
\end{figure}

The number of real graphs is already quite large, and we do not show the
corresponding figures. They are obtained by adding one final state gluon to
the Born flavour configuration, and by replacing one of the initial lines with
a gluon, adding a corresponding quark of opposite flavour to the final
state. Here we focus upon the bare flavour configuration $u \bar{u}
\rightarrow u \bar{d} \bar{u} d g$. The corresponding full flavour
configuration trees are depicted in fig.~\ref{fig:WWrealtrees}.

We will now label the gluon as $7$, and keep the same labels used in the Born
case for all other particles. The $P$ factors are now
\begin{eqnarray*}
  P^1_r & = & \frac{M_W^4}{(s_{34} - M_W^2)^2 + \Gamma_W^2 M_W^2} \times
  \frac{M_W^4}{(s_{56} - M_W^2)^2 + \Gamma_W^2 M_W^2}\,,\\
  P^2_r & = & \frac{M_W^4}{(s_{347} - M_W^2)^2 + \Gamma_W^2 M_W^2} \times
  \frac{M_W^4}{(s_{56} - M_W^2)^2 + \Gamma_W^2 M_W^2}\,,\\
  P^3_r & = & \frac{M_W^4}{(s_{34} - M_W^2)^2 + \Gamma_W^2 M_W^2} \times
  \frac{M_W^4}{(s_{567} - M_W^2)^2 + \Gamma_W^2 M_W^2}\,,\\
  P^4_r & = & \frac{M_Z^4}{(s_{35} - M_Z^2)^2 + \Gamma_Z^2 M_Z^2} \times
  \frac{M_Z^4}{(s_{46} - M_Z^2)^2 + \Gamma_Z^2 M_Z^2} \,,\\
  P^5_r & = & \frac{M_Z^4}{(s_{357} - M_Z^2)^2 + \Gamma_Z^2 M_Z^2} \times
  \frac{M_Z^4}{(s_{46} - M_Z^2)^2 + \Gamma_Z^2 M_Z^2}\,,\\
  P^6_r & = & \frac{M_Z^4}{(s_{35} - M_Z^2)^2 + \Gamma_Z^2 M_Z^2} \times
  \frac{M_Z^4}{(s_{467}^{} - M_Z^2)^2 + \Gamma_Z^2 M_Z^2}\,.
\end{eqnarray*}
The singular regions $\alpha_r$ are displayed in tab.~\ref{tab:alr-regions-example}.
\begin{table}[tbh]
  \centering
  \begin{tabular}{|c|c|c|c|}
    \hline
    $\alpha_r$ & $f_r$ & emitter & $d^{- 1} (\alpha_r)$\\
    \hline
    1 & 1 & 0 & $d^{- 1}_7$\\
    \hline
    2 & 2 & 3 & $d^{- 1}_{37,2}$\\
    \hline
    3 & 2 & 4 & $d^{- 1}_{47,2}$\\
    \hline
    4 & 3 & 5 & $d^{- 1}_{57,3}$\\
    \hline
    5 & 3 & 6 & $d^{- 1}_{67,3}$\\
    \hline
    6 & 4 & 0 & $d^{- 1}_7$\\
    \hline
    7 & 5 & 3 & $d^{- 1}_{37,5}$\\
    \hline
    8 & 5 & 4 & $d^{- 1}_{57,5}$\\
    \hline
    9 & 6 & 5 & $d^{- 1}_{47,6}$\\
    \hline
    10 & 6 & 6 & $d^{- 1}_{67,6}$\\
    \hline
  \end{tabular}
  \caption{\label{tab:alr-regions-example}}
\end{table} 
Notice that the final state radiation $d$ factors carry in the subscript the
position of the two partons that become collinear, and, after a comma, an index
specifying the resonance history. We are in fact assuming that the $d$
factors are computed in the frame of the resonance that owns the two
collinear partons. Notice also that the standard (non resonance aware) \verb!POWHEG!
implementation would have found 5 regions, one for the initial state
radiation, and 4 for final state radiation, corresponding to a gluon being
emitted by each final state parton.

It is interesting to see how the singular part of the cross section is shared
among the various resonance histories. We consider as an example the gluon
emission from particle 3, carrying the $d^{- 1}_{37}$ singularity. In the
standard \verb!POWHEG! formulation this region corresponds to a single $\alpha_r$. On
the other hand, in our resonance aware extension, that singularity is shared
by the $\alpha_r$ number 2 and 7. The one of the two that is more enhanced by
resonant propagators (i.e. by its $P$ factor) will dominate over the other. We
have
\begin{eqnarray}
  D_r & = & P^1_r d^{- 1}_7 + P^2_r (d^{- 1}_{37 ,2} + d^{- 1}_{47 ,2}) +
  P^3_r (d^{- 1}_{57 ,3} + d^{- 1}_{67 ,3}) \nonumber\\
  & + & P^4_r d^{- 1}_7 + P^5_r (d^{- 1}_{37 ,5} + d^{- 1}_{57 ,5}) + P^6_r
  (d^{- 1}_{47 ,6} + d^{- 1}_{67 ,6}), \\
  R_2 & = & \frac{P^2_r d^{- 1}_{37 ,2} }{D_r} R, \\
  R_7 & = & \frac{P^5_r d^{- 1}_{37 ,5} }{D_r} R. 
\end{eqnarray}
Notice that near the 3,7 collinear singularity, we have
\begin{eqnarray}
  R_2 & = & \frac{P^2_r d^{- 1}_{37 ,2} }{D_r} R \; \cong \; \frac{P^2_r d^{-
  1}_{37 ,2}}{P^2_r d^{- 1}_{37 ,2} + P^5_r d^{- 1}_{37 ,5}} \, R \; \cong
  \; \frac{P^2_r}{P^2_r + P^5_r} R, \\
  R_7 & = & \frac{P^5_r d^{- 1}_{37 ,5} }{D_r} R \; \cong \; \frac{P^5_r d^{-
  1}_{37 ,5}}{P^2_r d^{- 1}_{37 ,2} + P^5_r d^{- 1}_{37 ,5}} R \; \cong \;
  \frac{P^5_r}{P^2_r + P^5_r} R, 
\end{eqnarray}
where the last equality follows from our requirement that the $d$ factors are
Lorentz invariant in the collinear limit. We thus see that in the collinear
limit the collinear contribution is distributed among the 2 and 5 resonance
histories, favouring the one that is nearer the resonance peaks.

The underlying Born corresponding to the real flavour configurations $f_r \in
\{ 1, 2, 3 \}$ is the Born flavour configuration $f_b = 1$. In the limit of
vanishing momentum of the additional radiation in leg 7, i.e. when the
momenta on the legs 1--6 in the real diagrams can be mapped to a given set of
momenta of its underlying Born diagram, all the $P_r$ factors of the real flavour
configuration reduce to the $P_b$ factors of the corresponding underlying Born
contributions. In our case:
\begin{equation}
  P_r^1 \rightarrow P^1_b, \hspace{1em} P_r^2 \rightarrow P^1_b, \hspace{1em}
  P_r^3 \rightarrow P^1_b ; \hspace{2em} P_r^4 \rightarrow P^2_b, \hspace{1em}
  P_r^5 \rightarrow P^2_b, \hspace{1em} P_r^6 \rightarrow P^2_b .
\end{equation}
This implies that in the same limit:
\begin{equation}
  D_r = P^1_b (d^{- 1}_7 + d^{- 1}_{37 ,2} + d^{- 1}_{47 ,2} + d^{- 1}_{57
  ,3} + d^{- 1}_{67 ,3}) + P^2_b (d^{- 1}_7 + d^{- 1}_{37 ,5} + d^{- 1}_{47
  ,5} + d^{- 1}_{57 ,6} + d^{- 1}_{67 ,6}),
\end{equation}
so that, in the soft limit, for example
\begin{equation}
  R_2 \; \cong \; \frac{P^1_b d^{- 1}_{37 ,2} }{P^1_b (d^{- 1}_7 + d^{-
  1}_{37 ,2} + d^{- 1}_{47 ,2} + d^{- 1}_{57 ,3} + d^{- 1}_{67 ,3}) +
  P^2_b (d^{- 1}_7 + d^{- 1}_{37 ,5} + d^{- 1}_{47 ,5} + d^{- 1}_{57 ,6} +
  d^{- 1}_{67 ,6})},
\end{equation}
and a similar relation holds for all other $\alpha_r$ contributions.
As shown in eq.~(\ref{eq:dijsoftlim}) we cannot drop the resonance history
dependence in the $d_{ij}$ factors. Thus,
unlike the case of collinear singularities, in the soft limit a full
factorization of the $P$ and $d^{- 1}$ factors does not hold in general.
We will see in the
following sections that this fact leads to a minor complication in the
evaluation of the soft contribution.

\section{Soft-collinear contributions}\label{sec:softcoll}

In the \tmverbatim{V2} implementation of narrow resonance decays, the soft
collinear contributions (to be added to the virtual one) are computed assuming
that no interference terms arise between the different resonances (or between
a resonance and the direct production). In the finite width case we are
considering now, this restriction has to be removed, because such interference
terms do arise. Furthermore, the soft-collinear contributions depend upon the
adopted subtraction procedure, and we are now departing from the default one
used in the \tmverbatim{POWHEG BOX}. We thus need to discuss in detail and
compute the soft-collinear contributions in the present framework.

As specified previously, each singular region $\alpha_r$ is
associated with a single full flavour structure $f_r (\alpha_r)$. The
treatment of initial state singularities remain the same as in the standard
case, since no resonance decays are involved in ISR (initial state radiation).
We thus focus upon FSR (final state radiation). Thus, from now on, the
singular region $\alpha_r$ corresponds to final state particles $i$ and $j$
becoming collinear. Furthermore, the soft singularity is associated with
particle $i$ becoming soft. This is the case when $i$ is a gluon and $j$ is a
quark. If also $j$ is a gluon, in \tmtexttt{POWHEG,} the $R_{\alpha_r}$
contribution is multiplied by a factor of the form
 $2 h (E_j) / (h (E_i) + h(E_j))$,
where $h$ is typically a power. This factor damps the soft
singularity when particle $j$ becomes soft. Since the cross section is
symmetric in the exchange of the two gluons, this procedure leads to the
correct result.

Given the singular region, \tmtexttt{POWHEG} selects a
phase space mapping from the Born phase space $\Phi_B$ and the three radiation
variables $\xi_i$, $y_{i j}$ and $\phi$ to the full real emission phase space.
In the $\alpha_r$ region, partons $i$ and $j$ will arise from the same
resonance $k_{\tmop{res}}$. The phase space mapping will thus be chosen in
such a way that only the momenta of the decay product of the resonance will be
affected. For example, in case of partons $i$ and $j$ corresponding to a gluon
and a $b$ quark arising from top decay, the phase space mapping will build the
radiation phase space starting from the momenta of the top decay products
(i.e. the $b$ and the $W^+$), maintaining fixed all remaining momenta together
with the top four-momentum. The mapping procedure will correspond to the
prescription described in the FNO paper (ref.~{\cite{Frixione:2007vw}}),
applied to the top decay product in the rest frame of the top. It is the same
procedure that is applied in ref.~{\cite{Campbell:2014kua}} for the case of $t
\bar{t}$ production and decay in the factorized approach.

Following the notation of the FNO paper, we thus write the
phase space as
\begin{eqnarray}
  \mathd \Phi_{n + 1} & = & (2 \pi)^d \delta^d  \hspace{-0.17em}
  \hspace{-0.17em} \left( k_{\oplus} + k_{\ominus} - \sum_{l = 1}^{n + 1} k_l
  \right) \hspace{-0.17em}  \left[ \prod_{l = 1}^{n + 1} \mathd \Phi_l
  \right], \nonumber\\
  \mathd \Phi_l & = & \frac{\mathd^{d - 1} k_l}{2 k_l^0 (2 \pi)^{d - 1}} . 
  \label{eq:ddimphasespace}
\end{eqnarray}
where $d = 4 - 2 \epsilon$ is the dimensionality of spacetime. Furthermore, we
introduce the parametrization
\begin{equation}
  \mathd \Phi_i = \frac{\mathd^{d - 1} k_i}{2 k_i^0 (2 \pi)^{d - 1}} =
  \frac{(k_{\tmop{res}}^2)^{1 - \epsilon}}{(4 \pi)^{3 - 2 \epsilon}} 
  \hspace{0.17em} \xi^{1 - 2 \epsilon} \mathd \xi \hspace{0.17em} \mathd
  \Omega^{3 - 2 \epsilon},
\end{equation}
\[ \  \]
where $\xi$ is defined as
\[ \xi = \frac{2 k^0_i}{\sqrt{k_{\tmop{res}}^2}} \hspace{0.17em}, \]
computed in the rest frame of resonance $k_{\tmop{res}}$. In the soft limit,
the phase space becomes
\begin{equation}
  \mathd \Phi_{n + 1} \Rightarrow \mathd \Phi_B  \frac{(k_{\tmop{res}}^2)^{1 -
  \epsilon}}{(4 \pi)^{3 - 2 \epsilon}}  \hspace{0.17em} \xi^{1 - 2 \epsilon}
  \mathd \xi \hspace{0.17em} \mathd \Omega^{3 - 2 \epsilon}
\end{equation}
where, following the notation of FNO, the underlying Born kinematics is expressed
in terms of the barred variables, and $\mathd \Phi_B$ is the
underlying Born phase space.

The $\alpha_r$ contribution to the cross section can be written as
\begin{equation}
  \int R_{\alpha_r} \mathd \Phi_{n + 1} .
\end{equation}
We now introduce the expansion
\begin{equation}
  \xi^{- 1 - 2 \epsilon} = - \frac{1}{2 \epsilon} \delta (\xi) + \xi^{- 1 - 2
  \epsilon}_+,
\end{equation}
where
\begin{equation}
  \xi^{- 1 - 2 \epsilon}_+ = \left( \frac{1}{\xi} \right)_+ - 2 \epsilon
  \left( \frac{\log \xi}{\xi} \right)_+ + \ldots,
\end{equation}
i.e. is defined as a distribution with a vanishing integral between 0 and 1.
We get
\begin{equation}
  \int R_{\alpha_r} \mathd \Phi_{n + 1} = I_{s, \alpha_r} + I_{+, \alpha_r},
\end{equation}
with
\begin{eqnarray}
  I_{s, \alpha_r} & = & - \frac{1}{2 \epsilon} \int \mathd \Phi_B
  \frac{(k_{\tmop{res}}^2)^{1 - \epsilon}}{(4 \pi)^{3 - 2 \epsilon}} \mathd
  \Omega^{3 - 2 \epsilon} \lim_{\xi \rightarrow 0}  [\xi^2 R_{\alpha_r}], \\
  I_{+, \alpha_r} & = & \int \mathd \Phi_{n + 1} \frac{\xi^{- 1 - 2
  \epsilon}_+}{\xi^{- 1 - 2 \epsilon}_{}} R_{\alpha_r}, 
\end{eqnarray}
where the meaning of the second equation is simply to replace $\xi^{- 1 - 2
\epsilon}$ with \ $\xi^{- 1 - 2 \epsilon}_+$ in the real cross section
integral, since the corresponding $\delta (\xi)$ contribution has been
subtracted out.

\subsection{Soft terms}
We now discuss explicitly the computation of the soft term $I_{s, \alpha_r}$.
In the standard treatment, by summing over all singular regions one recovers
the full $R$, that can be approximated in the soft limit by the eikonal
formula. We cannot follow this procedure now, since it
requires that the soft limit is taken in the same frame for all $\alpha_r$,
which is not our case. In order to deal with this complication, we employ the
following trick. We use the identity
\begin{equation}
  \int_0^{\infty} \mathd \xi \xi^{- 1 - 2 \epsilon} e^{- \xi} = \Gamma (- 2
  \epsilon) = \frac{\Gamma (1 - 2 \epsilon)}{- 2 \epsilon},
\end{equation}
to rewrite $I_{s, \alpha_r}$ as
\begin{eqnarray}
  I_{s, \alpha_r} & = & \frac{1}{\Gamma (1 - 2 \epsilon)} \int \mathd \Phi_B
  \int_0^{\infty} \mathd \xi \xi^{- 1 - 2 \epsilon} e^{- \xi}
  \frac{(k_{\tmop{res}}^2)^{1 - \epsilon}}{(4 \pi)^{3 - 2 \epsilon}} \mathd
  \Omega^{3 - 2 \epsilon} \lim_{\xi \rightarrow 0}  [\xi^2 R_{\alpha_r}]
  \nonumber\\
  & = & \frac{1}{\Gamma (1 - 2 \epsilon)}  \int \mathd \Phi_B \int \mathd
  \Phi_i e^{- \xi}  \tilde{R}_{\alpha_r} \nonumber\\
  & = & \frac{1}{\Gamma (1 - 2 \epsilon)}  \int \mathd \Phi_B \int \mathd
  \Phi_i e^{- \frac{2 k_i \cdot k_{\tmop{res}}}{k_{\tmop{res}}^2}} 
  \tilde{R}_{\alpha_r} \,, \label{eq:isoftres}
\end{eqnarray}
where with $\tilde{R}_{\alpha_r}$ we denote the Taylor expansion of
$R_{\alpha_r}$ in the soft limit for $k_{_i}$:
\begin{equation}
  \tilde{R}_{\alpha_r} = \frac{1}{\xi^2} \lim_{\xi \rightarrow 0}  [\xi^2
  R_{\alpha_r}] .
\end{equation}
We notice that $\tilde{R}_{\alpha_r}$ is obviously independent from the frame
used to define $\xi$. It is in fact obtained from $R_{\alpha_r}$ by
linearizing it in the $k_i$ momentum.

In formula (\ref{eq:isoftres}) the frame dependence of the soft contribution
is all contained in the exponential, the rest of the expression being fully
Lorentz invariant. In order to perform the integral we proceed as follows. We
rewrite (\ref{eq:isoftres}) as
\begin{eqnarray}
  I_{s, \alpha_r} & = & I_{s, \alpha_r}^{(1)} \noplus + I^{(2)}_{s, \alpha_r}
  \nonumber\\
  I_{s, \alpha_r}^{(1)} & = & \frac{1}{\Gamma (1 - 2 \epsilon)} \int \mathd
  \Phi_{\Beta}  \int \mathd \Phi_i  \tilde{R}_{\alpha_r}  \left\{ \exp \left[
  - \frac{2 k_i \cdot k_{\tmop{res}}}{k_{\tmop{res}}^2} \right] - \exp \left[
  - \frac{2 k_i \cdot m}{m^2} \right] \right\} \nonumber\\
  I^{(2)}_{s, \alpha_r} & = & \frac{1}{\Gamma (1 - 2 \epsilon)} \int \mathd
  \Phi_{\Beta}  \int \mathd \Phi_i  \tilde{R}_{\alpha_r} \exp \left[ - \frac{2
  k_i \cdot m}{m^2} \right],  \label{eq:softsplit}
\end{eqnarray}
where $m$ is an arbitrary timelike momentum. For definiteness, we choose $m =
q$, the total four-momentum of the final state particles. The $I_{s,
\alpha_r}^{(1)}$ term in eq.~(\ref{eq:softsplit}) is infrared finite. The
$I^{(2)}_{s, \alpha_r}$ term can now be integrated in any frame we like. We
then just pick a common frame for all $\alpha_r$ that have the same underlying
Born bare flavour structure $F_b$, and sum over all of them. We get
\begin{equation}
  \sum_{F_b (\alpha_r) = F_b} I^{(2)}_{s, \alpha_r} = \int \mathd \Phi_{\Beta}
  \frac{1}{\Gamma (1 - 2 \epsilon)} \int \mathd \Phi_i \exp \left[ - \frac{2
  k_i \cdot m}{m^2} \right] \sum_{F_b (\alpha_r) = F_b} \tilde{R}_{\alpha_r} .
  \label{eq:softint0}
\end{equation}
In the sum of $\tilde{R}_{\alpha_r}$, all dependencies upon the partition of
the resonance regions and upon the $d^{- 1}$ factors cancel out, yielding the
full $\tilde{R}$ for a soft gluon emission with an underlying Born flavour
configuration equal to $F_b$. In fact, the bare flavour structure of the
$\alpha_r$ such that $F_b (\alpha_r) = F_b$ consist of the same flavour
assignment $F_b$ plus one (soft) gluon. Thus, from
eq.~(\ref{eq:realpartition}), it follows that in the sum in
eq.~(\ref{eq:softint0}) all resonance history and $d^{- 1}$ factors simplify. We
thus have
\begin{equation}
\sum_{F_b (\alpha_r) = F_b} \tilde{R}_{\alpha_r} = 4 \pi \alpha_S
   {\mu}_R^{2 \epsilon}  \left[ \sum_{l m} B_{l m}^{(F_b)} \frac{k_l \cdot
   k_m}{(k_l \cdot k_i ) (k_m \cdot k_i)} - B^{(F_b)} \sum_l \frac{k_l^2}{(k_l
   \cdot k_i)^2} C_l \right] .
\end{equation}
Performing the integration in the rest frame of $m$, we can use again the
replacement
\begin{equation}
 \xi^{- 1 - 2 \epsilon} e^{- \xi} \Rightarrow \frac{\Gamma (1 - 2
   \epsilon)}{- 2 \epsilon} \delta (\xi),
\end{equation}
that leads to the standard calculation of the soft contributions, regardless
of their resonance assignment. The corresponding result is reported in
Appendix A.1 of the \tmtexttt{POWHEG BOX} paper (ref.~{\cite{Alioli:2010xd}}).

At this stage, we can split the Born terms again in terms of their full
flavour structures, and thus compute each contribution using the
appropriate (importance sampled according to the resonance structure) phase
space.

The treatment of the $I_{s, \alpha_r}^{(1)}$ term of eq.~(\ref{eq:softsplit})
requires some care, since although the soft singularity is no longer there, it
has still a collinear singularity corresponding to the $\alpha_r$ region. We
evaluate the $I_{s, \alpha_r}^{(1)}$ integral in the CM rest frame. Other
choices are possible, but there is no reason to make more complex choices,
since the resonance virtualities in $\tilde{R}_{\alpha_r}$ do not depend upon
the soft momentum $k_i$, and thus no particular importance sampling is needed
in the $k_i$ integration. We thus write
\begin{eqnarray}
  I^{(1)}_{s, \alpha_r} & = & \int \mathd \Phi_{\Beta}  \frac{1}{\Gamma (1 - 2
  \epsilon)} \int \frac{s^{1 - \epsilon}}{(4 \pi)^{3 - 2 \epsilon}} 
  \hspace{0.17em} \xi^{1 - 2 \epsilon}  (1 - y^2)^{- \epsilon} \mathd \xi
  \hspace{0.17em} \mathd y \hspace{0.17em} d \Omega^{2 - 2 \epsilon}
  \nonumber\\
  & \times & \tilde{R}_{\alpha_r}  \left\{ \exp \left[ - \frac{2 k_i \cdot
  k_{\tmop{res}}}{k_{\tmop{res}}^2} \right] - \exp [- \xi] \right\}, 
\end{eqnarray}
where
\begin{equation}
  y = \cos \theta_{i j},
\end{equation}
and $j$ is the emitter associated with the $\alpha_r$ region. By expanding
\begin{eqnarray}
  (1 - y^2)^{- \epsilon} & = & (1 + y)^{- \epsilon}  (1 - y)  (1 - y)^{- 1 -
  \epsilon} \nonumber\\
  & = & (1 + y)^{- \epsilon}  (1 - y) \left[ - \frac{2^{-
  \epsilon}}{\epsilon} \delta (1 - y) + \left( \frac{1}{1 - y} \right)_+ +
  \mathcal{O} (\epsilon) \right] \nonumber\\
  & = &  (1 - y) \left[ - \frac{2^{- 2 \epsilon}}{\epsilon} \delta (1 - y) +
  \left( \frac{1}{1 - y} \right)_+ + \mathcal{O} (\epsilon) \right], 
  \label{eq:ydistrib}
\end{eqnarray}
we can write
\begin{equation}
  I_{s, \alpha_r}^{(1)} = I^{(1)}_{s +, \alpha_r} + I_{s \delta,
  \alpha_r}^{(1)},
\end{equation}
where
\begin{eqnarray}
  I^{(1)}_{s +, \alpha_r} & = & \int \mathd \Phi_{\Beta} \int \frac{s^{}
  \xi}{(4 \pi)^3}  \left( \frac{1}{1 - y} \right)_+ \times (1 - y)
  \tilde{R}_{\alpha_r}  \left\{ e^{- \frac{2 k_i \cdot
  k_{\tmop{res}}}{k_{\tmop{res}}^2} } - e^{- \xi} \right\} \mathd \xi
  \hspace{0.17em} \mathd y \hspace{0.17em} \mathd \phi, \phantom{aaaa} \\
  I_{s \delta, \alpha_r}^{(1)} & = & - \int \mathd \Phi_{\Beta} 
  \frac{1}{\Gamma (1 - 2 \epsilon)} \int \frac{s^{1 - \epsilon}}{(4 \pi)^{3 -
  2 \epsilon}}  \hspace{0.17em} \xi^{1 - 2 \epsilon} \frac{2^{- 2
  \epsilon}}{\epsilon} \delta (1 - y) \mathd \xi \hspace{0.17em} \mathd y
  \hspace{0.17em} \mathd \Omega^{2 - 2 \epsilon} \nonumber\\
  & \times & \lim_{y \rightarrow 1} [(1 - y) \tilde{R}_{\alpha_r}]  \left\{
  e^{- \frac{2 k_i \cdot k_{\tmop{res}}}{k_{\tmop{res}}^2}} - e^{- \xi}
  \right\} . 
\end{eqnarray}
The $I^{(1)}_{s +, \alpha_r}$ term has to to be computed numerically. It has
no analogue in the previous \tmtexttt{POWHEG BOX} implementation.

We now work through the $I_{s \delta, \alpha_r}^{(1)}$. We have
\begin{equation}
  \lim_{y \rightarrow 1} [(1 - y) \tilde{R}_{\alpha_r}] = \frac{32 \pi
  \alpha_s {\mu}^{2 \epsilon} C_{j (f_b)}}{s \xi^2} B_{f_b (\alpha_r)},
  \label{eq:collapprox}
\end{equation}
where $C_j$ is the Casimir invariant associated with the particle that
underwent the splitting for the region $\alpha_r$. Observe that in deriving
the identity (\ref{eq:collapprox}) {\tmem{we have assumed that in the
collinear limit the $d^{- 1}$ factors associated with a given pair of
final state particles all coincide, irrespective of the resonance structure}},
as we have remarked earlier. Using
\begin{equation}
  \mathd \Omega^{2 - 2 \epsilon} = \frac{2 \pi^{1 - \epsilon}}{\Gamma (1 -
  \epsilon)}
\end{equation}
we get
\begin{eqnarray}
  I_{s \delta, \alpha_r}^{(1)} & = & - \int \mathd \Phi_{\Beta} 
  \frac{1}{\Gamma (1 - 2 \epsilon)} \int \frac{s^{1 - \epsilon}}{(4 \pi)^{3 -
  2 \epsilon}}  \hspace{0.17em} \xi^{1 - 2 \epsilon} \frac{2^{- 2
  \epsilon}}{\epsilon} \delta (1 - y) \mathd \xi \hspace{0.17em} \mathd y
  \hspace{0.17em} \mathd \Omega^{2 - 2 \epsilon} \nonumber\\
  & \times & \frac{32 \pi \alpha_s {\mu}^{2 \epsilon} C_{j (f_b)}}{s
  \xi^2} B_{f_b (\alpha_r)}  \left\{ \exp \left[ - \frac{2 k \cdot
  k_{\tmop{res}}}{k_{\tmop{res}}^2} \right] - \exp \left[ - \frac{2 k \cdot
  m}{m^2} \right] \right\} \nonumber\\
  & = & - \frac{(4 \pi)^{\epsilon}}{\Gamma (1 - 2 \epsilon) \Gamma (1 -
  \epsilon)} \left( \frac{{\mu}^2}{s} \right)^{\epsilon}  \frac{\alpha_s
  C_{j (f_b)}}{\pi} \int \mathd \Phi_{\Beta} B_{f_b (\alpha_r)}
  \frac{1}{\epsilon} \nonumber\\
  & \times & \int \mathd \xi \xi^{- 1 - 2 \epsilon} \left\{ \exp \left[ - \xi
  \frac{\sqrt{s}  \bar{k}_j \cdot k_{\tmop{res}}}{\bar{k}_j^0
  k_{\tmop{res}}^2} \right] - \exp [- \xi] \right\} \,,
\end{eqnarray}
where we have written, in the collinear limit
\begin{equation}
  k_i = \frac{k_i^0}{\bar{k}^0_j} \bar{k}_j = \frac{\xi \sqrt{s}}{2
  \bar{k}^0_j}  \bar{k}_j,
\end{equation}
and $\bar{k}_j$ is the momentum of the emitter in the soft limit, i.e. at
the underlying Born level. Performing the $\xi$ integration (from 0 to
$\infty$), we get
\begin{eqnarray}
  I_{s \delta, \alpha_r}^{(1)} & = & - \frac{(4 \pi)^{\epsilon}}{\Gamma (1 - 2
  \epsilon) \Gamma (1 - \epsilon)} \left( \frac{{\mu}^2}{s}
  \right)^{\epsilon}  \frac{\alpha_s C_{j (f_b)}}{\pi} \int \mathd
  \Phi_{\Beta} B_{f_b (\alpha_r)} \frac{1}{\epsilon} \nonumber\\
  & \times & \left[ \left( \frac{\sqrt{s}  \bar{k}_j \cdot
  k_{\tmop{res}}}{\bar{k}_j^0 k_{\tmop{res}}^2} \right)^{2 \epsilon} \Gamma (-
  2 \epsilon) - \Gamma (- 2 \epsilon) \right] \nonumber\\
  & = & \frac{\mathcal{N}}{2 \epsilon^2} \left( \frac{Q^2}{s}
  \right)^{\epsilon}  \frac{\alpha_s C_{j (f_b)}}{\pi} \int \mathd
  \Phi_{\Beta} B_{f_b (\alpha_r)} \left[ \left( \frac{\sqrt{s}  \bar{k}_j
  \cdot k_{\tmop{res}}}{\bar{k}_j^0 k_{\tmop{res}}^2} \right)^{2 \epsilon} - 1
  \right] \nonumber\\
  & = & \frac{\mathcal{N}}{2 \epsilon^2}  \frac{\alpha_s C_{j (f_b)}}{\pi}
  \int \mathd \Phi_{\Beta} B_{f_b (\alpha_r)} \left[ \left( \frac{Q \bar{k}_j
  \cdot k_{\tmop{res}}}{\bar{k}_j^0 k_{\tmop{res}}^2} \right)^{2 \epsilon} -
  \left( \frac{Q^2}{s} \right)^{\epsilon} \right] \nonumber\\
  & = & \mathcal{N}  \frac{\alpha_s C_{j (f_b)}}{\pi} \int \mathd
  \Phi_{\Beta} B_{f_b (\alpha_r)} \nonumber\\
  & \times & \left[ \frac{1}{\epsilon} \log
  \frac{\sqrt{s}  \bar{k}_j \cdot k_{\tmop{res}}}{\bar{k}_j^0
  k_{\tmop{res}}^2} + \left( \log^2  \frac{Q \bar{k}_j \cdot
  k_{\tmop{res}}}{\bar{k}_j^0 k_{\tmop{res}}^2} - \log^2 \frac{Q}{\sqrt{s}}
  \right) \right] \,,
\end{eqnarray}
where we have introduced the common normalization factor
\begin{equation}
  \mathcal{N} = \frac{(4 \pi)^{\epsilon}}{\Gamma (1 - \epsilon)} \left(
  \frac{{\mu}^2}{Q^2} \right)^{\epsilon} . \label{eq:normfact}
\end{equation}
The normalization factor in (\ref{eq:normfact}) should be the same one adopted
in the computation of the virtual term, as defined in the FNO and
\tmverbatim{POWHEG BOX} papers. If this is the case, the $1 / \epsilon$
singularities in the virtual and soft virtual contributions cancel exactly,
and one needs only to retain the finite terms.

\subsection{Collinear terms}

We now turn to the collinear integral
\begin{eqnarray}
  I_{+, \alpha_r} & = & \int \mathd \Phi_{n + 1} \frac{\xi^{- 1 - 2
  \epsilon}_+}{\xi^{- 1 - 2 \epsilon}_{}} R_{\alpha_r} . 
\end{eqnarray}
We are considering the region where $i, j$ become collinear, and where, as
discussed previously, $k_j$ does not lead to a soft singularity. This is the
case for $g q$ pairs $i$ should correspond to $g$. In the $g g$ case, the
$d^{- 1}_{i j}$ factors are supplemented with an energy damping factor
\begin{equation}
  \frac{2 h (E_{j_{}})}{h (E_{i_{}}) + h (E_{j_{}})},
\end{equation}
where, as in ref.~{\cite{Frixione:2007vw}}, $h$ is typically defined to be a
simple power law. Since there is no soft singularity in $k_j$, in the
following we can assume that we have a lower cutoff on the $k_j$ energy, that
can be smoothly removed at the end of the calculation, so that in our
manipulation we can always assume that $k_j$ is not vanishingly small. We now
write the $I_{+, \alpha_r}$ term as
\begin{eqnarray}
  I_{+, \alpha_r} & = & \int (2 \pi)^d \delta^d  \hspace{-0.17em}
  \hspace{-0.17em} \left( k_{\oplus} + k_{\ominus} - \sum_{l = 1}^{n + 1} k_l
  \right) \hspace{-0.17em}  \left[ \prod_{l \neq j, i} \mathd \Phi_l \right]
  \mathd \Phi_j \mathd \Phi_i \frac{\xi^{- 1 - 2 \epsilon}_+}{\xi^{- 1 - 2
  \epsilon}_{}} R_{\alpha_r}, 
\end{eqnarray}
where we remind that $\xi$ is evaluated in the frame of the resonance to which
$i$ and $j$ belong. We introduce for $\Phi_i$ the phase space (always
defined in the rest frame of the resonance)
\begin{equation}
  \mathd \Phi_i = \frac{(k^0_i)^{1 - 2 \epsilon}}{2 (2 \pi)^{3 - 2 \epsilon}}
  \mathd k^0_i (1 - y^2)^{- \epsilon} \hspace{0.17em} \mathd y \hspace{0.17em}
  \mathd \Omega_i^{2 - 2 \epsilon},
\end{equation}
where the angular integration is done with respect to the $j$ direction, i.e.
$y = 1 - \cos \theta_{i j}$. We separate out the collinear divergent term by
using eq.~(\ref{eq:ydistrib}), that yields
\begin{eqnarray}
  I_{+, \alpha_r} & = & I_{+ \delta, \alpha_r} + I_{+ +, \alpha_r},
  \nonumber\\
  I_{+ +, \alpha_r} & = & \int \mathd \Phi_{n + 1} (1 - y) \left( \frac{1}{1 -
  y} \right)_+ \xi \left( \frac{1}{\xi} \right)_+ R_{\alpha_r}, 
  \label{eq:Iplusplus}
\end{eqnarray}
and
\begin{eqnarray}
  I_{+ \delta, \alpha_r} & = & \int (2 \pi)^d \delta^d  \hspace{-0.17em}
  \hspace{-0.17em} \left( k_{\oplus} + k_{\ominus} - \sum_{l = 1}^{n + 1} k_l
  \right) \hspace{-0.17em}  \left[ \prod_{l \neq j, i} \mathd \Phi_l \right]
  \nonumber\\
  & \times & \mathd \Phi_j  \left[ - \frac{2^{- 2 \epsilon}}{\epsilon} \delta
  (1 - y) \right]  \frac{(k^0_i)^{1 - 2 \epsilon}}{2 (2 \pi)^{3 - 2 \epsilon}}
  \mathd k^0_i \mathd y \mathd \Omega_i^{2 - 2 \epsilon}  \frac{\xi^{- 1 - 2
  \epsilon}_+}{\xi^{- 1 - 2 \epsilon}_{}} \lim_{y \rightarrow 1} [(1 - y)
  R_{\alpha_r}] . \phantom{aaaa}
\end{eqnarray}
Eq.~(\ref{eq:Iplusplus}) should be interpreted as following: take the $\int
\mathd \Phi_{n + 1} R_{\alpha_r}$ expression, work it out in $y$ and $\xi$
variables, and replace $1 / (1 - y)$ and $1 / \xi$ by the corresponding $+$
distributions.

We now write the $\Phi_j$ integral in terms of angular and radial variables,
and introduce a variable $k^0 = k_j^0 + k_i^0$
\begin{eqnarray}
  I_{+ \delta, \alpha_r} & = & \int (2 \pi)^d \delta^d  \hspace{-0.17em}
  \hspace{-0.17em} \left( k_{\oplus} + k_{\ominus} - \sum_{l = 1}^{n + 1} k_l
  \right) \hspace{-0.17em}  \left[ \prod_{l \neq j, k} \mathd \Phi_l \right]
  \nonumber\\
  & \times &  \frac{(k^0_j)^{1 - 2 \epsilon}}{2 (2 \pi)^{3 - 2 \epsilon}}
  \mathd k^0_j \mathd \Omega_j^{3 - 2 \epsilon} \mathd k^0 \delta (k^0 - k_i^0
  - k_j^0) \nonumber\\
  & \times &  \left[ - \frac{2^{- 2 \epsilon}}{\epsilon} \delta (1 - y)
  \right]  \frac{(k^0_i)^{1 - 2 \epsilon}}{2 (2 \pi)^{3 - 2 \epsilon}} \mathd
  k^0_i \mathd y \mathd \Omega_i^{2 - 2 \epsilon}  \frac{\xi^{- 1 - 2
  \epsilon}_+}{\xi^{- 1 - 2 \epsilon}_{}} \lim_{y \rightarrow 1} [(1 - y)
  R_{\alpha_r}] . \phantom{aaaa}
\end{eqnarray}
Because of the $\delta (1 - y)$ factor we have now that $k_j$ and $k_i$ are
proportional, and thus $\Omega_j$ represents their common direction. Defining
\begin{equation}
  z = 1 - \frac{k_i^0}{E} \nocomma \nocomma,
\end{equation}
and defining
\begin{equation}
  k = k_i + k_j,
\end{equation}
performing the $\mathd k^0_j$ integration using the energy $\delta$ function
we get $k^0_j = z k^0$, and trading $k^0_i$ for $z$ we get
\begin{eqnarray}
  I_{+ \delta, \alpha_r} & = & \int (2 \pi)^d \delta^d  \hspace{-0.17em}
  \hspace{-0.17em} \left( k_{\oplus} + k_{\ominus} - \sum_{l = 1}^{n + 1} k_l
  \right) \hspace{-0.17em}  \left[ \prod_{l \neq j, k} \mathd \Phi_l \right]
  \nonumber\\
  & \times &  \frac{(k^0)^{1 - 2 \epsilon}}{2 (2 \pi)^{3 - 2 \epsilon}}
  \mathd k^0 \mathd \Omega_j^{3 - 2 \epsilon} \nonumber\\
  & \times &  \left[ - \frac{2^{- 2 \epsilon}}{\epsilon} \right] 
  \frac{(k^0)^{2 - 2 \epsilon}}{2 (2 \pi)^{3 - 2 \epsilon}} z^{1 - 2 \epsilon}
  (1 - z)^{1 - 2 \epsilon} \mathd z \mathd \Omega_i^{2 - 2 \epsilon} 
  \frac{\xi^{- 1 - 2 \epsilon}_+}{\xi^{- 1 - 2 \epsilon}_{}} \lim_{y
  \rightarrow 1} [(1 - y) R_{\alpha_r}] .  \phantom{aaaa}
\end{eqnarray}
We notice that the expression on the second line corresponds to the phase
space of the parton into which $i$ and $j$ have merged, i.e. the $\bar{k}_j$
phase space. The whole expression thus becomes
\begin{eqnarray}
  && I_{+ \delta, \alpha_r} = \int \mathd \Phi_B  \left[ - \frac{2^{- 2
  \epsilon}}{\epsilon} \right]  \frac{(k^0)^{2 - 2 \epsilon}}{2 (2 \pi)^{3 - 2
  \epsilon}} z^{1 - 2 \epsilon} (1 - z)^{1 - 2 \epsilon} \mathd z \left[
  \frac{2 \pi^{1 - \epsilon}}{\Gamma (1 - \epsilon)} \right]  \frac{\xi^{- 1 -
  2 \epsilon}_+}{\xi^{- 1 - 2 \epsilon}_{}} \lim_{y \rightarrow 1} [(1 - y)
  R_{\alpha_r}] \nonumber \\
  && = - \frac{1}{\epsilon} \frac{(4 \pi)^{\epsilon}}{\Gamma (1 - \epsilon)}
  \frac{2^{- 2 \epsilon}}{8 \pi^2} \int \mathd \Phi_B  (\bar{k}^0_j)^{2 - 2
  \epsilon} \mathd z z^{1 - 2 \epsilon} (1 - z)^{1 - 2 \epsilon}  \frac{\xi^{-
  1 - 2 \epsilon}_+}{\xi^{- 1 - 2 \epsilon}_{}} \lim_{y \rightarrow 1} [(1 -
  y) R_{\alpha_r}]\,,
\end{eqnarray}
where we have replaced $k^0$ with $\bar{k}^0_j$, that is the energy of the
underlying Born emitter. We have
\begin{equation}
  \xi = \xi_{\max}  (1 - z), \hspace{2em} \xi_{\max} = \frac{2
  \bar{k}^0_j}{\sqrt{k_{\tmop{res}}^2}},
\end{equation}
and
\begin{equation}
  \xi^{- 1 - 2 \epsilon}_+ = \xi^{- 1 - 2 \epsilon} + \frac{1}{2 \epsilon}
  \delta (\xi) = \xi_{\max}^{- 1 - 2 \epsilon} \times \left[ (1 - z)^{- 1 - 2
  \epsilon} + \frac{\xi_{\max}^{2 \epsilon}}{2 \epsilon} \delta (1 - z)
  \right],
\end{equation}
so that
\begin{eqnarray}
  I_{+ \delta, \alpha_r} & = & - \frac{1}{\epsilon} \frac{(4
  \pi)^{\epsilon}}{\Gamma (1 - \epsilon)}  \frac{2^{- 2 \epsilon}}{8 \pi^2}
  \int \mathd \Phi_B  (\bar{k}^{_{} 0}_j)^{2 - 2 \epsilon} \mathd z z^{1 - 2
  \epsilon}  \nonumber\\
  & \times & \left[ (1 - z)^{- 1 - 2 \epsilon} + \frac{\xi_{\max}^{2
  \epsilon}}{2 \epsilon} \delta (1 - z) \right] \times \lim_{y \rightarrow 1}
  [(1 - z)^2 (1 - y) R_{\alpha_r}] . 
\end{eqnarray}
The Altarelli-Parisi approximation in $4 - 2 \epsilon$ dimension yields
\begin{eqnarray}
  \lim_{y \rightarrow 1} [(1 - y) R_{\alpha_r}] & = & \frac{8 \pi \alpha_s
  {\mu}^{2 \epsilon}}{2 (\bar{k}^{_{} 0}_j)^2 z (1 - z)} P_{\alpha_r} (z)
  B_{f_b (\alpha_r)}, 
\end{eqnarray}
where with $P_{\alpha_r} (z)$ we mean
\begin{eqnarray}
  P_{g \rightarrow g g} (z) & = & C_A  \left( \frac{z}{1 - z} + \frac{1 -
  z}{z} + z (1 - z) \right) \frac{2 h (z)}{h (z) + h (1 - z)}, \\
  P_{g \rightarrow q \bar{q}} (z) & = & T_F  \frac{(1 - z)^2 + z^2 -
  \epsilon}{1 - \epsilon}, \\
  P_{q \rightarrow q g} (z) & = & C_F \left( \frac{1 + z^2}{1 - z} - \epsilon
  (1 - z) \right) . 
\end{eqnarray}
We thus get
\begin{eqnarray}
  I_{+ \delta, \alpha_r} & = & - \frac{\mathcal{N}}{\epsilon} 
  \frac{\alpha_s}{2 \pi^{}} \int \mathd \Phi_B B_{f_b (\alpha_r)}  \left(
  \frac{Q}{2 \bar{k}^{_{} 0}_j} \right)^{2 \epsilon}  \int \mathd z z^{- 2
  \epsilon}  \nonumber\\
  & \times & \left[ (1 - z)^{- 1 - 2 \epsilon} + \frac{\xi_{\max}^{2
  \epsilon}}{2 \epsilon} \delta (1 - z) \right] \times (1 - z) P_{\alpha_r}
  (z) . 
\end{eqnarray}
Let us begin by evaluating the integrals
\begin{eqnarray}
  \int_0^1 \mathd z z^{- 2 \epsilon}  (1 - z)^{- 2 \epsilon} \times P_{g
  \rightarrow g g} (z) & = & \int_0^1 \mathd z z^{- 2 \epsilon}  (1 - z)^{- 2
  \epsilon} \times \frac{P_{g \rightarrow g g} (z) + P_{g \rightarrow g g} (1
  - z)}{2} \nonumber\\
  & = & C_A \int_0^1 \mathd z z^{- 2 \epsilon}  (1 - z)^{- 2 \epsilon} \left(
  \frac{z}{1 - z} + \frac{1 - z}{z} + z (1 - z) \right) \nonumber\\
  & = & C_A \int_0^1 \mathd z z^{- 2 \epsilon}  (1 - z)^{- 2 \epsilon} \left(
  \frac{2 z}{1 - z} + z (1 - z) \right) \nonumber\\
  & = & C_A \int_0^1 \mathd z z^{- 2 \epsilon}  (1 - z)^{- 2 \epsilon} \left(
  \frac{2}{1 - z} - 2 + z (1 - z) \right) \nonumber\\
  & = & C_A \left[ \int_0^1 \mathd z (1 - z)^{- 2 \epsilon}  \frac{2}{1 - z}
  - 4 \epsilon \int_0^1 \mathd z \frac{\log (z)}{1 - z} \right. \nonumber\\
  & + & \left. \int_0^1 \mathd z (- 2 + z (1 - z)) (1 - 2 \epsilon \log [z (1
  - z)]) \right] \nonumber\\
  & = & C_A \left[ \frac{- 2}{2 \epsilon} + \epsilon \frac{2 \pi^2}{3} -
  \epsilon \frac{67}{9} - \frac{11}{6} \right] \nonumber\\
  & = & \frac{- 2 C_A}{2 \epsilon} - \epsilon \left( \frac{67}{9} - \frac{2
  \pi^2}{3} \right) C_A - \frac{11 C_A}{6}, 
\end{eqnarray}
where in the first two steps we have used twice the $z \rightarrow 1 - z$
symmetry of the integral.
For the $q\to qg$ case we have
\begin{eqnarray}
  && \int_0^1 \mathd z z^{- 2 \epsilon}  (1 - z)^{- 2 \epsilon} \times P_{q
  \rightarrow q g} (z) \nonumber \\
  & = & \int_0^1 \mathd z z^{- 2 \epsilon}  (1 - z)^{- 2
  \epsilon} \times C_F \left( \frac{1 + z^2}{1 - z} - \epsilon (1 - z) \right)
  \nonumber\\
  & = & C_F \left[ \int_0^1 \mathd z z^{- 2 \epsilon}  (1 - z)^{- 2 \epsilon}
  \times \left( \frac{2}{1 - z} - (1 + z) - \epsilon (1 - z) \right) \right]
  \nonumber\\
  & = & C_F \left[ \int_0^1 \mathd z (1 - z)^{- 2 \epsilon}  \frac{2}{1 - z}
  - 4 \epsilon \int_0^1 \mathd z \frac{\log (z)}{1 - z} \right. \nonumber\\
  &&\; + \; \left. \int_0^1 \mathd z (- (1 + z) - \epsilon (1 - z)) (1 - 2
  \epsilon \log [z (1 - z)]) \right] \nonumber\\
  & = & C_F \left[ \frac{- 2}{2 \epsilon} + \epsilon \frac{2 \pi^2}{3} -
  \frac{3}{2} - \epsilon \frac{13}{2} \right] 
  \; = \; \frac{- 2 C_F}{2 \epsilon} - \epsilon \left[ \frac{13}{2} - \frac{2
  \pi^2}{3} \right] C_F - \frac{3 C_F}{2} . 
\end{eqnarray}
Finally, for the $g\to q\bar{q}$ case:
\begin{eqnarray}
  \int_0^1 \mathd z z^{- 2 \epsilon}  (1 - z)^{- 2 \epsilon} \times P_{g
  \rightarrow q \bar{q}} (z) & = & \int_0^1 \mathd z z^{- 2 \epsilon}  (1 -
  z)^{- 2 \epsilon} \times T_F  \frac{(1 - z)^2 + z^2 - \epsilon}{1 -
  \epsilon} \nonumber\\
  & = & \frac{2 T_F }{3} \noplus + \epsilon \frac{23 T_F }{9} \,.
\end{eqnarray}
We now define, as usual
\begin{equation}
  \begin{array}{lllllll}
    \gamma_g & = & \frac{11 C_A - 4 T_F n_F}{6}\;, &  & \gamma_g' & = & \left(
    \frac{67}{9} - \frac{2 \pi^2}{3} \right) C_A - \frac{23}{9} T_F n_F   \;,\\
    \gamma_q & = & \frac{3}{2} C_F\;, &  & \gamma'_q & = & \left( \frac{13}{2} -
    \frac{2 \pi^2}{3} \right) C_F \;,
  \end{array}
\end{equation}
and find
\begin{equation}
  \sum_{\alpha_r \in \alpha_r (f_b)} I_{+ \delta, \alpha_r} = \mathcal{N} 
  \frac{\alpha_s}{2 \pi^{}} \int \mathd \Phi_B B_{f_b}  \left( \frac{Q}{2
  \bar{k}^{_{} 0}_j} \right)^{2 \epsilon}  \left[ \frac{1 - \xi_{\max}^{2
  \epsilon}}{\epsilon^2} C_{j (f_b)} + \frac{\gamma_{j (f_b)}}{\epsilon} +
  \gamma'_{j (f_b)} \right], 
\end{equation}
where by $j (f_b)$ we mean the flavour of the $j^{\tmop{th}}$ parton in the
$f_b$ flavour structure. We get
\begin{equation} \hspace{-0.4cm}
  \sum_{\alpha_r \in \alpha_r (f_b)} I_{+ \delta, \alpha_r} = \mathcal{N} 
  \frac{\alpha_s}{2 \pi^{}} \int \mathd \Phi_B B_{f_b}  \left(
  \frac{Q^2}{k^2_{\tmop{res}}} \right)^{\epsilon} \xi_{\max}^{- 2 \epsilon}
  \left[ \frac{1 - \xi_{\max}^{2 \epsilon}}{\epsilon^2} C_{j (f_b)} +
  \frac{\gamma_{j (f_b)}}{\epsilon} + \gamma'_{j (f_b)} \right],
\end{equation}
where we have used for $\xi_{\max}$ the covariant expression
\begin{equation}
  \xi_{\max} = \frac{2 \bar{k}_j \cdot k_{\tmop{res}}}{k_{\tmop{res}}^2} .
\end{equation}
We now expand it as
\begin{eqnarray}
  &&  \hspace{-1cm} \sum_{\alpha_r \in \alpha_r (f_b)} I_{+ \delta, \alpha_r} \; = \; \mathcal{N} 
  \frac{\alpha_s}{2 \pi^{}} \int \mathd \Phi_B B_{f_b}  \left(
  \frac{Q^2}{k^2_{\tmop{res}}} \right)^{\epsilon}  \left[ \frac{- 2 \log
  \xi_{\max} + 2 \epsilon \log^2 \xi_{\max}}{\epsilon} C_{j (f_b)} \right.
  \nonumber\\
  && \phantom{aaaaaaaaa}
   \; + \; \left.  \frac{\gamma_{j (f_b)}}{\epsilon} (1 - 2 \epsilon \log
  \xi_{\max}) + \gamma'_{j (f_b)} \right] \nonumber\\
  && \; = \;  \mathcal{N}  \frac{\alpha_s}{2 \pi^{}} \int \mathd \Phi_B B_{f_b}
  \left[ \frac{- 2 \log \xi_{\max} + 2 \epsilon \log^2 \xi_{\max} - 2 \epsilon
  \log \xi_{\max} \log (Q^2 / k_{\tmop{res}}^2)}{\epsilon} C_{j (f_b)} \right.
  \nonumber\\
  &&\quad + \; \left. \frac{\gamma_{j (f_b)}}{\epsilon} + \gamma_{j (f_b)} \log
  \frac{Q^2}{k_{\tmop{res}}^2} - 2 \gamma_{j (f_b)} \log \xi_{\max} +
  \gamma'_{j (f_b)} \right] \nonumber\\
  && \; = \; \mathcal{N}  \frac{\alpha_s}{2 \pi^{}} \int \mathd \Phi_B B_{f_b}
  \left[ \frac{- 2 \log \xi_{\max}}{\epsilon} C_{j (f_b)} \noplus +
  \frac{\gamma_{j (f_b)}}{\epsilon}  \right. \nonumber\\
  && \quad + \; \left. 2 \log \xi_{\max} \left( \log \xi_{\max} - \log
  \frac{Q^2}{k_{\tmop{res}}^2} \right) C_{j (f_b)}  \right. \nonumber \\
  && \quad + \left. \left( \log
  \frac{Q^2}{k_{\tmop{res}}^2} - 2 \log \xi_{\max} \right) \gamma_{j (f_b)} +
  \gamma'_{j (f_b)} \right]\,.
\end{eqnarray}
We now combine this term with the $I_{s \delta, \alpha_r}^{(1)}$ integral:
\begin{eqnarray}
  I_{A, \alpha_r} & = & I_{+ \delta, \alpha_r} + I_{s \delta, \alpha_r}^{(1)}
  \nonumber\\
  & = & \mathcal{N}  \frac{\alpha_s}{2 \pi^{}} \int \mathd \Phi_B B_{f_b}
  \left[ \frac{2}{\epsilon} \log \frac{\sqrt{s}}{2 \bar{k}^0_j} C_{j (f_b)} +
  \frac{\gamma_{j (f_b)}}{\epsilon} \right. \nonumber\\
  & + & 2 \left( \log \frac{\sqrt{s}}{2 \bar{k}^0_j} \noplus + \log
  \xi_{\max} \right) \left( \log \frac{\sqrt{s}}{2 \bar{k}^0_j } \noplus +
  \log \xi_{\max} + \log \frac{Q^2}{s} \right) C_{j (f_b)}\\
  & + & \left. 2 \log \xi_{\max} \left( \log \xi_{\max} - \log
  \frac{Q^2}{k_{\tmop{res}}^2} \right) C_{j (f_b)} + \left( \log
  \frac{Q^2}{k_{\tmop{res}}^2} - 2 \log \xi_{\max} \right) \gamma_{j (f_b)} +
  \gamma'_{j (f_b)} \right], \nonumber
\end{eqnarray}
where $f_b$ stands for $f_b (\alpha_r)$, and $j$ is the emitter for the region
$\alpha_r$. Notice also that now $\bar{k}^0_j$ represents the energy of the
emitter in our common reference frame, while earlier, with the same symbol we
denoted its energy in the resonance frame. Notice also that $\xi_{\max}$ is
now frame dependent. We will denote as $I_A^{(0)}$ the finite part of $I_A$.

\subsection{Summary}
We now summarize the real and soft-collinear terms that need to be included in
the calculation:
\begin{enumeratealphacap}
  \item Real integral:
  \begin{eqnarray}
    I_{+ +, \alpha_r} & = & \int \mathd \Phi_{n + 1} (1 - y) \left( \frac{1}{1
    - y} \right)_+ \xi \left( \frac{1}{\xi} \right)_+ R_{\alpha_r}, 
    \label{eq:Iplusplusfinal}
  \end{eqnarray}
  \item Collinear terms:
\begin{multline}
\hspace{-1cm}    I_{A, \alpha_r}^{(0)} = \frac{\alpha_s}{2 \pi^{}}  \int \mathd \Phi_B
    B_{f_b} \left[ 2 \left( \log \frac{\sqrt{s}}{2 \bar{k}^0_j} \noplus + \log
    \xi_{\max} \right) \left( \log \frac{\sqrt{s}}{2 \bar{k}^0_j } \noplus +
    \log \xi_{\max} + \log \frac{Q^2}{s} \right) C_{j (f_b)}  \right.\\
\hspace{-1cm}      + \left. 2 \log \xi_{\max} \left( \log \xi_{\max} - \log
    \frac{Q^2}{k_{\tmop{res}}^2} \right) C_{j (f_b)} + \left( \log
    \frac{Q^2}{k_{\tmop{res}}^2} - 2 \log \xi_{\max} \right) \gamma_{j (f_b)}
    + \gamma'_{j (f_b)} \right]_{},
  \end{multline}
  where $j$ is the emitter and $k_{\tmop{res}}$ is the momentum of the
  resonance that contains the emitter for the region $\alpha_r$. With
  $\bar{k}^0_j$ we denote the energy of the emitter in our common reference
  frame, that is the CM frame of the final state. On the other hand,
  $\xi_{\max}$ is computed in the resonance frame.
  \item Soft terms remain the same as in the standard treatment, and are
  reported in Appendix A.1 of the \tmtexttt{POWHEG BOX} paper
  (ref.~{\cite{Alioli:2010xd}}).
  \item Soft mismatch:
  \begin{eqnarray}
    I^{(1)}_{s +, \alpha_r} & = & \int \mathd \Phi_{\Beta} \int_0^{\infty}
    \mathd \xi \int_{- 1}^1 \mathd y \int_0^{2 \pi} \mathd \phi \frac{s^{}
    \xi}{(4 \pi)^3} \times \Biggl\{ \tilde{R}_{\alpha_r} \left[ e^{- \frac{2
    k_i \cdot k_{\tmop{res}}}{k_{\tmop{res}}^2} } - e^{- \xi} \right]
    _{\alpha_r} \nonumber\\
    & - & \nobracket \nobracket \frac{32 \pi \alpha_s C_{j (f_b)}}{s
    \xi^2} B_{f_b (\alpha_r)} \frac{\left[ 
    e^{- \frac{2 \bar{k}^{}_j \cdot k_{\tmop{res}}}{k_{\tmop{res}}^2} \frac{k^0_i}{\bar{k}^0_j} } - e^{- \xi}
    \right]_{\alpha_r} }{1 - \cos \theta} \Biggr\}, 
  \end{eqnarray}
  to be summed over all the $\alpha_r$ with the emitter belonging to a
  resonance.

  The integration phase space is defined in the partonic CM frame, with the
  third axis pointing along the direction of the emitter $j$.
  \item Collinear terms related to initial state radiation remain the same as
  in the standard treatment.
\end{enumeratealphacap}
In tab.~\ref{tab:newstuffinbox}, we list the terms that needed to be newly
implemented in the new, resonance aware version of the \tmverbatim{POWHEG BOX}
that we are presenting here.
\begin{table}[h]
    \centering
{\small \begin{tabular}{|l|l|l|l|}
    \hline
    Term & Description & Already present & New\\
    \hline
    $I^{(2)}_{s, \alpha_r}$ & Soft terms & $\mathcal{I}_{i j}$ in the
    \tmverbatim{POWHEG BOX} & \\
    \hline
    $I^{(1)}_{s +, \alpha_r}$ & Soft mismatch &  & Should be done
    numerically\\
    \hline
    $I_{+ +, \alpha_r}$ & Real integral & Already present (resonance
    extension) & \\
    \hline
    $I^{(0)}_{A, \alpha_r}$ & Collinear terms & To be deleted & To be added\\
    \hline
  \end{tabular} } 
  \caption{The terms that had to be added or modified in the {\tt POWHEG BOX} in
  order to implement the subtraction scheme of the present
  publication.\label{tab:newstuffinbox}}
\end{table}

\subsection{Soft $\log \Gamma$ terms}

In the procedure that we have illustrated, collinear singular regions arise
only among partons produced in the decay of the same resonance. This property
arises because, in the separation of the singular regions, we restrict ourselves
to singular structures that are compatible with the resonance history. While
this feature guarantees a smooth cancellation of the collinear logarithms in
the subtraction procedure, we cannot expect a corresponding cancellation of
all soft, non collinear logarithms. There are in fact two sources of soft
radiation with a lower or upper cut off of the order of the resonance
virtualities:
\begin{itemizedot}
  \item Soft radiation arising from the interference of soft emissions from
  coloured partons belonging to different resonances. These terms have an
  {\tmem{upper}} cut-off of the order of the resonance width.
  
  \item Soft emission involving amplitudes with radiation arising from the
  resonances internal lines. These terms have a {\tmem{lower}} cut off of the
  order of the resonance width.
\end{itemizedot}
We thus expect that in our procedure $\log \Gamma$ terms will arise in the
integration of the real cross section. The virtual corrections will also have
corresponding $\log \Gamma$ terms, that cancel the real ones when summed together.

In this section we discuss the structure of these soft terms. As we will see,
it is possible, in principle, to remove them from the integration of the real
cross section, and include them in the soft term, in such a way that their
cancellation takes place in the soft-virtual contribution. However, we have
not attempted to implement this in the \tmverbatim{POWHEG BOX}. In view of the
relatively large size of the resonance widths in the typical processes that we
consider, it is unlikely that they may cause problems in practical NLO
calculations. Furthermore, as far as NLO+PS implementations are concerned,
these terms are in fact properly treated in our resonance-aware \tmverbatim{POWHEG}
framework, and do not require any further action.

\begin{figure}[t]
\begin{center}
\includegraphics[width=0.48\textwidth]{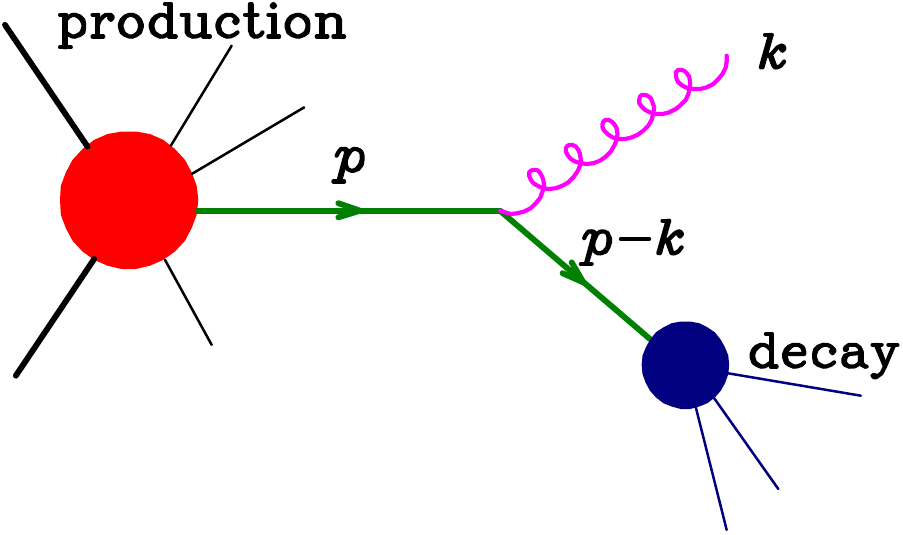}
\caption{\label{fig:softlogG}
Insertion of a soft gluon in an internal resonance propagator.}
\end{center}
\end{figure}

We now discuss the structure of the soft logarithms in the presence of narrow
resonances. For simplicity, we assume that all the resonances have comparable
widths of order $\Gamma_0$. We consider two regions:
\begin{itemizedot}
  \item Region $a$: is characterized by soft emissions with energy $\omega$
  larger than $\Gamma_0$. In this region $\Gamma_0$ plays the role of an
  infrared cutoff. The dominant region of integration has $\log \omega$
  uniformly distributed between $\log \Gamma_0$ (lower cut-off) and the log of
  some hard scale in the process (high cut-off), typically of the order of the
  mass of the resonances.
  
  \item Region $b$: is characterized by soft emissions with energy $\omega$
  less than $\Gamma_0$. The lower limit in this region is regulated in the
  usual technical ways (like dimensional
  regularization). Its upper cut-off is $\Gamma_0$.
\end{itemizedot}
In region $a$, since $\Gamma_0$ acts as an infrared cutoff, the emissions from
resonances internal lines near their mass shell should also be considered as
soft. In fig.~\ref{fig:softlogG}
we illustrate the insertion of a soft emission in an internal resonance line.
The product of the resonance propagators will be given by
\begin{multline}
  \frac{1}{p^2 - M^2 + i \Gamma M} \times \frac{1}{(p - k)^2 - M^2 + i \Gamma
  M}  \\
= \frac{1}{2 p \cdot k} \left[ \frac{1}{(p - k)^2 - M^2 + i \Gamma M} -
  \frac{1}{p^2 - M^2 + i \Gamma M} \right] .
\end{multline}
Under the assumption that $\omega = k^0 > \Gamma$, the two denominators cannot
be near their mass shell at the same time. When the first term in the square
bracket is near its mass shell, the process corresponds
to the resonance radiating during production.
In fact, in this case the $p$ momentum is far from the mass shell by a scale
of order $k^0$, while the $p-k$ momentum is near the mass shell by a scale of
order $\Gamma$. In coordinate space, this means that the line carrying
momentum $p$ has a length of order $1/k^0$, much shorter than the length
of order $1/\Gamma$ of the $p-k$ line. Conversely, if the second
term is on-shell, radiation is taking place during decay. When squaring the
amplitude, interference between these two terms is suppressed, since the two
propagators cannot be on-shell at the same time, and the
integration is effectively cut off by at a scale of order $\Gamma$, leaving
no phase space for soft logarithms to build up. For the same reason,
interference from emissions arising at production with emissions from
resonance decay, as well as from emissions arising from the decay of different
resonances, do not yield soft logarithms, since they also lead to
propagators off the resonance peaks in the interfering amplitudes.

Reasoning in terms of radiation and decay times,  by assuming $\omega >
\Gamma_0$ we are assuming that radiation time is shorter than the resonances
lifetimes. Thus, soft radiation in production cannot interfere with radiation
in decays, since they happen at different times, and for the same reason
radiation from different resonances cannot interfere. So, as far as soft
singularities are concerned, the process can be though of as the product of
independent production and decay processes, each one of them with resonances
appearing only as initial or final state particles, but not as internal lines.
For all these independent components, soft emissions is given by the usual
eikonal formula applied only to initial and final state particles, that in
this case can also be unstable resonances.

The structure of the soft singularity in region $b$ is determined by the
initial and final state particles after the decay of all resonances. The
resonances are considered as off-shell particles, as far as soft emissions are
concerned, and interference terms from emissions arising from different
resonances are not suppressed by small Breit-Wigner weights, since the
emission energy is below the resonance widths. In terms of time, this is the
case when the time for soft radiation is longer than the resonance widths, so
that only particles that live longer than the resonances can contribute.

The form of the soft subtraction term in the $b$ region is the same one that
we have adopted in the present method. However, that in
our present treatment we are considering unrestricted emissions, while the $b$
region is defined to involve soft energies below $\Gamma_0$.
When considering a given underlying Born resonance history, we will thus have
the following cases:
\begin{itemizedot}
  \item In the emissions from pairs of coloured massless partons belonging to
  the same resonance, the terms in regions $a$ and $b$ will combine, yielding
  an unrestricted soft energy integral, and no $\log \Gamma$ terms.
  
  \item In the emission from pairs of coloured massless partons belonging to
  different resonances, only the terms in the $b$ region will be present.
  These contributions will be cut-off at energies above $\Gamma_0$, since for
  larger energies they will push one of the two resonances
  out of its mass shell, thus damping the cross section. They will
  thus lead to $\log \Gamma$ contributions to the cross section.
  
  \item In any emission from an internal resonance leg, the emission energy
  will have a lower infrared cutoff $\Gamma_0$, and will yield other $\log
  \Gamma$ terms.
\end{itemizedot}

It is conceivable that our method may be modified, by adding further soft
subtraction terms to the real cross section and corresponding integrated soft
terms to the soft-virtual cross section, in such a way that the $\log \Gamma$
terms cancel within the soft-virtual contribution. This procedure may make
the NLO calculation more convergent in the zero width limit. However, it would
have no effect in the generation of radiation according to the
\tmverbatim{POWHEG} method. In \tmverbatim{POWHEG}, the cancellation of the
$\log \Gamma$ terms takes place numerically in the calculation of the
$\tilde{B}$ function, between the real and the soft-virtual integral. In the
generation of radiation, the cross section is unitarized by construction, so
that no further $\log \Gamma$ terms arise in inclusive quantities. We thus did
not attempt to implement such an improvement in the present work.

\section{Code organization}

The implementation of the subtraction scheme described in the present paper
has required an extensive rewriting of several parts of the \tmverbatim{POWHEG
BOX} framework. While we postpone writing a full documentation for the new
code, we will describe in the present section the structures that are used to
describe the various components of the cross section in terms of flavour and
resonance histories.

The flavour structures used to implement our subtraction scheme are organized
as follows. The process specific code provides the flavour structure in terms
of arrays carrying the flavour of the particles involved in the process,
including intermediate resonances. We have the arrays
\begin{alltt}

flst_born(1:flst_bornlength(iborn),iborn), iborn=1...flst_nborn;
flst_bornres(1:flst_bornlength(iborn),iborn), iborn=1...flst_nborn;

flst_real(1:flst_reallength(ireal),ireal), ireal=1...flst_nreal;
flst_realres(1:flst_reallength(ireal),ireal), ireal=1...flst_nreal.

\end{alltt}
These arrays are set in the user process routines. The following arrays are
also set by the user process routines:
\begin{alltt}
flst_bornresgroup(1:flst_nborn);
flst_nbornresgroup.
\end{alltt}
{\noindent}These have the purpose of grouping together the Born full flavour
configurations that have similar resonance structure so that they can be
integrated together. Thus, the value of
\tmverbatim{flst\_bornresgroup(iborn)} (an integer from 1 to
\tmverbatim{flst\_nbornresgroup}) labels the resonance structure group of the
born full flavour structure \tmverbatim{iborn}. This is needed because the
\tmverbatim{POWHEG} \tmverbatim{BOX} groups together flavour structures with
similar resonance histories when performing the integration, since these
configuration can be integrated with the same importance sampling.

In the user process arrays, each flavour structure can appear only once,
following the traditional approach of FNO. In the
case of resonances, this leads to a non-trivial subtlety that we describe now.
Consider the flavour structure for the subprocess $q \bar{q} \rightarrow e^+
e^- e^+ e^-$. According to the traditional approach of the \tmverbatim{POWHEG
BOX} there is only one flavour structure associated with this process, i.e.
only a single ordering of the final state electron and positron will appear.
When resonance structures are considered, we realize that we have two ways of
pairing the electrons--anti-electrons to build an intermediate $Z$ boson. These
two pairings are fully equivalent up to permutations of the final state
particles, and thus only one resonance assignment will appear for them. We
should not forget, however, that the contribution with a given resonance
assignment carries a resonance projection factor. Assigning the ordering 1 to
8 to the $q \bar{q} \rightarrow Z Z \rightarrow e^+ e^- e^+ e^-$, assuming
that the $Z$ in position 3 decays into the $e^{\noplus +} e^-$ pair in
position 5-6, and assuming that we have only these two resonance histories, we
will have a factor of the form
\begin{equation}
  \frac{P (5, 6 ; 7, 8)}{P (5, 6 ; 7, 8) + P (5, 8 ; 7, 6)},
\end{equation}
where, for example,
\begin{equation}
  P (5, 6 ; 7, 8) = \frac{M_Z^4}{(s_{56} - M_Z^2)^2 + \Gamma_Z^2 M_Z^2} \times
  \frac{M_Z^4}{(s_{78} - M_Z^2)^2 + \Gamma_Z^2 M_Z^2} .
\end{equation}
It is clear now that we should supply a factor of 2 for this graph, since by
assigning the resonances we break part of the exchange symmetry for final
state identical particles. Thus, the user process should provide the symmetry
factor appropriate for the given final state irrespective of resonance
assignment. The \tmverbatim{POWHEG} machinery will take care of supplying the
appropriate factors arising from the resonance assignment specification. Thus,
the user process list is built from the list of flavour structures for all
distinct final states (where distinct means that final state differing by
permutations are not allowed). For each final state, the list will be expanded
by assigning all possible resonance histories. But, again, full flavour
structure differing only by a permutation of the resonance histories will not
be allowed. The \tmverbatim{POWHEG BOX} checks explicitly that no full flavour
structures equivalent up to a permutation will appear.

Notice that the factor of two will lead to a total resonance factor of
\begin{equation}
  \frac{2 P (5, 6 ; 7, 8)}{P (5, 6 ; 7, 8) + P (5, 8 ; 7, 6)},
\end{equation}
that is not 1. However, by symmetry, an analysis of generated events that does
not distinguish among identical final state particles will lead to the same
results as if we included both weights

\tmverbatim{}
\begin{equation}
  \frac{P (5, 6 ; 7, 8)}{P (5, 6 ; 7, 8) + P (5, 8 ; 7, 6)} + \frac{P (5, 8 ;
  7, 6)}{P (5, 6 ; 7, 8) + P (5, 8 ; 7, 6)} = 1 .
\end{equation}

\

Given the real, the \tmverbatim{POWHEG BOX} finds all singular regions
associated with the real graphs, and builds the corresponding arrays
\begin{alltt}

flst_alr(1:flst_alrlength(alr),alr), alr=1...flst_nalr;
flst_alrres(1:flst_alrlength(alr),alr), alr=1...flst_nalr.

\end{alltt}
{\noindent}It furthermore fills the arrays
\tmverbatim{flst\_emitter(1:flst\_nalr)} with the emitter of the given
singular region, and the array \tmverbatim{flst\_alrmult(1:flst\_nalr)} with
the multiplicity of the singular region. A multiplicity factor can arise if we have
identical partons in the final state. For example, if we have several gluons
and a quark in the final state, there will be regions associated with each
gluon being collinear to the quark, and the program will find as many regions
of this type as there are gluon. It will recognize that all these regions are
equivalent, and it will emit a single region with a multiplicity factor equal
to the number of equivalent regions.

If there are real contributions that do not have any singular region, they are
collected into the "regular" arrays
\begin{alltt}

flst_regular(1:flst_regularlength(ireg),ireg), ire=1...flst_nregular;
flst_regularres(1:flst_regularlength(ireg),ireg), ireg=1...flst_nregular.

\end{alltt}
{\noindent}An array \tmverbatim{flst\_regularmult(1:flst\_nregular)} is
provided also in this case.

The task of finding out the singular and regular contributions is carried out
by the subroutine \tmverbatim{genflavreglist}, in the file
\tmverbatim{find\_regions.f}. In the traditional \tmverbatim{POWHEG BOX}
implementation, at this stage, for each \tmverbatim{alr} a list of competing
singular regions was found. These were needed since each \tmverbatim{alr}
contribution is obtained by multiplying the corresponding real graph by the
ratio of the $d^{- 1}_{\alpha_r}$ factor divided by the sum of all the $d^{-
1}_{\alpha_r}$ associated with the other competing singular regions. In the
current implementation, we also need to list together the competing resonance
histories that lead to the same final state, in order to compute the
corresponding resonance projection factor. So, for either the alr, the regular
or the Born terms, we have an array of pointers
\begin{alltt}
flst_XXXnumrhptrs(1:flst_nXXX), flst_XXXrhptrs(maxreshists,flst_nXXX),
\end{alltt}
{\noindent}where \tmverbatim{XXX} stands for either of \tmverbatim{alr},
\tmverbatim{regular} or \tmverbatim{born}. The integer \
\tmverbatim{flst\_XXXnumrhptrs(iXXX)} stores the number of resonance histories
associated with the \tmverbatim{iXXX} full flavour structure. The integers
\tmverbatim{flst\_XXXrhptrs(1:flst\_XXXnumrhptrs(iXXX),iXXX)}, in case
\tmverbatim{XXX} is either \tmverbatim{alr} or \tmverbatim{regular}, are
indices in the arrays
\begin{alltt}

flst_allrhlength(maxreshists),
flst_allrh(nlegreal,maxreshists), 
flst_allrhres(nlegreal,maxreshists),

\end{alltt}
{\noindent}that represent the full flavour structure of the competing
resonance histories in the real graphs, and in case \tmverbatim{XXX} is
\tmverbatim{born}, are indices in the arrays
\begin{alltt}

flst_allbornrhlength(maxreshists),
flst_allbornrh(nlegborn,maxreshists), 
flst_allbornrhres(nlegborn,maxreshists),

\end{alltt}
{\noindent}representing the full flavour structure of the competing resonance
histories for the Born graphs. We stress that we cannot use
\tmverbatim{flst\_real} and \tmverbatim{flst\_born} in place of
\tmverbatim{flst\_allrh} and \tmverbatim{flst\_allbornrh}, because in the
latter also configurations that differ by a permutation of the intermediate
resonances are included. The \tmverbatim{rh} arrays described above are filled
by the subroutine \tmverbatim{fill\_res\_histories,} in the
\tmverbatim{fill\_res\_histories.f} file, that also computes the multiplicity
factor associated with alternative resonance histories that differ only by a
permutation of the resonances from the contribution being considered. These
arrays are required for the computation of the weights needed to project out a
given resonance history contribution from the real and Born amplitudes.
Besides these, in the case of the \tmverbatim{alr} contributions, we also need
to know the singular regions associated with a given resonance history. This
information is contained in the arrays
\begin{alltt}

flst_allrhnumreg(maxreshists),
flst_allrhreg(1:2,maxregions,maxreshists).

\end{alltt}
{\noindent}For each entry of the \tmverbatim{realrh} arrays,
\tmverbatim{flst\_allrhnumreg} gives the number of singular regions, and
\tmverbatim{flst\_allrhreg} gives the indices of the two particles becoming
collinear in the corresponding singular region.

\section{The example of single top, $t$-channel production}

We consider the Born level process
$b q \rightarrow b e^+ \nu_e q'$. In the
following we will label for conciseness as $q$ and $q'$ all light quarks
or antiquarks (excluding the $b$) with the appropriate flavour structure
that can appear in the
process, and we imply also the presence of the corresponding processes with
exchanged initial state particles (i.e. $q b$ in this case).
This process is dominated by the single top production process
$b q \rightarrow (t \rightarrow b (W^+ \rightarrow e^+ \nu_e)) q'$,
such that the top quark is not produced at rest in 
the partonic centre-of-mass. Therefore this process is relevant for testing
our formalism. In fact, the standard momentum
mapping leading to the underlying Born configuration, in case of collinear
radiation from the $b$ quark arising from top decay, would conserve the
incoming partons 4-momentum by adjusting the 3-momenta of the $b$ and the $W^+ q'$
systems with appropriate boosts. This procedure would preserve the mass of the
$W^+ q'$ system, but not the mass of the top.

At the Born level, it is enough to consider a single
resonance history, namely
$b q \rightarrow (t \rightarrow b (W^+ \rightarrow e^+ \nu_e)) q'$.
Alternatively, one may consider two different resonance histories:
\begin{eqnarray}
  b q & \rightarrow & (t \rightarrow b (W^+ \rightarrow e^+ \nu_e)) q', \\
  b q & \rightarrow & b (W^+ \rightarrow e^+ \nu_e) q' . 
  \label{eq:secondhist}
\end{eqnarray}
The second one is actually not needed, since treating the $b W^+$ system as a
resonance (i.e. preserving its mass in the underling Born mapping) rather than
preserving the mass of the $W^+ q'$ system, as the \tmverbatim{POWHEG BOX}
would do for the resonance history of eq.~(\ref{eq:secondhist}), does not lead
to any inaccuracy. We did however include this resonance assignment as
an option, and used it to test that our setup works also in the case when
more than one resonance history is present at the Born level.

We are considering the following real processes:
$b q\to e^+  \nu_e q' g$ and $b g \to e^+  \nu_e q q'$. We do not consider
processes of the form $q g \to  b e^+  \nu_e q' \bar{b}$, that include
also $s$-channel contributions. This is adequate for the purposes
of the present paper, where we would like to present and validate a method,
rather then provide a realistic simulation of single top production.

We will now list the resonance histories for the real contributions
corresponding to the choice of a single resonance history at the Born level:
\begin{eqnarray}
  b q & \rightarrow & g (t \rightarrow b (W^+ \rightarrow e^+ \nu_e)) q', \\
  b q & \rightarrow & (t \rightarrow b g (W^+ \rightarrow e^+ \nu_e)) q', \\
  b g & \rightarrow & (t \rightarrow b g (W^+ \rightarrow e^+ \nu_e)) q q', \\
  b g & \rightarrow & b (Z / \gamma \rightarrow (W^+ \rightarrow e^+ \nu_e)
  (W^- \rightarrow q q'))  \label{eq:regular-bsing}\\
  b g & \rightarrow & (t \rightarrow b g (W^+ \rightarrow e^+ \nu_e))  (W^-
  \rightarrow q q') 
\end{eqnarray}
Notice that the last two processes are really regular ones, since for them no
collinear singularity can arise by pairing particles belonging to the same
resonance.

The Born (together with the colour correlated Born) and real matrix elements
for the process were easily generated using the \tmverbatim{MadGraph4-POWHEG}
interface described in ref.~{\cite{Campbell:2012am}}. The virtual contribution
was extracted by hand from code
generated using \verb!MadGraph5_aMC@NLO!~\cite{Alwall:2014hca}.

\subsection{Test at the NLO level}
We first tested our method (that we refer to as \verb!POWHEG-BOX-RES!)
by comparing its NLO level results
with a (traditional) \verb!POWHEG-BOX-V2! implementation of the same
process. More specifically, we implemented the $bq\to b e^+ \nu+e q'$ process
in the \verb!POWHEG-BOX-V2! framework, without the inclusion of any resonance information,
and using exactly the same matrix elements used with the new method.
 The comparison was carried out by using exactly the same choice of
scales and parton density functions.

We observe that the process~(\ref{eq:regular-bsing}) has initial state collinear
singularities, due to a contribution arising
from an initial state $g \rightarrow b \bar{b}$ splitting followed
by the subprocess $b \bar{b} \rightarrow (Z / \gamma \rightarrow (W^+
\rightarrow e^+ \nu_e) (W^- \rightarrow q q'))$. This process is not
subtracted in our procedure, since we do not have a corresponding underlying
Born contribution. In order to avoid the associated collinear
divergence, and only in the framework of the NLO tests that
we are discussing in this section,
we supplied our cross section with a damping factor of the form
\begin{equation} \label{eq:damping}
\frac{p_{t,b}^2}{p_{t,b}^2 + 20}\;,
\end{equation}
that damps low transverse momentum $b$ emissions. It is applied
to all terms of the cross section, as a function of the corresponding $b$ quark
kinematics.
We stress that this damping factor is not used when doing phenomenological calculations.
It is needed here only to guarantee that the NLO cross sections computed with the traditional
\verb!POWHEG BOX! implementation is finite and agrees formally with the one
computed with the new method.

In order to perform the NLO test we did not need the virtual contributions, since
they are the same in the two approaches. In the "traditional" implementation
the Born phase space was generated with importance sampling on the dominant
decay chain $b q \rightarrow (t \rightarrow b (W^+ \rightarrow e^+ \nu_e))
q'$. It was found that the new implementation yielded better convergence,
speeding up the calculation by about a factor of 2 or more. Very high
statistics runs of both implementations were performed, and full numerical
agreement was found for both the total cross section and the differential
distributions.

\subsection{Results and comparisons at the full shower level}
As mentioned earlier, the process we are considering is singular when final
state $b$ quarks have very small transverse momenta. Thus, event generation
requires a generation cut or a Born suppression factor~\cite{Alioli:2010qp}.
We adopt the latter method in the present work, using a suppression factor of
the form
\begin{equation}
\frac{p_{t,b}^2}{p_{t,b}^2 + 20}\;.
\end{equation}
As a result, less and less events are generated as the $b$ transverse momentum becomes
small, but the event weight is increased correspondingly, thus yielding a potentially
divergent cross section for observables that do not suppress the contribution of
small transverse momentum $b$ quarks.

The shower was performed using \verb!Pythia8! version 81.85
\cite{Sjostrand:2014zea,Sjostrand:2007gs,Sjostrand:2006za}.
In all cases, \verb!Pythia8! was run with its default initialization, supplemented
with the following calls:
\begin{verbatim}
    pythia.readString("SpaceShower:pTmaxMatch = 1");
    pythia.readString("TimeShower:pTmaxMatch = 1");

    pythia.readString("SpaceShower:QEDshowerByQ = off"); // From quarks.        
    pythia.readString("SpaceShower:QEDshowerByL = off"); // From Leptons.       
    pythia.readString("TimeShower:QEDshowerByQ = off"); // From quarks.         
    pythia.readString("TimeShower:QEDshowerByL = off"); // From Leptons.

    pythia.readString("PartonLevel:MPI = off");
\end{verbatim}
The first two calls cause \verb!Pythia8! to veto emissions harder than \verb!scalup!, if arising
at production level, and to allow unrestricted emissions from resonance decays.
Furthermore, $b$ hadron decays were switched off with calls of the following kind
\begin{verbatim}
  pythia.readString("521:mayDecay = off");
  pythia.readString("-521:mayDecay = off");
\end{verbatim}
for all $b$ flavoured mesons and baryons.

In order to test our generator, we generated four samples of one
million of events each, and compared the relative output. The
samples are obtained as follows:
\begin{itemize}
\item \verb!NORES! Sample. This is obtained using the traditional \verb!POWHEG-BOX-V2!
implementation for the process.
The events are fed to \verb!Pythia8!, with the setting listed earlier.
\verb!Pythia8! is required to veto radiations at scales harder than the value of
the Les Houches variable \verb!scalup!, set equal to the
transverse momentum of the \verb!POWHEG! generated radiation for each Les Houches event.
\item \verb!RES-HR! Sample. This sample is obtained using the
\verb!POWHEG-BOX-RES! implementation of the process.
The Les Houches events include the hardest radiation generated by \verb!POWHEG!
(the \verb!HR! in \verb!RES-HR! stands for ``hardest radiation'').
Since \verb!POWHEG! is generating the hardest radiation,
besides vetoing radiation in production with the usual \verb!scalup! mechanism,
we also forbid any
\verb!Pythia8! radiation from top decays harder than \verb!scalup!.
We do this by explicitly examining the showered events.
If a radiation generated by \verb!Pythia8! in top decay has a transverse
momentum greater than \verb!scalup!, the program discards it, and runs \verb!Pythia8! again
on the same Les Houches partonic event. This procedure is repeated indefinitely,
thus explicitly vetoing any event with radiation harder than \verb!scalup!.
\item \verb!RES-AR! Sample.
This sample is obtained using the \verb!POWHEG-BOX-RES! implementation of the process.
However, rather than keeping only the hardest radiation, we kept both
the hardest radiation in top decay and the hardest radiation in
production (the \verb!AR! in \verb!RES-AR!  stands for ``all
radiation'') . These radiations are composed into a single event,
using the usual \verb!POWHEG! mapping mechanism.  In this case, besides
the normal \verb!scalup! veto for radiation {\em in production}, we must
forbid \verb!Pythia8! radiation in top decays if harder than the \verb!POWHEG!
generated one {\em in top decay}. There are thus two different veto scales, one for production
(i.e. the \verb!scalup! value) and one for decay. There are no provisions in the
Les Houches Interface for User Processes to store the radiation scale for decaying
resonances. The program thus computes explicitly the transverse momentum of radiation
in top decay at the Les Houches level. If the showered event contains shower generated radiation
in top decay harder than this computed scale, the shower is discarded, and a
new shower is generated on the same Les Houches event, repeating the procedure
indefinitely until the event passes the required condition.

The \verb!RES-AR! implementation is fully analogous to the \verb!allrad! procedure
illustrated in ref.~\cite{Campbell:2014kua}. The method (and the software)
for vetoing  \verb!Pythia8! radiation is also borrowed from that reference.
\item \verb!ST-tch! Sample. This sample is generated using the \verb!ST-tch!
\verb!POWHEG! generator of ref.~\cite{Alioli:2009je}.
Radiation in decay is not included in this generator, and thus we let \verb!Pythia8! shower the event
according to its default setup, vetoing events with radiation in production harder
than \verb!scalup!, and with no veto in top decays.
In order to match more closely what we include in \verb!RES-HR!, we deleted in
the \verb!ST-tch!
the real processes initiated by a light (i.e. not $b$) quark and a gluon.
\end{itemize}

\subsection{Phenomenological analysis}
We have considered the LHC 8 TeV configuration for our phenomenological
runs. We have used throughout the MSTW2008 set~\cite{Martin:2009iq} at
NLO order. Other PDF sets, like those of
refs.~\cite{Lai:2010vv,Ball:2010de}, can be used as well, but we are
not interested in a PDF comparison in the present study. We only consider
the $b\, \mu^+ \nu_\mu$ final state (i.e. not the conjugate one). 

We set the top mass to $m_t=172.5$~GeV. For this value of the mass and PDF choice
the computed top width, including NLO strong corrections, is $1.3306$~GeV.
We use the same NLO value of the width also in the \verb!ST-tch! generator.
In this generator it only affects the top line-shape,
since the cross section is determined by the top cross section multiplied by
a user supplied branching fraction.
On the other hand, in our generator the width must be computed with the same Standard
Model parameters that are used in the matrix elements, since the cross section
will be proportional to a partial width (depending upon the couplings that are
used in the matrix elements) divided by the total width that appears in the denominator
of the top propagator.

Since we will compare generators that do not include top resonance information, our
analysis will be performed (unless explicitly stated otherwise)
without using ``Monte Carlo truth'' information
as far as the top particle is concerned,
thus relying solely upon a particle level reconstruction
of the top kinematics. We thus define the following objects:
\begin{itemize}
\item The lepton. This is the hardest $\mu^+$ in the event.
\item The neutrino. This is the hardest $\nu_\mu$ in the event.
\item The $W^+$. This is the system formed by the lepton and the neutrino.
\item The $b$ hadron. This is the hardest hadron with a $b$ quark content (not a $\bar{b}$!).
\item The $b$-jet. Jets are reconstructed using the anti-$k_t$ algorithm~\cite{Cacciari:2008gp},
as implemented in the \texttt{FastJet} package~\cite{Cacciari:2011ma}, with $R=0.5$.
The $b$-jet is the jet that contains the $b$ hadron.
\item The top quark. This is defined as the system comprising the $W$ and the $b$-jet.
      Only $b$-jets with a $p_t$ of at least 25~GeV and $|\eta|<4.5$ are accepted for
      this purpose. In case such a $b$-jet is not found, no top is reconstructed.
\end{itemize}
\subsection{{\tt RES-AR} and {\tt ST-tch} comparison}
\begin{figure}[t]
\begin{center}
\includegraphics[width=0.48\textwidth]{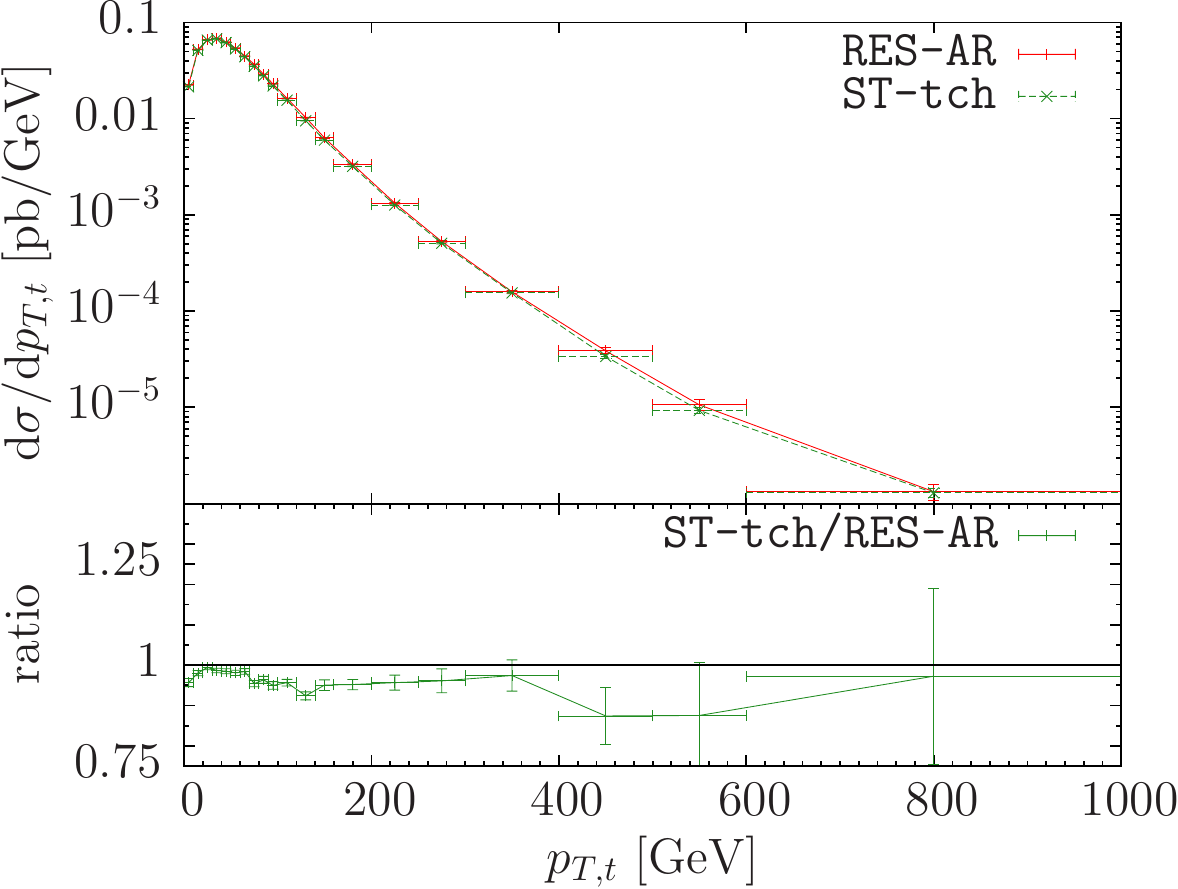}
\includegraphics[width=0.48\textwidth]{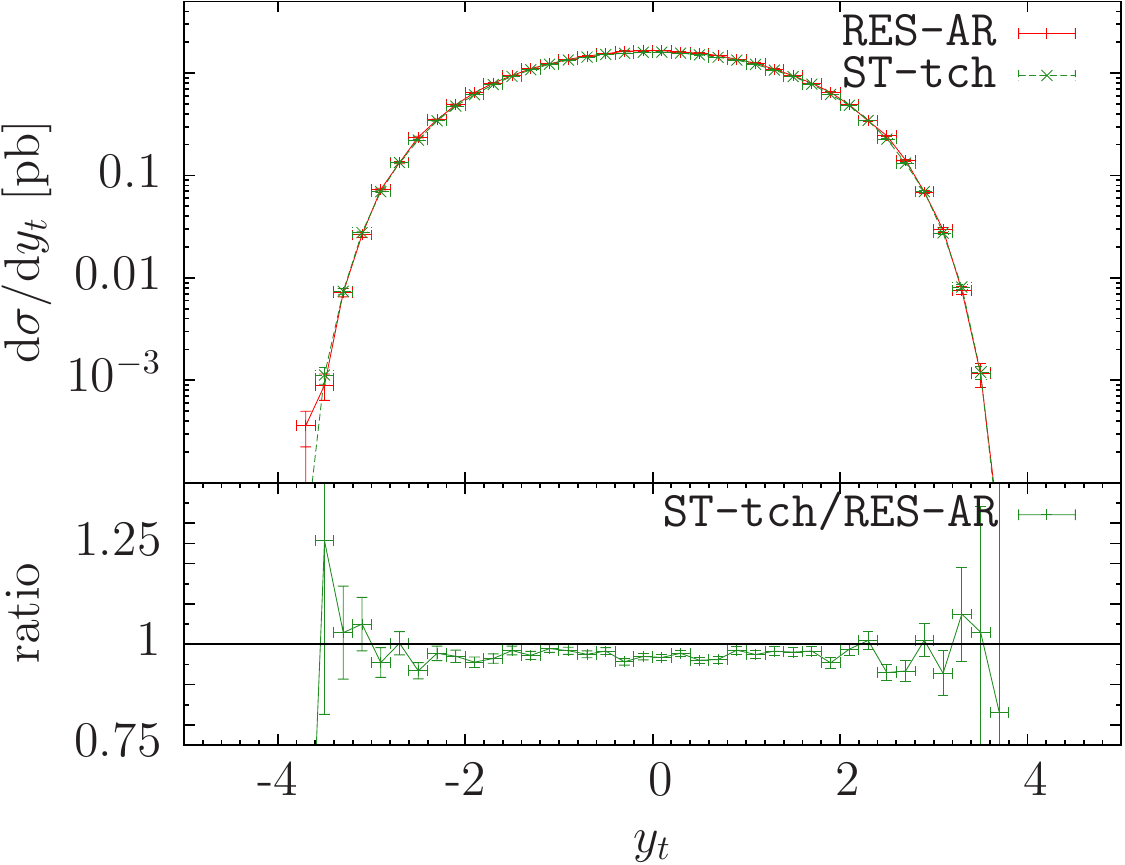}
\caption{\label{fig:Top-pt-RES-AR-ST} Transverse momentum distribution of the
top quark, obtained with the {\tt RES-AR} and the {\tt ST-tch} generators.}
\end{center}
\end{figure}

\begin{figure}[t]
\begin{center}
\includegraphics[width=0.48\textwidth]{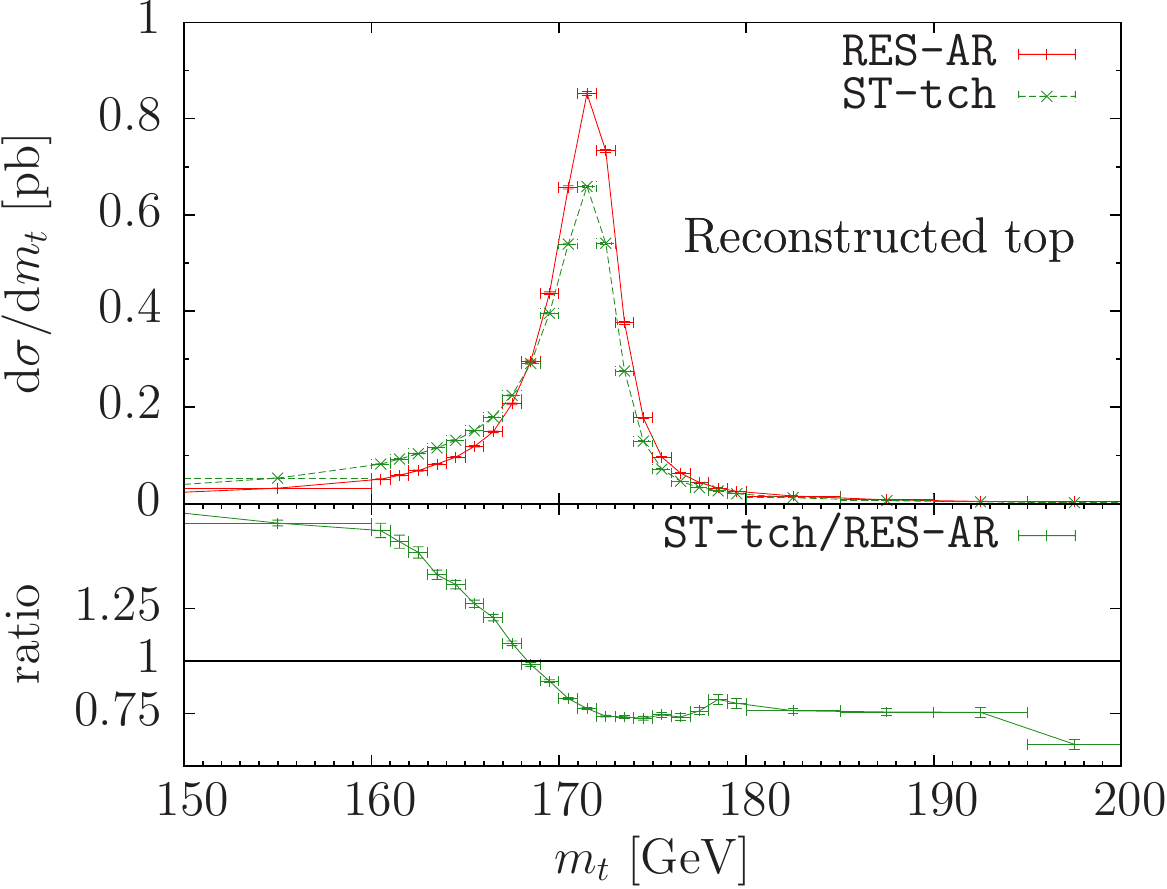}
\includegraphics[width=0.48\textwidth]{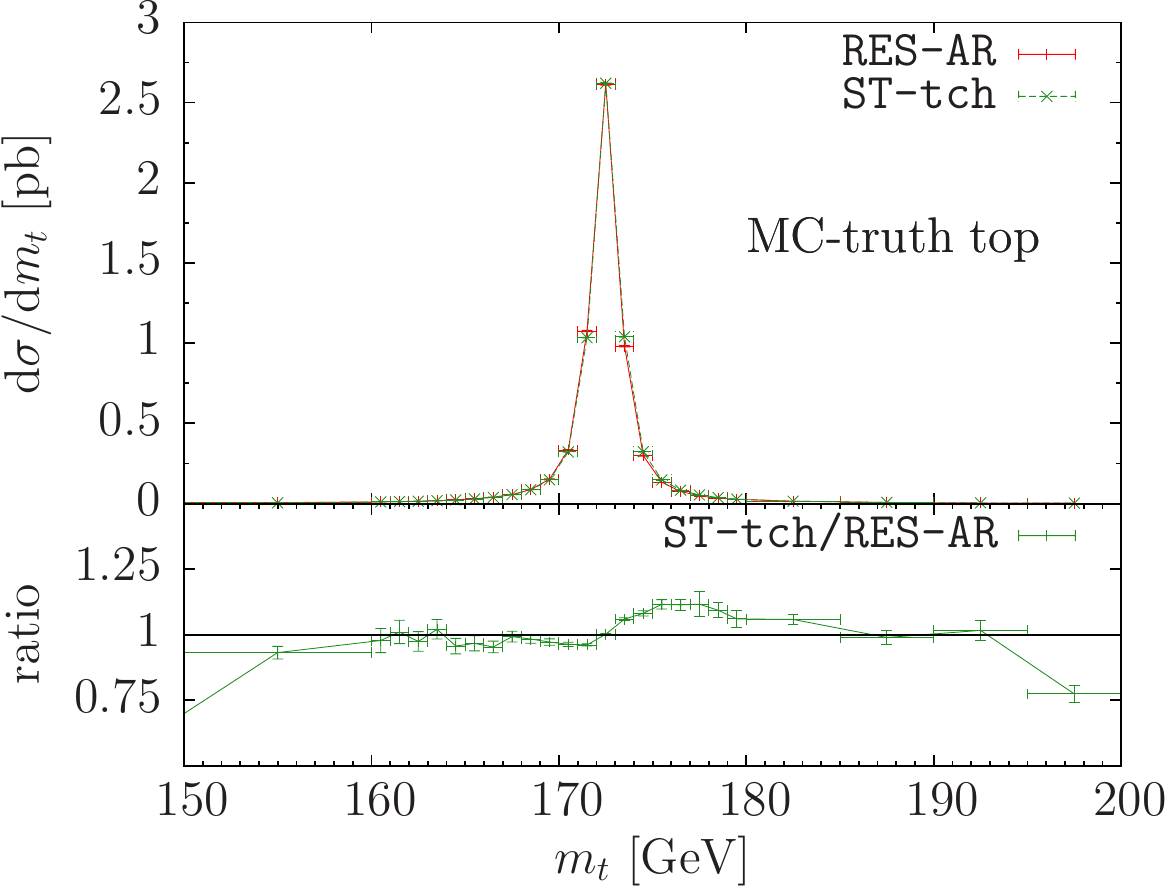}
\caption{\label{fig:Top-mass-RES-AR-ST} Invariant mass of the
top quark, obtained with the {\tt RES-AR} and the {\tt ST-tch} generators, at the
reconstructed level and at the MC-truth level.}
\end{center}
\end{figure}

\begin{figure}[t]
\begin{center}
\includegraphics[width=0.48\textwidth]{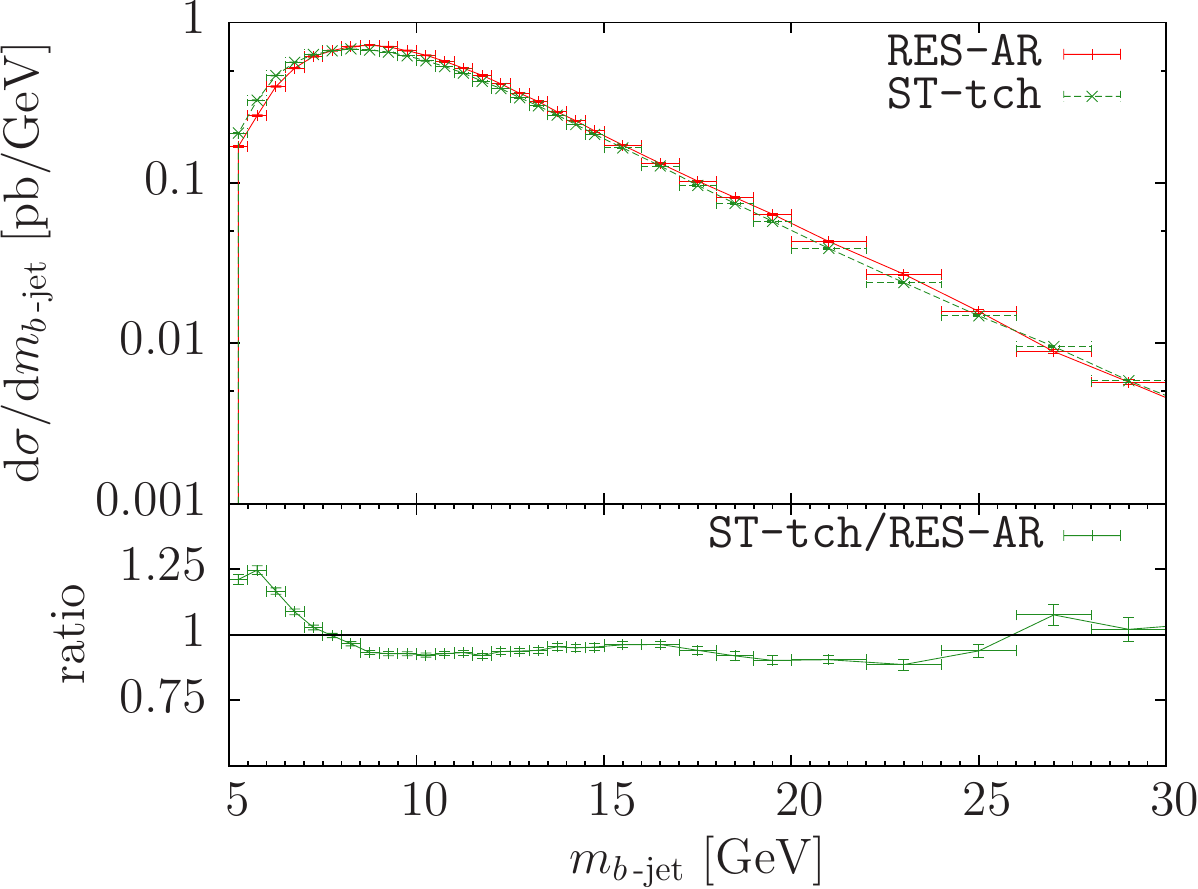}
\includegraphics[width=0.48\textwidth]{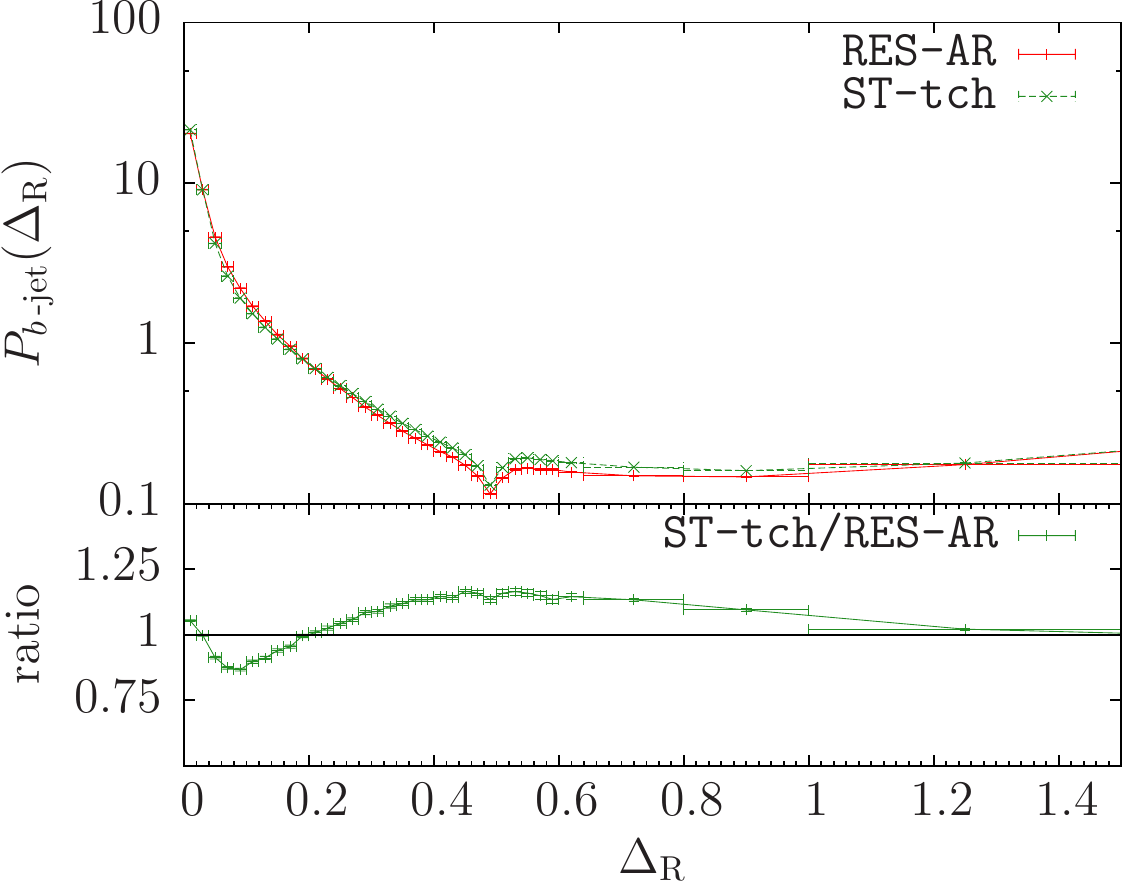}
\caption{\label{fig:bjet-RES-AR-ST} Mass and profile of the $b$ jet,
obtained with the {\tt RES-AR} and the {\tt ST-tch} generators.}
\end{center}
\end{figure}

\begin{figure}[t]
\begin{center}
\includegraphics[width=0.48\textwidth]{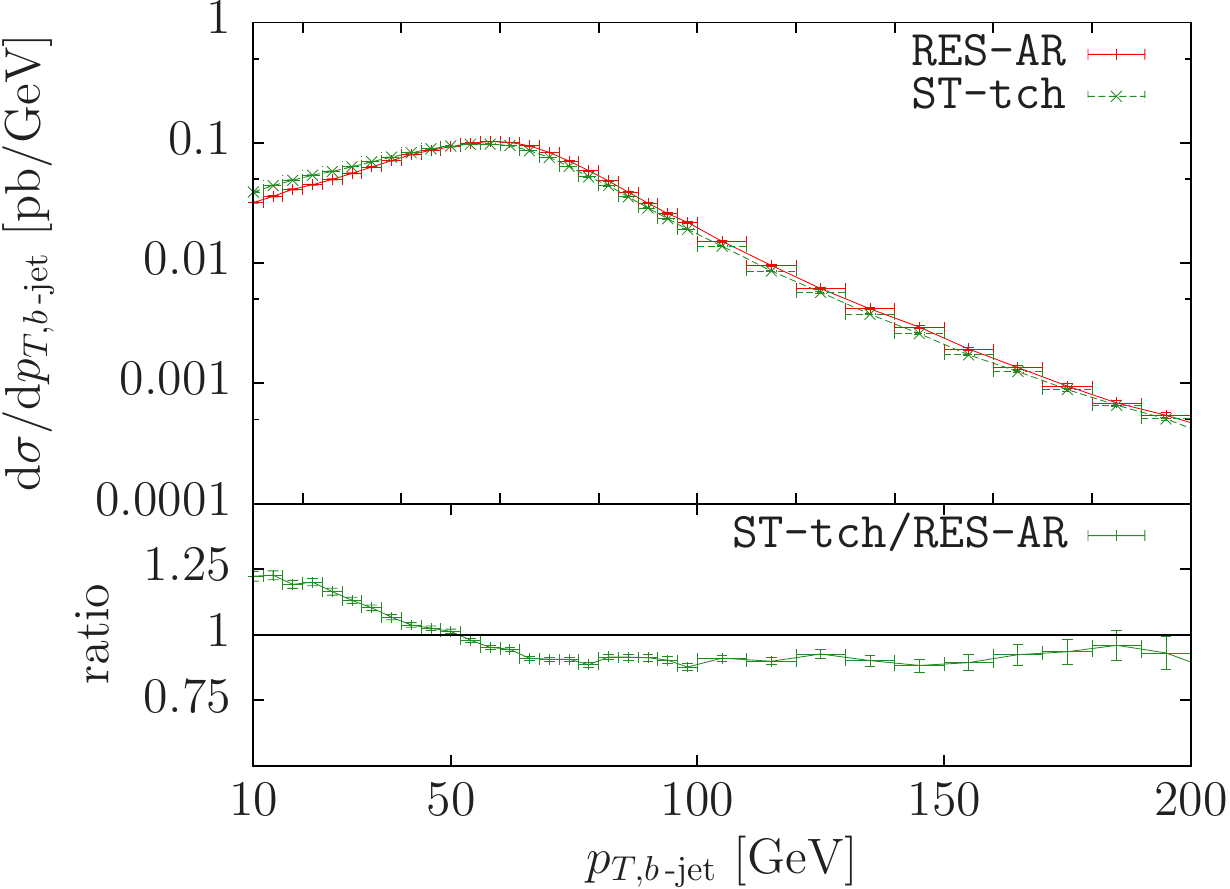}
\caption{\label{fig:bjet-pt-RES-AR-ST} Transverse momentum of the $b$ jet,
obtained with the {\tt RES-AR} and the {\tt ST-tch} generators.}
\end{center}
\end{figure}

We begin by comparing the \verb!RES-AR! and the \verb!ST-tch! results.
For this purpose (and only in this case) we have
deleted from the  \verb!RES-AR! generator the real amplitudes with the final non-$b$ light partons
in a colour singlet. This excludes in particular contributions with $tW^-$ associated production,
leading to a more meaningful comparison and to a better agreement on the total cross
sections,  since these contributions are not included in the \verb!ST-tch! generator.
In fig.~\ref{fig:Top-pt-RES-AR-ST}
we plot the transverse momentum and the rapidity distributions of the reconstructed top.
In these plots no requirement is made on the mass of the reconstructed top.
As one can see, reasonable agreement is found with these distributions.
In fig.~\ref{fig:Top-mass-RES-AR-ST}
we show the mass peak, both for the reconstructed top and for the top particle in the Monte Carlo
record (more specifically, we pick the last top in the Monte Carlo event record). It is apparent that
the line-shape of the reconstructed top are not in complete agreement. Assuming that this is due to
differences in the structure of the $b$-jet, we plot in fig.~\ref{fig:bjet-RES-AR-ST}
the $b$-jet mass and profile, for $b$-jets with transverse momentum above $15$~GeV and $|\eta|<4.5$.
The jet profile is defined as follows
\begin{equation}
P_{b\operatorname{-jet}}(\Delta_{\mathrm R})= N \int {\mathrm d}\sigma\;\frac{\sum_j p_{T,j} \times \delta(\Delta_{\mathrm R}^{(j,b\operatorname{-jet})}-\Delta_{\mathrm R})}%
{p_{T,b\operatorname{-jet}}},
\end{equation}
where $N$ is chosen in such a way that
\begin{equation}
\int_0^{0.5} P_{b\operatorname{-jet}}(\Delta_{\mathrm R})  \mathd \Delta_{\mathrm R} = 1\,,
\end{equation}
where $0.5$ is the $\Delta_{\mathrm R}$ value that defines the $b$-jet.
Thus, for $\Delta_{\mathrm R}<0.5$, $P_{b\operatorname{-jet}}(\Delta_{\mathrm R})$ is the fractional distribution
of the transverse momentum in the jet. 
In fig.~\ref{fig:bjet-pt-RES-AR-ST}
we show the transverse momentum of the $b$-jet.

We see that these plots show consistently that the $b$-jet is harder and more massive in the \verb!RES-AR!
case than in the \verb!ST-tch! one. In particular, the jet profile plot shows that there are more partons
sharing the jet momentum in the region with $\Delta_{\rm R}$ near 0.1 in the \verb!RES-AR! case. In the
\verb!ST-tch! case more momentum is concentrated at very small $\Delta_R$
(presumably due to the $b$ meson), and there
are more partons at larger values of $\Delta_R$, also outside of the jet cone. We interpret these
fact as consistent with the reconstructed top mass peak being slightly shifted to the
right in the \verb!RES-AR! case.

In order to quantify the shift in the mass extraction that one would
get using one or the other Monte Carlo, we define an observable
$M_{\rm trec}$, equal to the average value of the reconstructed top mass in a
window of $\pm 15$~GeV
around $m_t$, in order to mimic the typical experimental resolution on the reconstructed top mass.
We get $M_{\rm trec}=170.54(2)$~GeV for the \verb!RES-AR!, and $M_{\rm trec}=169.59(1)$~GeV for the
\verb!ST-tch! generator.
Thus, extracting the top mass with the \verb!ST-tch! generator we would get a value 1~GeV larger than
if we used the \verb!RES-AR! one.

As a further comment on our findings, we remind the reader that, as far as the reconstructed top line-shape
is concerned, the  \verb!RES-AR! and  \verb!ST-tch! generators differ mainly in the way that radiation
from the $b$ quark is treated. In the \verb!RES-AR! generator the hardest radiation from the $b$ quark
is always handled by \verb!POWHEG!, with \verb!Pythia8! handling the remaining radiation.
In the  \verb!ST-tch! generator, on the other hand, \verb!POWHEG! generates no radiation
from the decaying top. Thus, all radiation from the $b$ quark is handled by \verb!Pythia8!.
We must therefore ascribe the differences that we find to the different treatment of radiation
in \verb!POWHEG! and \verb!Pythia8!.
This issue was also discussed in ref.~\cite{Campbell:2014kua}. In view of its impact on the
top mass determination, this topic deserves a more detailed phenomenological study, that goes
beyond the scope of the present work.

\subsection{{\tt RES-AR} and {\tt RES-HR} comparison}
We now compare the \verb!RES-AR! and \verb!RES-HR! generators. As mentioned earlier, the two generators differ
in the way that the Les Houches record is formed after the stage of generation of radiation.
In the former, the hardest radiation is kept for both production and top decay independently.
So, events with up to two more partons with respect to the Born kinematics are stored in the
Les Houches record, and are passed to \verb!Pythia8! for showering. The shower in production is
limited to the hardness of the radiation in production, while the shower in top decay is limited
by the hardness of radiation in top decay. In the latter, only the hardest radiation of all is kept.
Shower radiation, whether from production or from decay, is limited by the scale of the hardest
radiation in \verb!POWHEG!.

\begin{figure}[t]
\begin{center}
\includegraphics[width=0.48\textwidth]{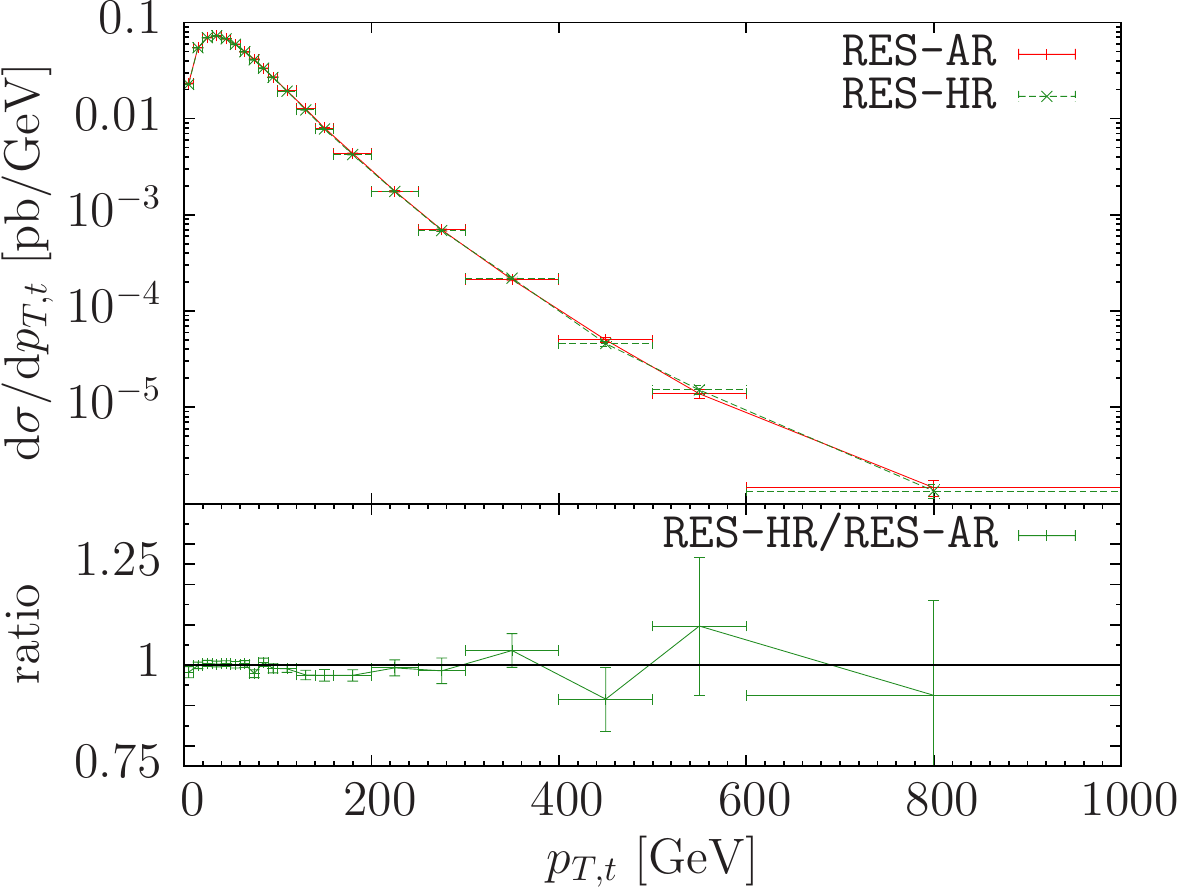}
\includegraphics[width=0.48\textwidth]{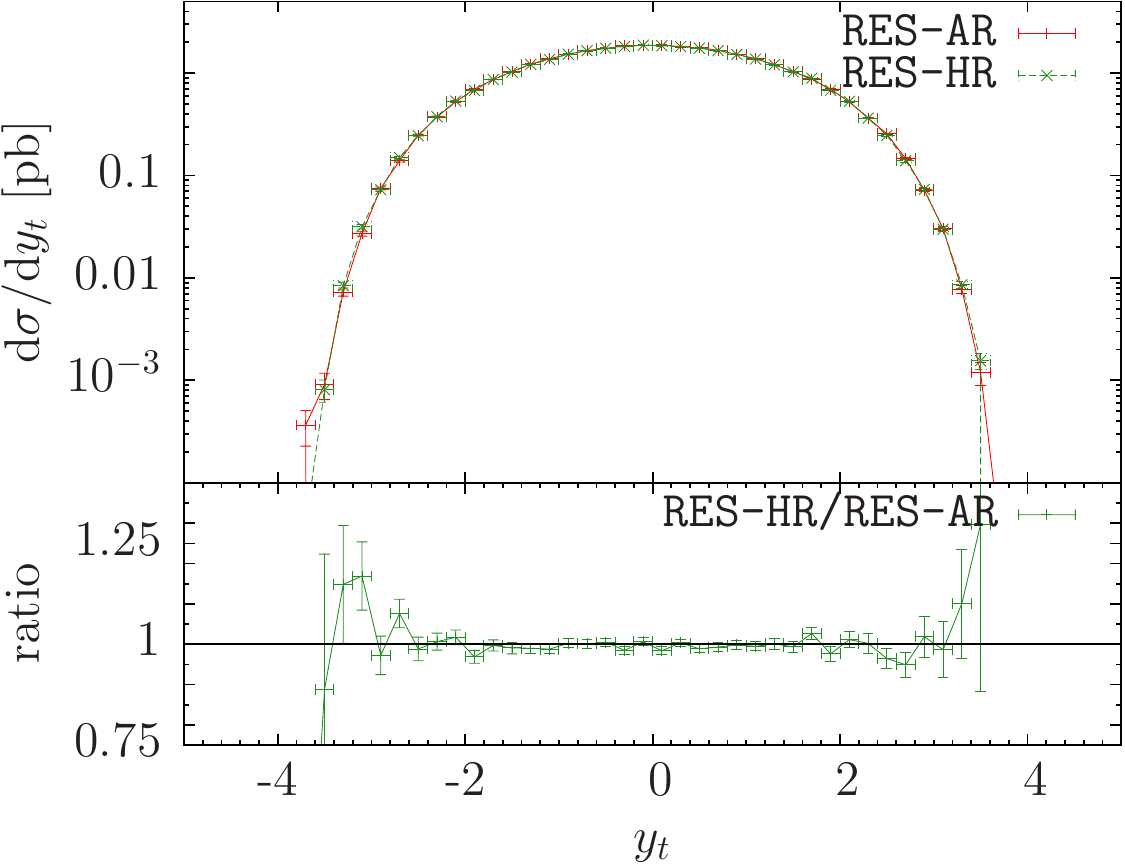}
\caption{\label{fig:Top-pt-RES-AR-RES-HR} Transverse momentum distribution of the
top quark, obtained with the {\tt RES-AR} and the {\tt RES-HR} generators.}
\end{center}
\end{figure}

\begin{figure}[t]
\begin{center}
\includegraphics[width=0.48\textwidth]{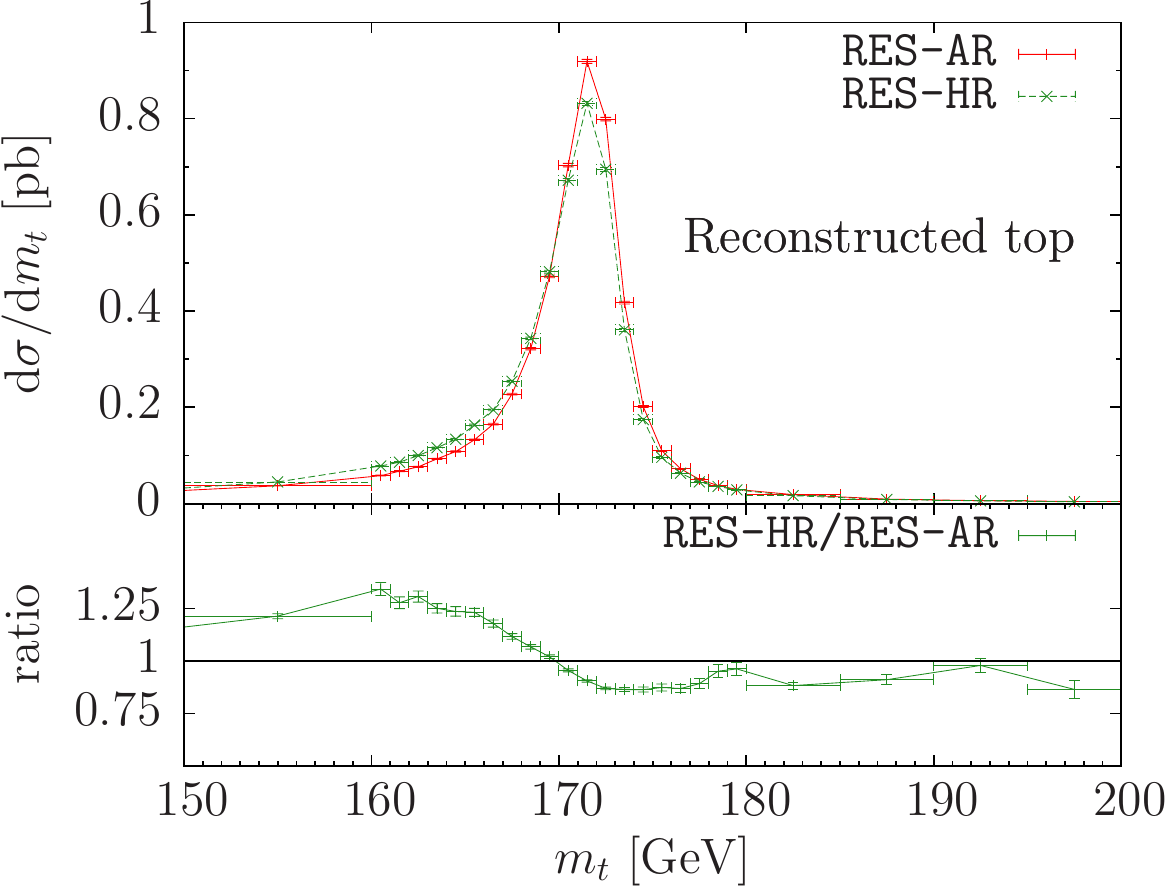}
\includegraphics[width=0.48\textwidth]{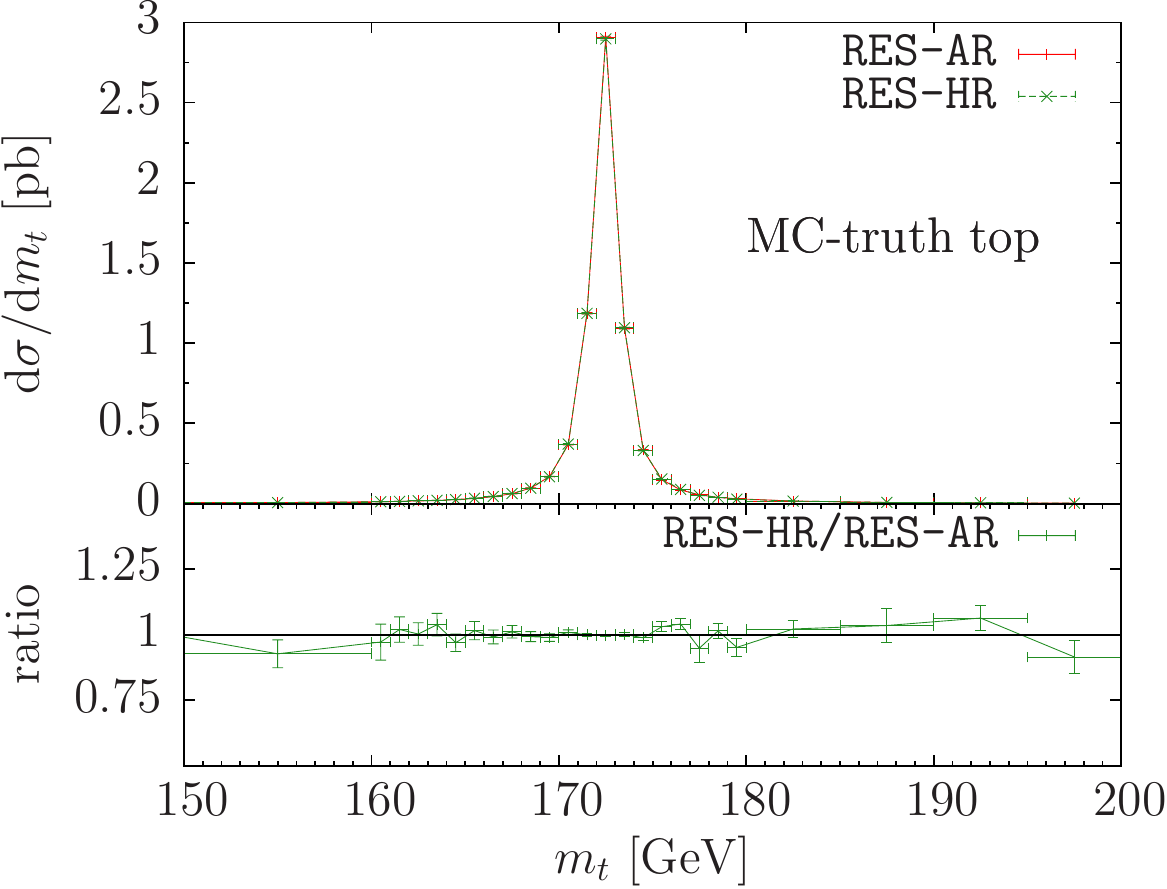}
\caption{\label{fig:Top-mass-RES-AR-RES-HR} Invariant mass of the
top quark, obtained with the {\tt RES-AR} and the {\tt RES-HR} generators, at the
reconstructed level and at the MC-truth level.}
\end{center}
\end{figure}

\begin{figure}[t]
\begin{center}
\includegraphics[width=0.48\textwidth]{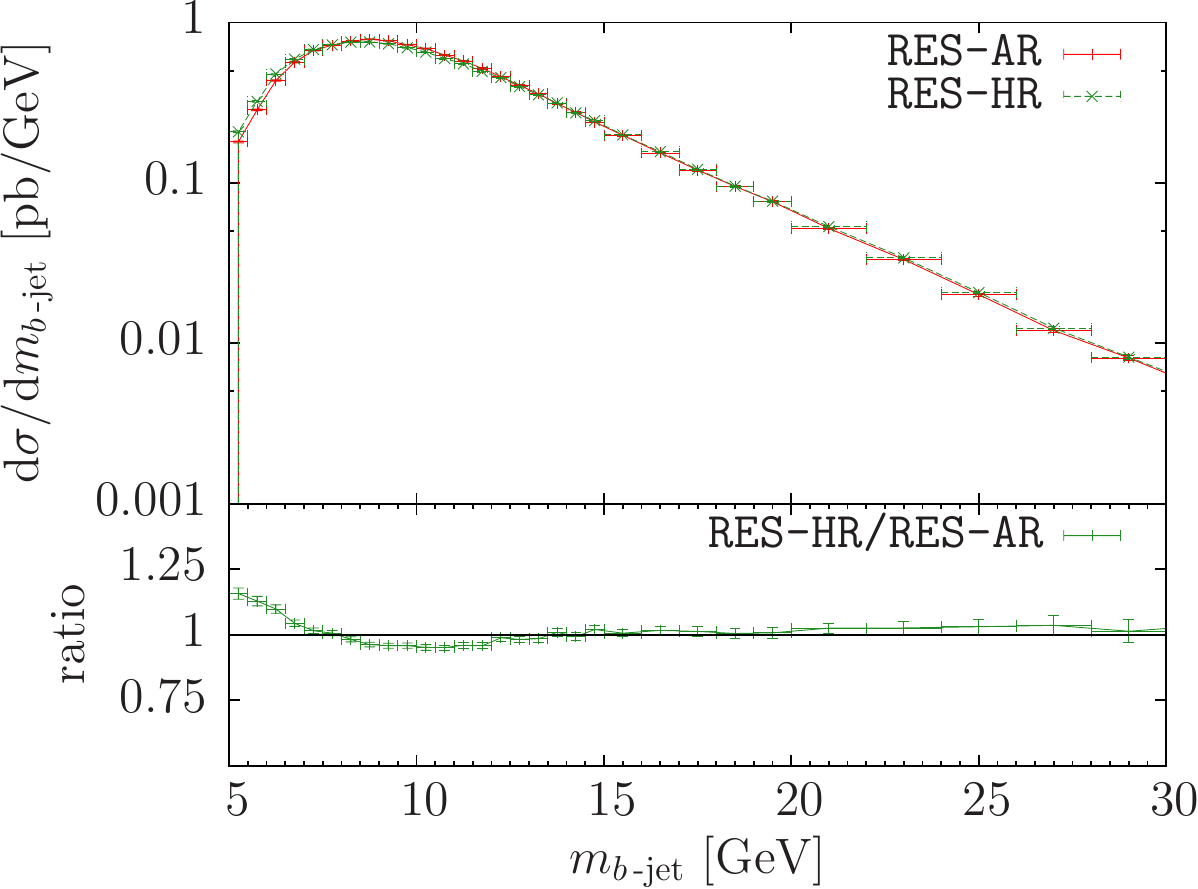}
\includegraphics[width=0.48\textwidth]{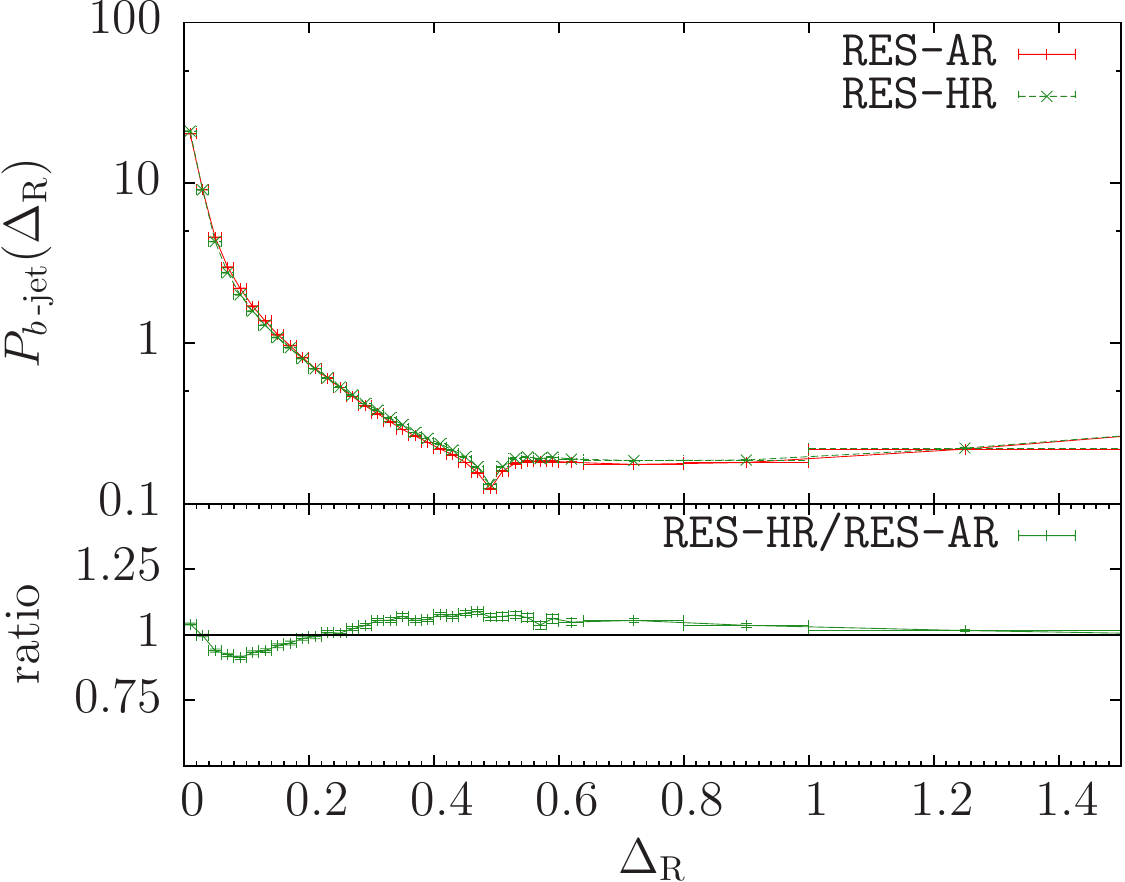}
\caption{\label{fig:bjet-RES-AR-RES-HR} Mass and profile of the $b$-jet,
obtained with the {\tt RES-AR} and the {\tt RES-HR} generators.}
\end{center}
\end{figure}

\begin{figure}[t]
\begin{center}
\includegraphics[width=0.48\textwidth]{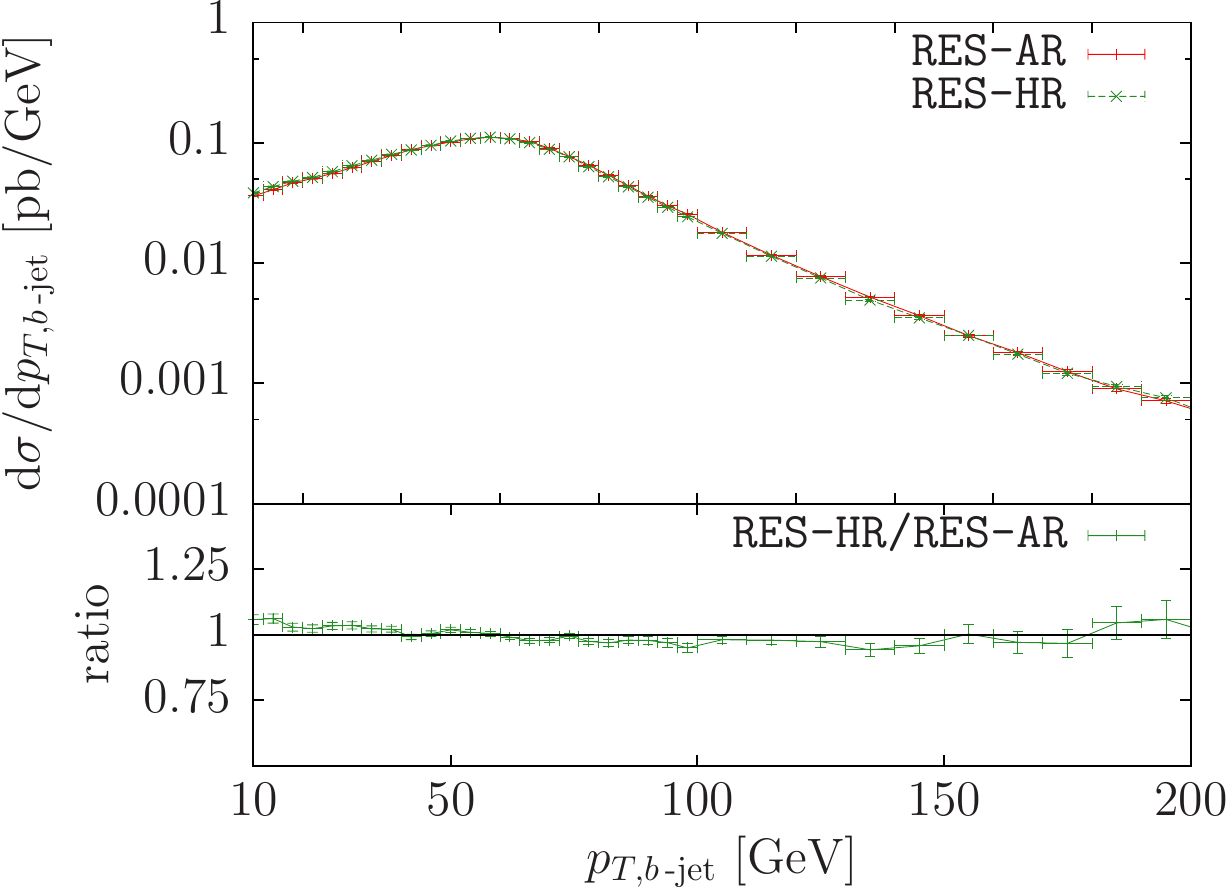}
\caption{\label{fig:bjet-pt-RES-AR-RES-HR} Transverse momentum of the $b$ jet,
obtained with the {\tt RES-AR} and the {\tt RES-HR} generators.}
\end{center}
\end{figure}

In fig.~\ref{fig:Top-pt-RES-AR-RES-HR}
we plot the transverse momentum and the rapidity distributions of the reconstructed top.
In fig.~\ref{fig:Top-mass-RES-AR-RES-HR}
we show the mass peak, both for the reconstructed top and for the top particle in the Monte Carlo
record. In fig.~\ref{fig:bjet-RES-AR-RES-HR}
 we plot the mass and profile of the $b$-jet,
while in fig.~\ref{fig:bjet-pt-RES-AR-RES-HR}
we plot its transverse momentum.

We find, as in the case of the \verb!RES-AR! and \verb!ST-tch! comparison, differences in the 
top lineshape. In this case, however, they are less pronounced. The comparison of observables
related to the $b$-jet also follow a similar pattern. They are qualitatively similar to the
previous case, but less pronounced, in particular for the case of
the $b$-jet transverse momentum distribution. We interpret these findings as being due to the fact
that in the \verb!RES-HR! generator the $b$-jet hardest radiation is in part controlled by \verb!POWHEG!
and in part by \verb!Pythia8!.
We remark that also in the \verb!RES-HR! case \verb!POWHEG! should correct the hardest radiation
from the $b$ quark to yield an NLO accurate result, at least for sufficiently hard radiation.

It is again interesting to quantify the shift in the mass extraction that one would get using one or the other
Monte Carlo. Computing, as before, our $M_{\rm trec}$ observable,
we get $M_{\rm trec}=170.06(3)$~GeV for the \verb!RES-HR! generator,
and $M_{\rm trec}=170.55(2)$~GeV for the \verb!RES-AR!. The small difference in the \verb!RES-AR! result
with respect to the one given in the previous subsection was due to the fact that there a set
of real contributions was left out, as explained earlier.

\subsection{{\tt NORES} and {\tt RES-HR} comparison}
We now compare the \verb!NORES! and \verb!RES-HR! generators. The purpose of this
comparison is to see if and how a generator that is not aware of resonance
structures can exhibit visible distortions.
In this case, MC resonance truth information is not available for the \verb!NORES!
generator. It is possible however to try to guess the resonance information from the
structure of the event, as we will detail later.

\begin{figure}[t]
\begin{center}
\includegraphics[width=0.48\textwidth]{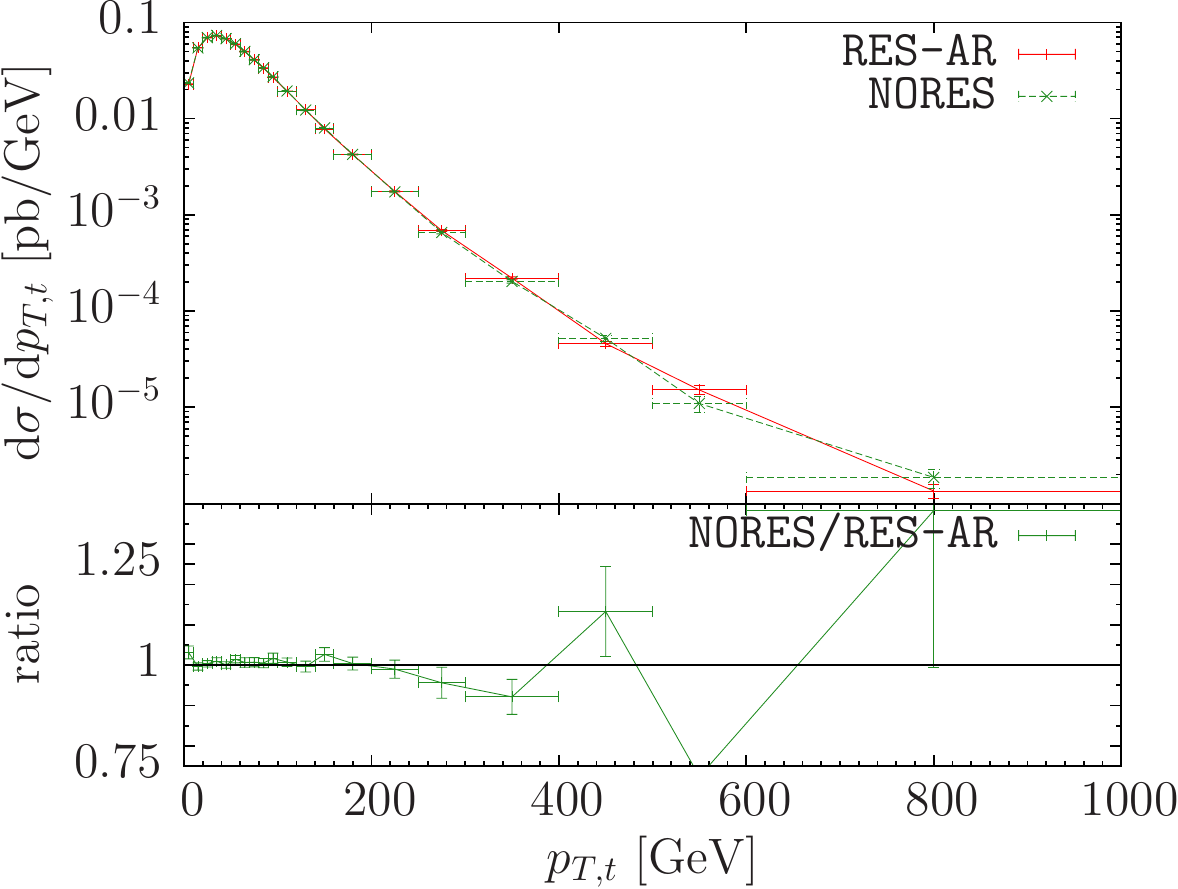}
\includegraphics[width=0.48\textwidth]{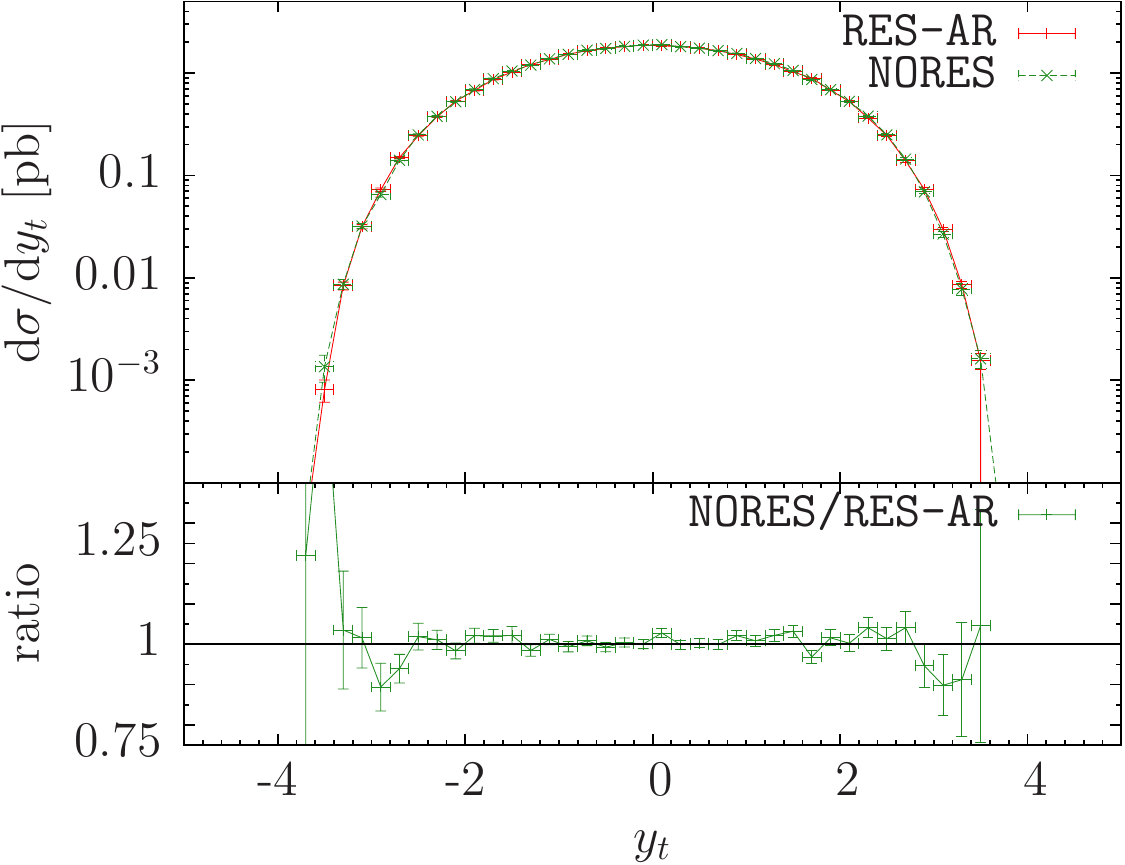}
\caption{\label{fig:Top-pt-NORES-RES-HR} Transverse momentum distribution of the
top quark, obtained with the {\tt NORES} and the {\tt RES-HR} generators.}
\end{center}
\end{figure}

\begin{figure}[t]
\begin{center}
\includegraphics[width=0.48\textwidth]{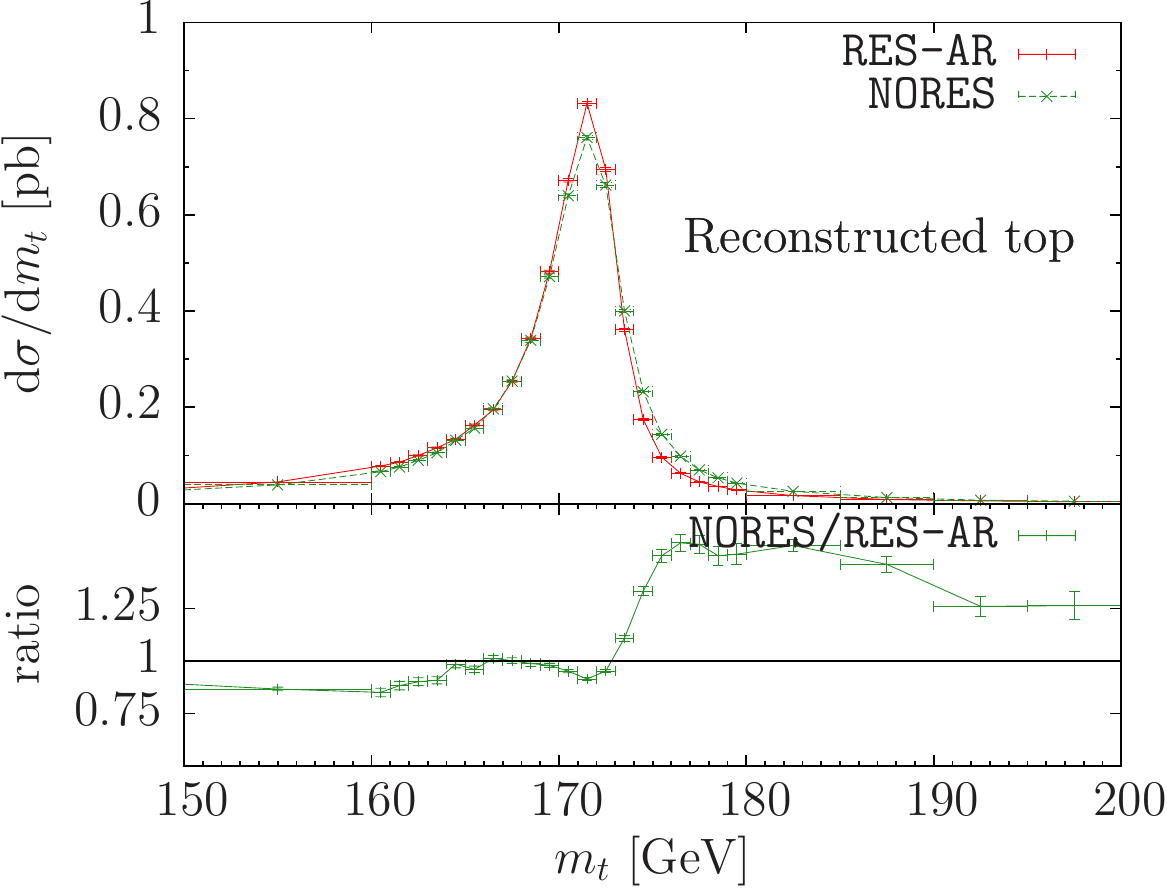}
\caption{\label{fig:Top-mass-NORES-RES-HR} Invariant mass of the
top quark, obtained with the {\tt NORES} and the {\tt RES-HR} generators, at the
reconstructed level.}
\end{center}
\end{figure}

\begin{figure}[t]
\begin{center}
\includegraphics[width=0.48\textwidth]{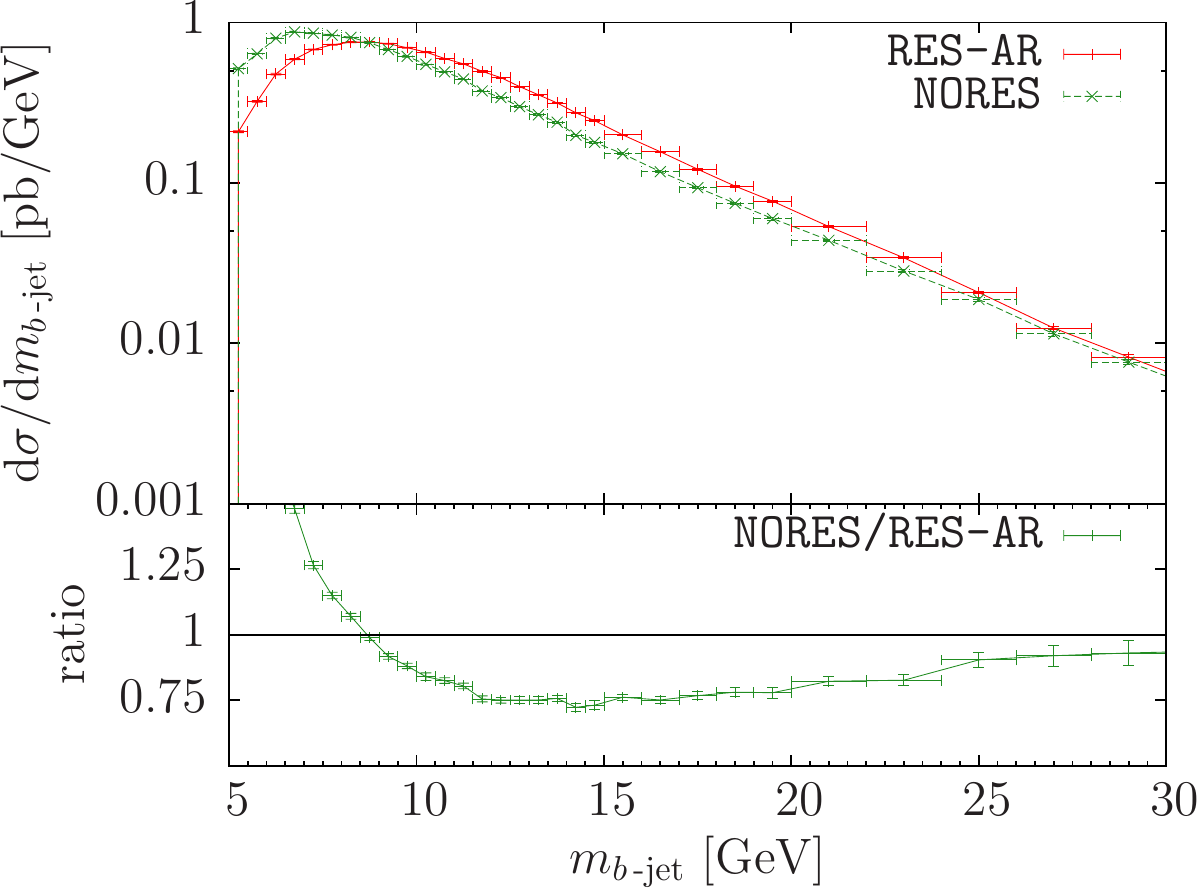}
\includegraphics[width=0.48\textwidth]{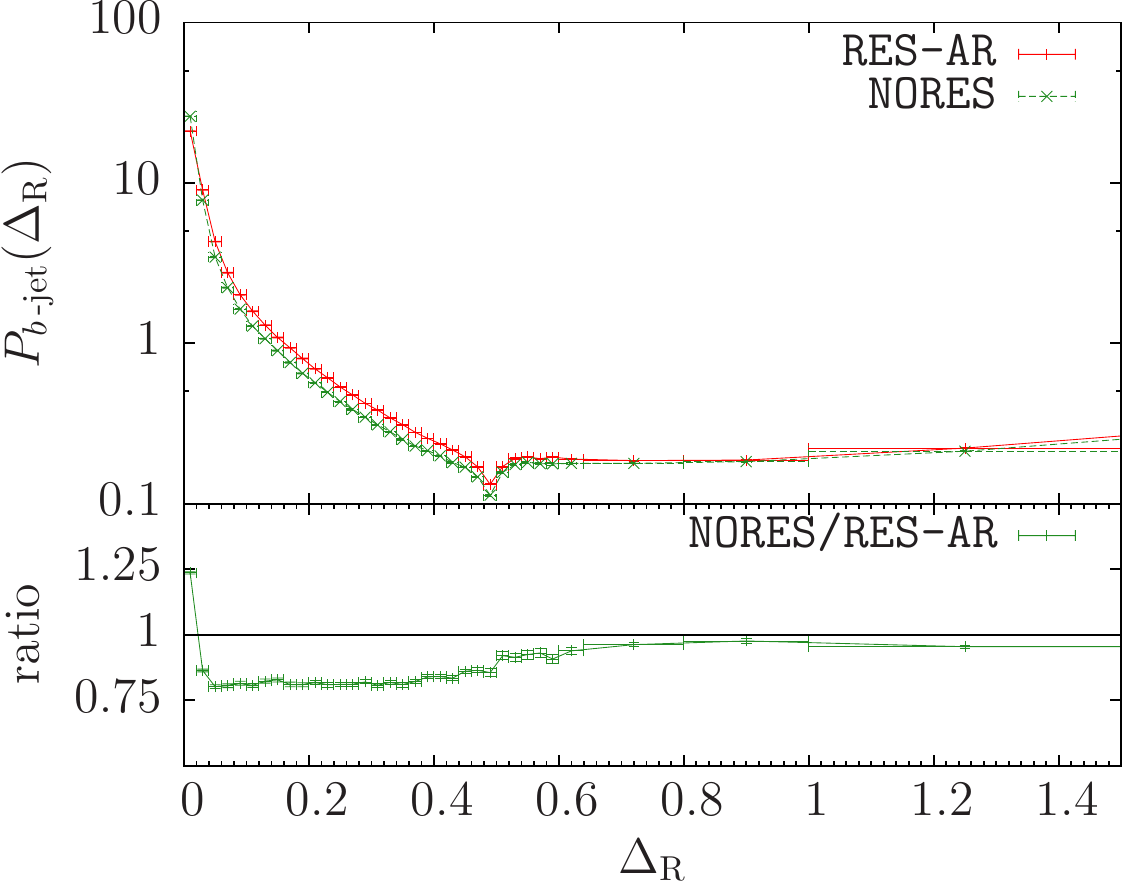}
\caption{\label{fig:bjet-NORES-RES-HR} Mass and profile of the $b$ jet,
obtained with the {\tt NORES} and the {\tt RES-HR} generators.}
\end{center}
\end{figure}

\begin{figure}[t]
\begin{center}
\includegraphics[width=0.48\textwidth]{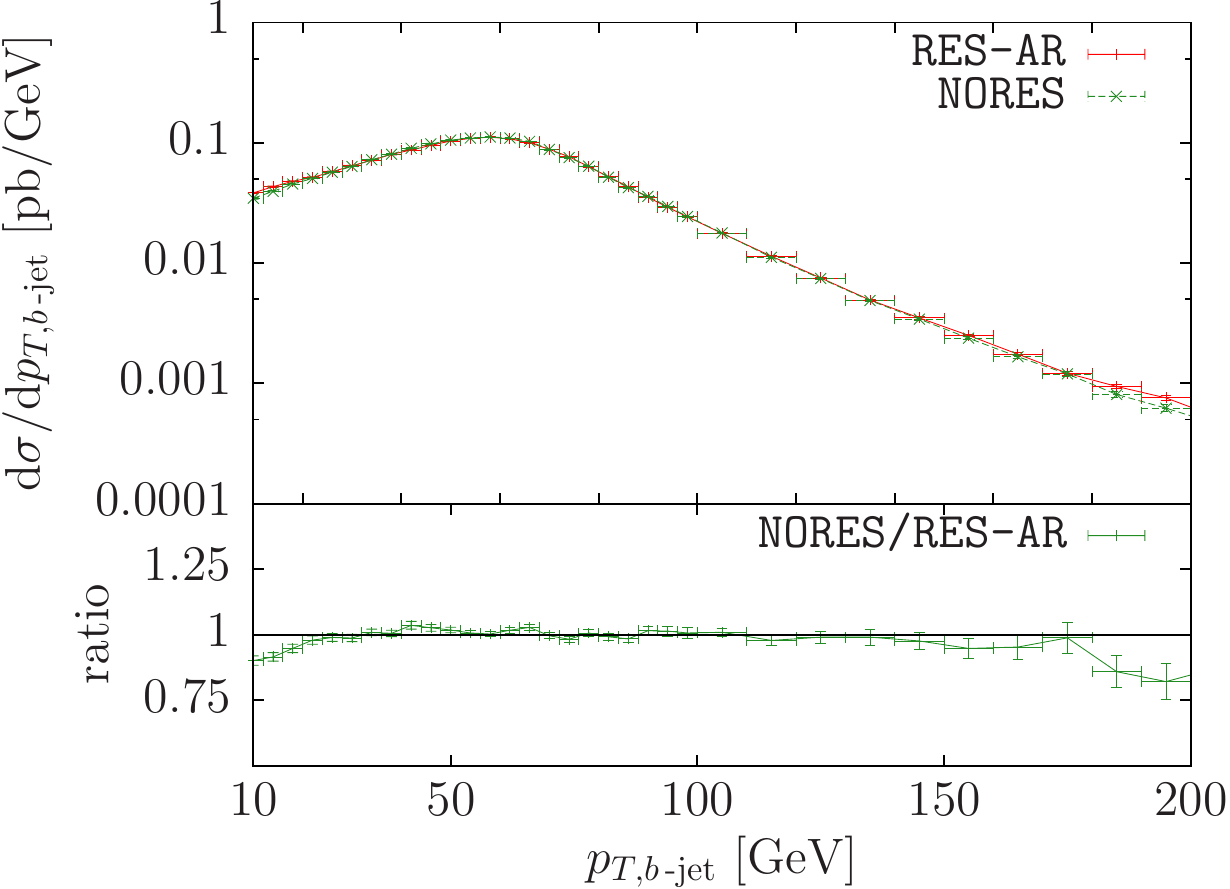}
\caption{\label{fig:bjet-pt-NORES-RES-HR} Transverse momentum of the $b$ jet,
obtained with the {\tt NORES} and the {\tt RES-HR} generators.}
\end{center}
\end{figure}

We begin by
showing results with the \verb!NORES! contribution without performing any
resonance assignment.
In fig.~\ref{fig:Top-pt-NORES-RES-HR}
we plot the transverse momentum and the rapidity distributions of the reconstructed top.
In fig.~\ref{fig:Top-mass-NORES-RES-HR}
we show the mass peak for the reconstructed top.
In fig.~\ref{fig:bjet-NORES-RES-HR}
we plot the $b$-jet mass and profile.
In fig.~\ref{fig:bjet-pt-NORES-RES-HR}
we show the $b$-jet transverse momentum.

We see marked distortions in the mass peak, in the
$b$-jet mass and profile, and in the transverse momentum distribution.
In this case we get 
$M_{\rm trec}=170.54(2)$~GeV for the \verb!NORES! and $M_{\rm trec}=170.06(3)$~GeV for the
\verb!RES-HR! generator.
We remark that in this case no MC-truth was available for the top quark mass in the
\verb!NORES! case, and thus the corresponding plot is missing.

\begin{figure}[t]
\begin{center}
\includegraphics[width=0.48\textwidth]{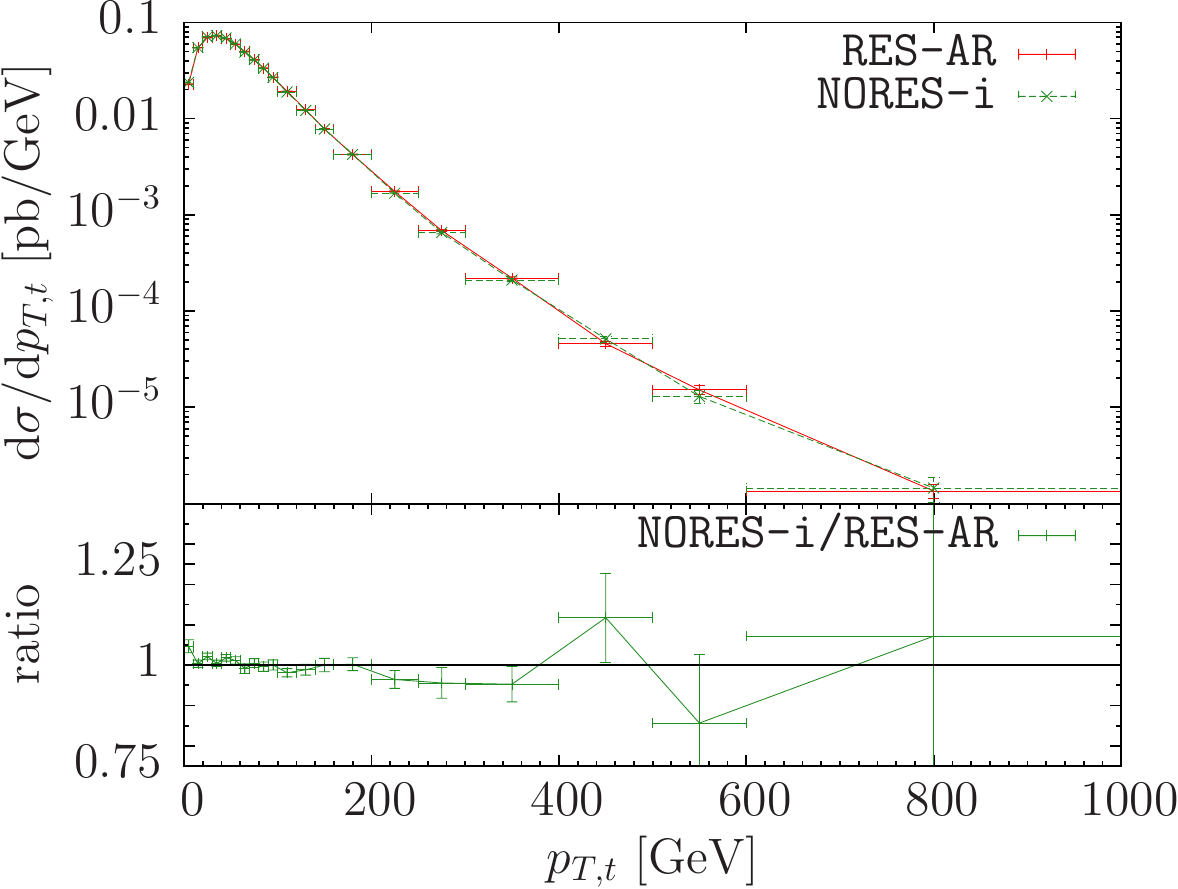}
\includegraphics[width=0.48\textwidth]{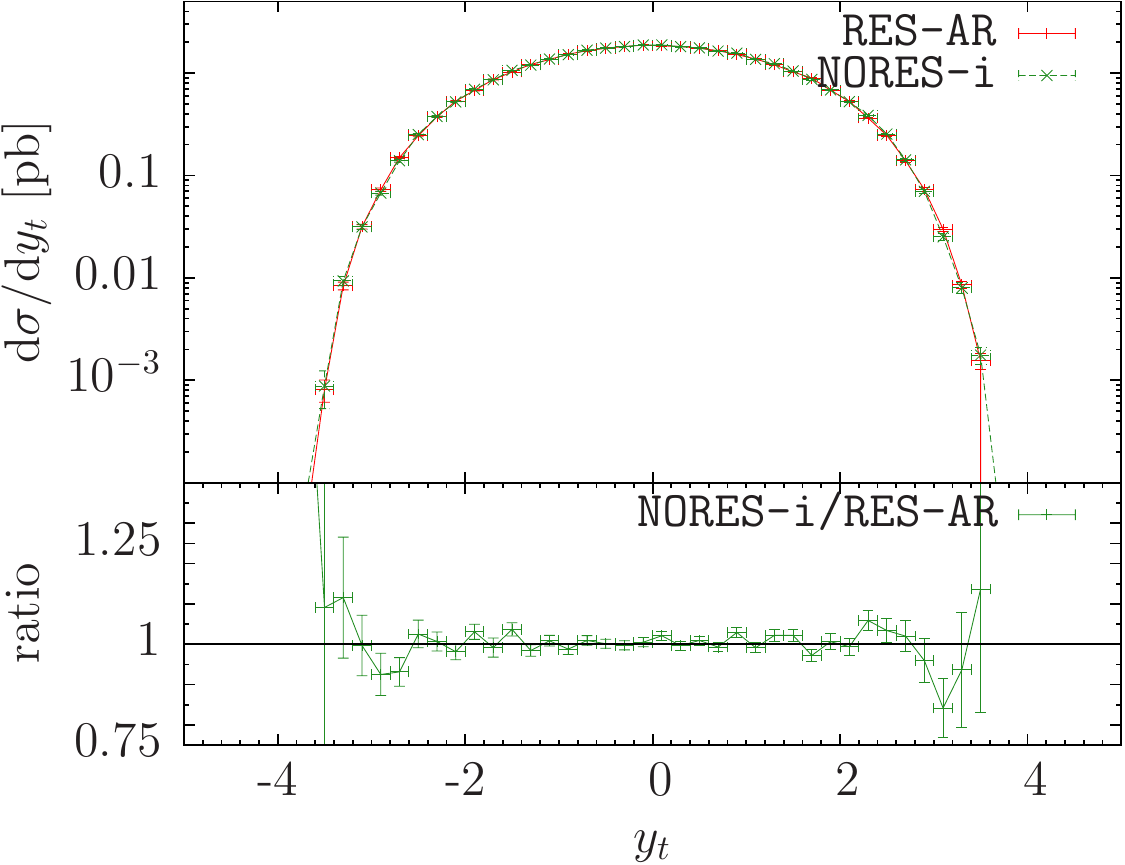}
\caption{\label{fig:Top-pt-NORES-i-RES-HR} Transverse momentum distribution of the
top quark, obtained with the {\tt NORES-i} and the {\tt RES-HR} generators.}
\end{center}
\end{figure}

\begin{figure}[t]
\begin{center}
\includegraphics[width=0.48\textwidth]{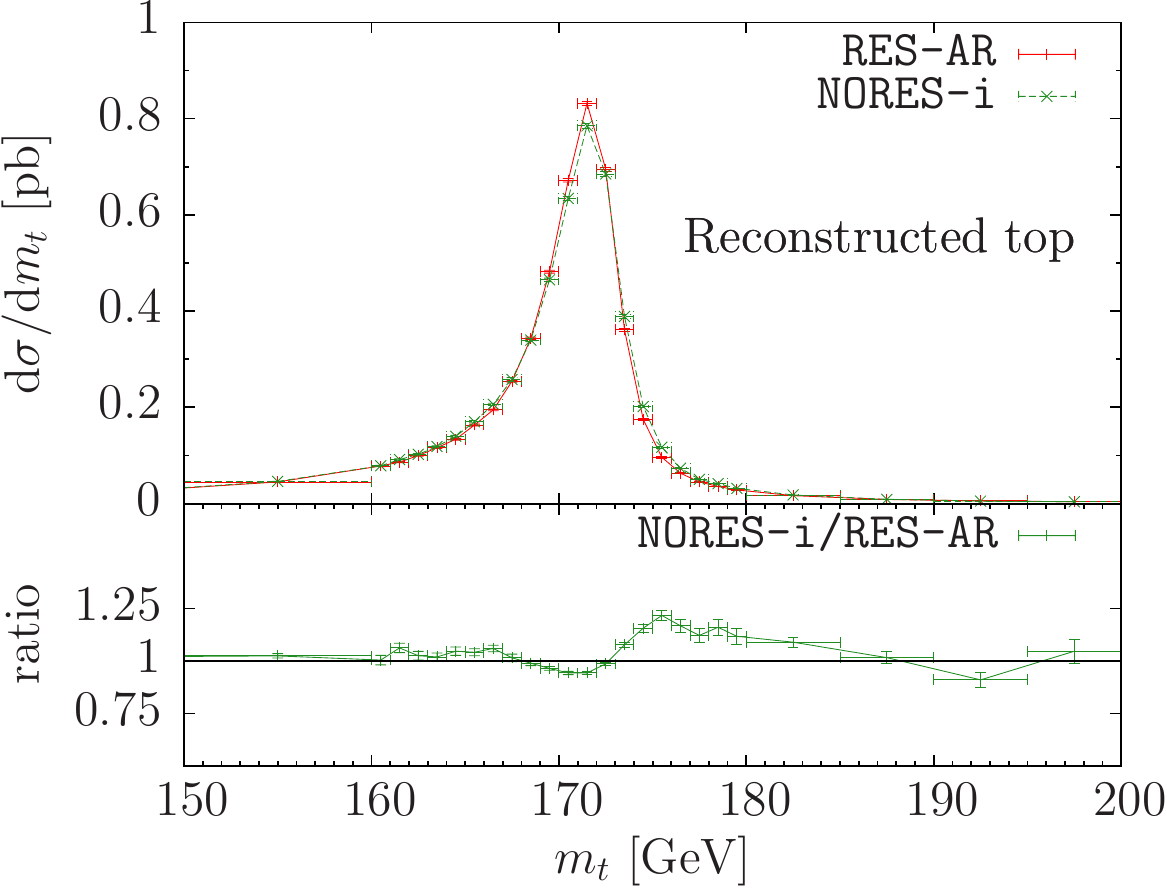}
\includegraphics[width=0.48\textwidth]{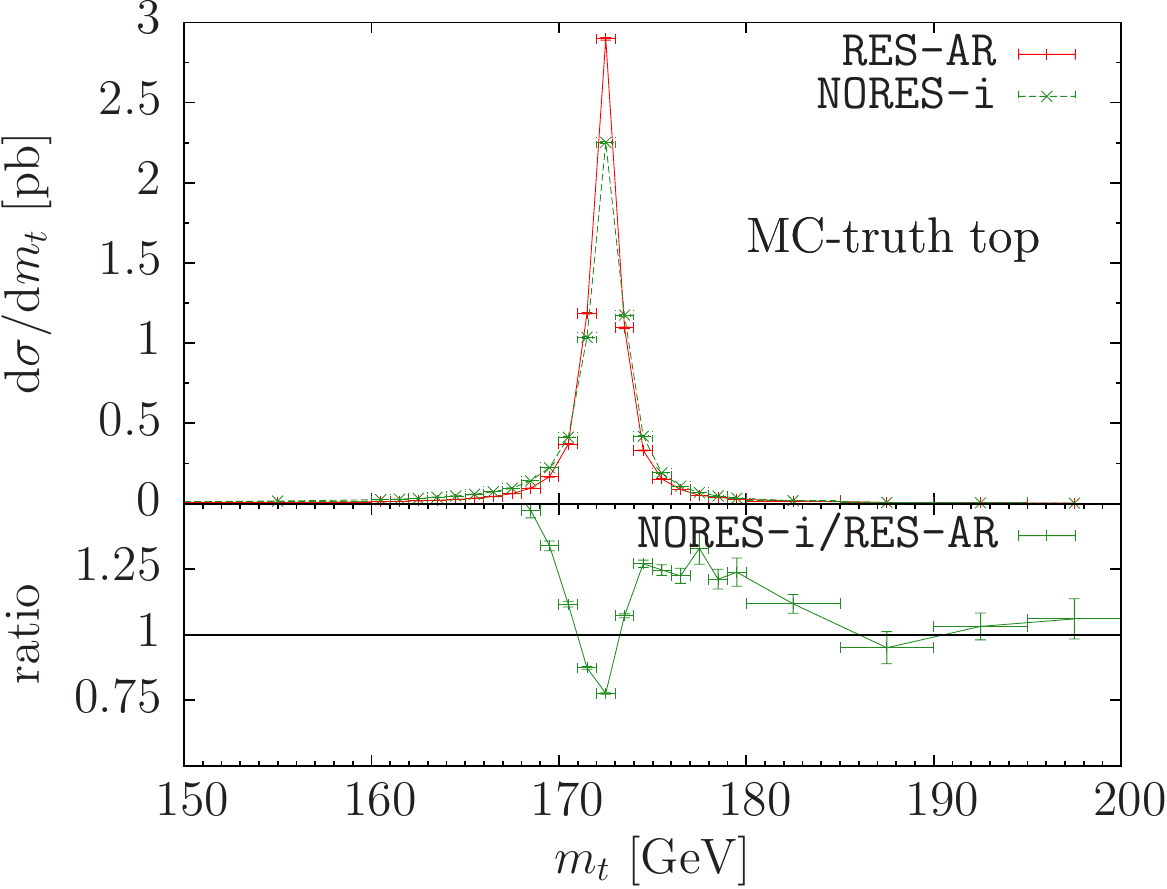}
\caption{\label{fig:Top-mass-NORES-i-RES-HR} Invariant mass of the
top quark, obtained with the {\tt NORES-i} and the {\tt RES-HR} generators, both at the
reconstructed and at the MC-truth level.}
\end{center}
\end{figure}

\begin{figure}[t]
\begin{center}
\includegraphics[width=0.48\textwidth]{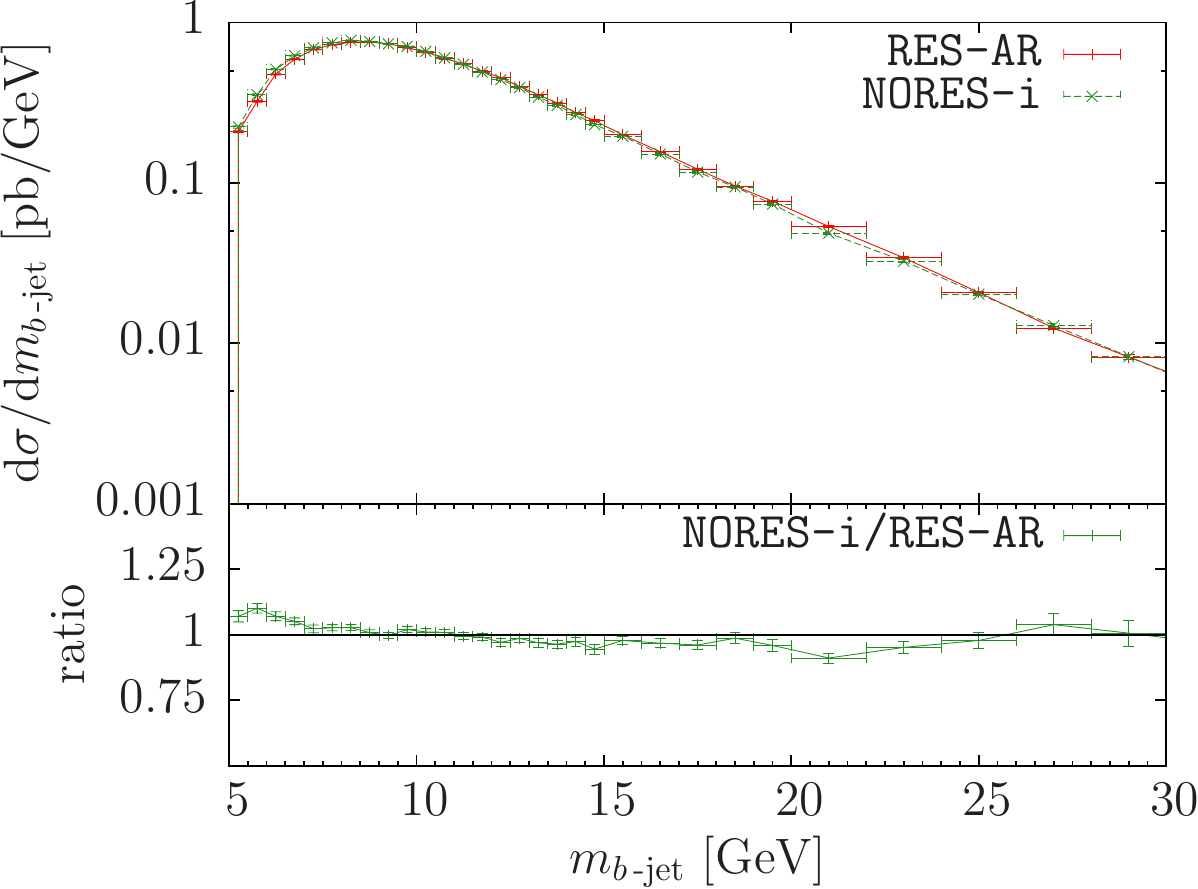}
\includegraphics[width=0.48\textwidth]{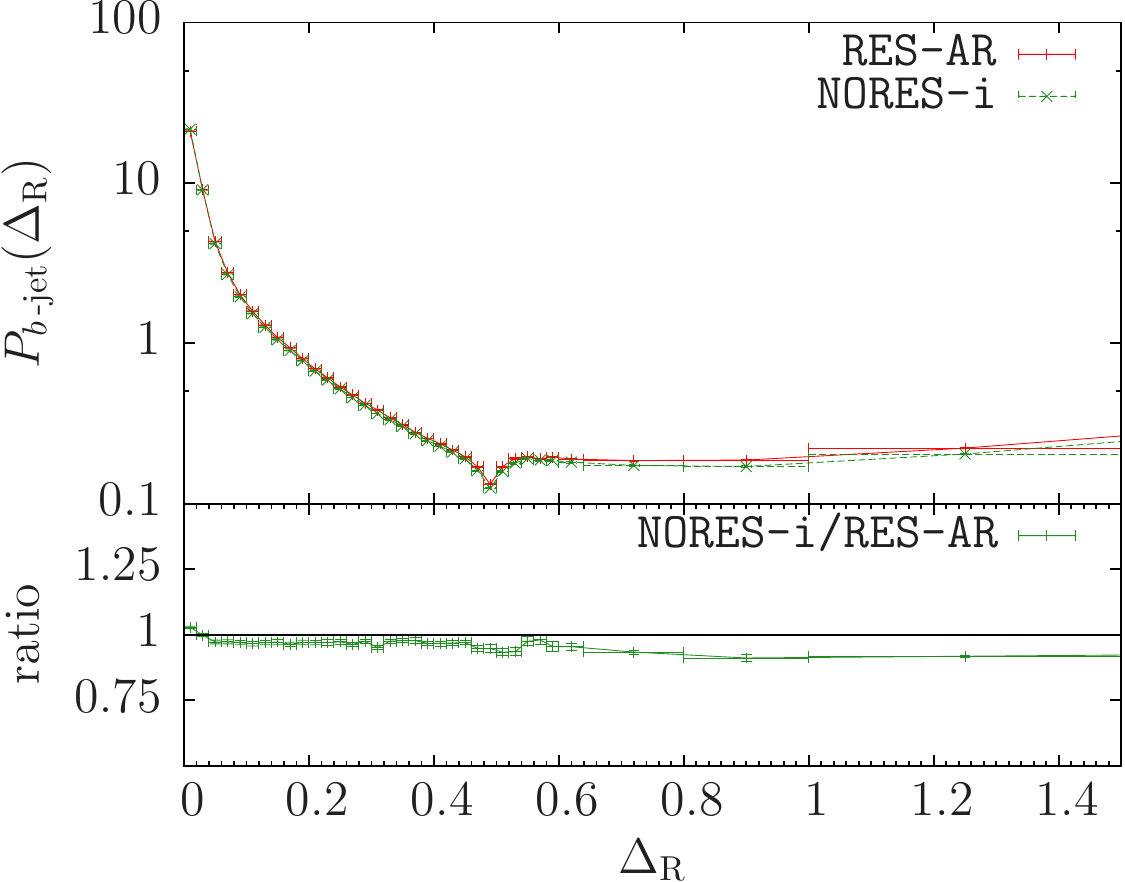}
\caption{\label{fig:bjet-NORES-i-RES-HR} Mass and profile of the $b$ jet,
obtained with the {\tt NORES-i} and the {\tt RES-HR} generators.}
\end{center}
\end{figure}

\begin{figure}[t]
\begin{center}
\includegraphics[width=0.48\textwidth]{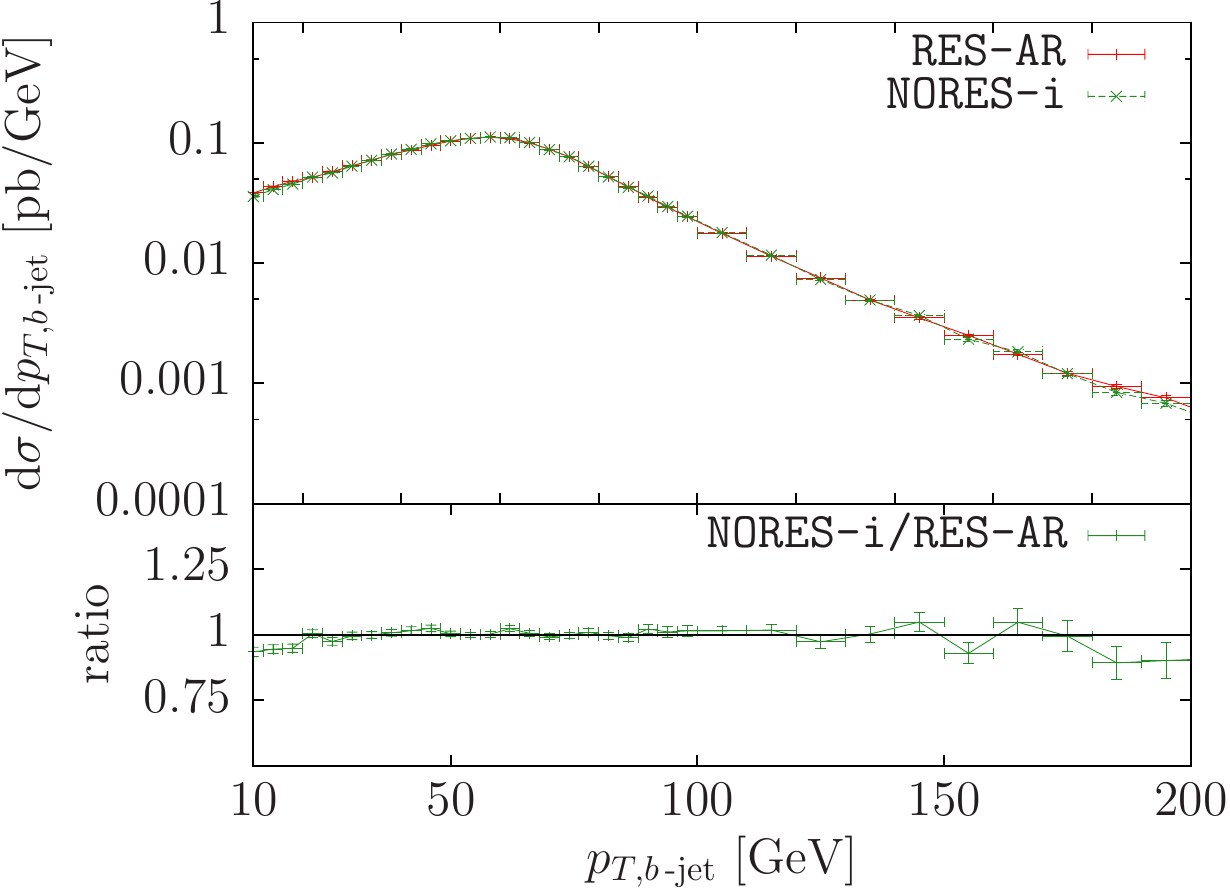}
\caption{\label{fig:bjet-pt-NORES-i-RES-HR} Transverse momentum of the $b$ jet,
obtained with the {\tt NORES-i} and the {\tt RES-HR} generators.}
\end{center}
\end{figure}

Finally, we performed a guess resonance assignment on the \verb!NORES! output record, in the
following way:
\begin{itemize}
\item The $b\, \mu^+\nu_\mu$ system is assigned to the top in all events.
\item If the radiated parton is a gluon, and the $g\, b$ system
      has the colour of a quark, then we compute
      the transverse momentum of the gluon relative to the beam axis,  $k_{T,{\rm isr}}$,
      relative to the final state light quark, $k_{T,{\rm fsr}}$,
      and relative to the $b$ quark in the $g\, b\, \mu^+\nu_\mu$ frame,  $k_{T,b}$.
      Furthermore we compute the quantities
      \begin{eqnarray}
        f_1 &=& \frac{1}{(s_{b\, \mu^+ \nu_\mu}-m_t^2)^2+(\Gamma_t m_t)^2}\,, \\
        f_2 &=& \frac{1}{(s_{g\, b\, \mu^+ \nu_\mu}-m_t^2)^2+(\Gamma_t m_t)^2}\,.
      \end{eqnarray}
      The gluon is assigned or not assigned to the top resonance with probabilities
      proportional to
      \begin{equation}
        \frac{f_2}{k_{T,b}^2}\quad {\rm and} \quad f_1\times \left(\frac{1}{k_{T,{\rm isr}}^2}
         +\frac{1}{k_{T,{\rm fsr}}^2}\right) \,.
      \end{equation}
\item If the radiated parton is not a gluon, it is not assigned to the top.
\end{itemize}
We label as \verb!NORES-i! ({\tt i} for ``improved'') the corresponding generator,
and show its comparison with the \verb!RES-HR! output in figs.~\ref{fig:Top-pt-NORES-i-RES-HR}
through~\ref{fig:bjet-pt-NORES-i-RES-HR}.
We see that now the differences are much less pronounced, and do not
seem to affect significantly the determination of the
top mass. In fact, the average value of our $M_{\rm trec}$ observable
is now $M_{\rm trec}=170.07(2)$~GeV for the \verb!NORES-i! and $M_{\rm trec}=170.06(3)$~GeV for the
\verb!RES-HR! generator. Mild distortions are observed for the remaining distributions.
On the other hand, we see that the MC-truth top line-shape seems to exhibit unphysical
features, presumably due to the way that resonance assignment was performed.

\section{Conclusions}
In this work we have presented a formalism for dealing with intermediate resonances
in NLO+PS generators built within the \verb!POWHEG! framework. We have
formulated a subtraction method such that no double logarithms of the
resonance's width arise separately in the integrated real and in the
soft-virtual term of the NLO calculation. Single logarithms of the widths do however
arise in the soft-virtual term (in fact in the virtual contribution) and in the
integrated real cross section, and cancel only when adding them up.
Thus, in the framework of a \verb!POWHEG! generator, these single log terms cancel
in the $\tilde{B}$ function, so that both double and
single logarithms of the resonance widths are absent there. In \verb!POWHEG!, the
generation of radiation is unitarized by construction. Therefore, all
soft divergences (including also those that are cut-off by the resonance width)
are properly regulated there by Sudakov form factors and/or by finite width effects.

Our formalism is fully general, and has been implemented in a general way in a modified
version of the \verb!POWHEG-BOX-V2!.
The framework is such that in order to implement a specific process, one must supply
the Born, including spin and colour correlated amplitudes, the virtual and the
real amplitudes. The framework takes care of everything else: it finds the possible
resonance histories and singular regions, it builds a Born phase space consistent with the resonance
histories, and it constructs the subtraction terms, the soft-virtual contributions and the
mismatch terms described in sec.~\ref{sec:softcoll}.
It then performs the various stages of the \verb!POWHEG!
event generation.

In the present work we have considered, as a test case,
the process $p p\to \mu^+ \nu_\mu j_b j$, that is dominated by single-top $t$-channel production,
in the 5-flavours scheme. Within this framework, we have examined the output of our generator,
with particular attention to observables that can have an impact on the top mass
measurements.

 We have compared two variants of our generators. In the first one we use
the traditional \verb!POWHEG! method, dubbed \verb!RES-HR! in this paper,
retaining only the hardest radiation, feeding the
corresponding partonic event to a shower Monte Carlo, and vetoing any
shower-generated radiation harder than the \verb!POWHEG! one. In the second one,
dubbed \verb!RES-AR!, 
the hardest radiations in production and in resonance decays are both kept and
combined in the final partonic event. In this case, the veto scales on the hardness
of the Shower radiation in production and in decays are different, being set
to the corresponding scales of radiation in \verb!POWHEG!.
We also compare our generator to the previous single-top, t-channel generator,
the \verb!ST-tch! process of ref.~\cite{Alioli:2009je}
in the \verb!POWHEG-BOX-V2!.
We can briefly summarize our findings as follows. We find differences in the
reconstructed top, mostly due to the
structure of the $b$-jets, and we ascribe these differences to the fact that the
hardest radiation in the $b$-jet is fully determined by the Shower Monte Carlo
in the  \verb!ST-tch! generator, it is in part determined by \verb!POWHEG! and
in part by the Shower Monte Carlo in the \verb!RES-HR! generator, and
it is fully determined by \verb!POWHEG! in the \verb!RES-AR! generator.

We have also considered the output of a generator using the same, full
matrix elements for the $p p\to \mu^+ \nu_\mu j_b j$ process that we
have used in our new generator, implemented however in the traditional
\verb!POWHEG-BOX-V2! framework, that is to say without resonance aware
formalism. In this context we have again considered two alternative
options. In the first one we pass the events to the Shower
Monte Carlo without any resonance information. In the second one we
reconstruct a most probable resonance history of the event based upon kinematics.
The aim of such test is to search for distortions in the generated
radiation due to the lack of proper treatment of resonance decays.
We have found that if no resonance information is passed to the
shower important differences are in fact observed. On the other hand,
if we make an educated guess of the event resonance history, and pass
it to the shower, smaller differences are present at the reconstructed
level, although the top line-shape at the MC-truth level exhibits unphysical
features.  

We remark that one comparison is still missing in the present work: the comparison to
a generator using the on-shell approximation (i.e. a single top generator analogous
to the \verb!ttb_NLO_dec! $t\bar{t}$ generator of ref.~\cite{Campbell:2014kua}), in order to check
if off-shell and non-resonant contributions at the radiation level
are relevant for an accurate simulation of the production of a top
resonance.

The relevant code for the present work is available in the \verb!POWHEG BOX! repository
at the url \url{http://powhegbox.mib.infn.it/POWHEG-BOX-RES-beta}. It has to be
considered as very preliminary at the present stage, since only one relatively simple
process has been implemented with it so far.

\section*{Acknowledgments}
We wish to thank Kirill Melnikov, Roberto Tenchini and Giulia Zanderighi for useful discussions.
We also wish to thank Carlo Oleari for critically examining parts of the paper.



\providecommand{\href}[2]{#2}\begingroup\raggedright\endgroup

\end{document}